\documentclass[useAMS,fleqn,usenatbib]{mnras}

\usepackage{newtxtext,newtxmath}

\usepackage[T1]{fontenc}
\usepackage{ae,aecompl}

\usepackage{graphicx}	
\usepackage{amsmath}	
\usepackage{amssymb}	
\usepackage{etoolbox}
\makeatletter
\patchcmd\@combinedblfloats{\box\@outputbox}{\unvbox\@outputbox}{}{%
	\errmessage{\noexpand\@combinedblfloats could not be patched}%
}%
\makeatother
\usepackage{color} 
\usepackage[usenames,dvipsnames]{xcolor}
\usepackage[normalem]{ulem} 
\usepackage{hyperref} 
\usepackage{wasysym}

\def\Msun{\hbox{$\rm\, M_{\odot}$}}
\def\Mstar{\hbox{$\rm\, M_{\star}$}}
\def\Htwo{{\mathrm{H}_2}}
\newcommand{\twoc}{_{\textrm{200c}}}
\newcommand{\lgal}{\textsc{L-galaxies}}
\newcommand{\eg}[0]{$\textnormal{e.g. }$}
\newcommand{\ie}[0]{$\textnormal{i.e. }$}

\newcommand{\Zsun}[0]{\,\textnormal{Z}_\odot}

\newcommand{\sub}[1]{_{\textnormal{#1}}}

\definecolor{newgreen}{rgb}{0.20, 0.75, 0.20}

\def\mnras{{MNRAS}}

\title [\lgal~2020] {L-GALAXIES 2020: Spatially resolved cold gas phases, star formation and chemical enrichment in galactic discs}

\author[Bruno M. B. Henriques et al.]  
{Bruno M. B. Henriques$^{1}$\thanks{E-mail:brunohenriques83@gmail.com},
Robert M. Yates$^{2}$, Jian Fu$^{3}$, Qi Guo$^{4}$,
\newauthor
Guinevere Kauffmann$^{2}$, Chaichalit Srisawat$^{5}$, Peter A. Thomas$^{6}$, Simon D. M. White$^{2}$ 
\vspace{0.4cm}\\
 {}$^{1}$Institute for Astronomy, ETH Zurich, CH-8093 Zurich, Switzerland\\
  {}$^{2}$Max-Planck-Institut f\"ur Astrophysik, Karl-Schwarzschild-Str. 1, D-85741 Garching b. M\"unchen, Germany\\ 
    {}$^{3}$Key Laboratory for Research in Galaxies and Cosmology, Shanghai Astronomical Observatory, Chinese Academy of Sciences, \\
  80 Nandan Road., Shanghai, 200030, China\\
   {}$^{4}$Partner Group of the Max-Planck-Institut f\"ur Astrophysik, National Astronomical Observatories, Chinese Academy of Sciences, \\
             ~Beijing, 100012, China\\
{}$^{5}$Center for Astrophysics and Cosmology, Science Institute, University of Iceland, Dunhagi 5, 107 Reykjavik, Iceland\\
  {}$^{6}$Astronomy Centre, University of Sussex, Falmer, Brighton BN1 9QH, UK\\}
 
\begin{document}

\date{Accepted February 2020}

\volume{491}
\pagerange{5795--5814} \pubyear{2020}

\maketitle

\label{firstpage}

\begin{abstract}
  We have updated the Munich galaxy formation model, \lgal, to follow the radial distributions of stars and atomic and molecular gas in galaxy discs. We include an H$_2$-based star-formation law, as well as a detailed chemical-enrichment model with explicit mass-dependent delay times for SN-II, SN-Ia and AGB stars. Information about the star formation, feedback and chemical-enrichment histories of discs is stored in 12 concentric rings.  The new model retains the success of its predecessor in reproducing the observed evolution of the galaxy population, in particular, stellar mass functions and passive fractions over the redshift range $0\leq z \leq 3$ and mass range $8\leq \log (M_*/\Msun)\leq 12$, the black hole-bulge mass relation at $z=0$, galaxy morphology as a function of stellar mass and the mass-metallicity relations of both stellar and gas components. In addition, its detailed modelling of the radial structure of discs allows qualitatively new comparisons with observation, most notably with the relative sizes and masses of the stellar, atomic and molecular components in discs. Good agreement is found with recent data. Comparison of results obtained for simulations differing in mass resolution by more than two orders of magnitude shows that all important distributions are numerically well converged even for this more detailed model. An examination of metallicity and surface-density gradients in the stars and gas indicates that our new model, with star formation, chemical enrichment and feedback calculated self-consistently on local disc scales, reproduces some but not all of the trends seen in recent many-galaxy IFU surveys.
  
\end{abstract}


\section{Introduction}
\label{sec:intro}

Rapid developments in telescope/detector technology and in computational power in recent decades have resulted in major advances in our understanding of galaxy formation from both theoretical and observational perspectives. Observations have been  able to fully characterize the stellar mass \citep[e.g.][]{Bell2003, Bundy2005, Faber2007, Ilbert2013, Muzzin2013} and star formation rate \citep[e.g.][]{Lilly1996, Madau1996, Madau2014} content of the Universe across most of cosmic history, while simulations have developed fair representations of the real universe in cosmological boxes while resolving galaxies smaller than the Small Magellanic Cloud (SMC) \citep[e.g.][]{Henriques2015, Vogelsberger2014, Schaye2015}.

Despite these successes, most observational galaxy formation work has, until recently, been focused on statistical studies of the distributions of global galaxy properties with little attention paid to how material is distributed within galaxies. However, this is changing swiftly with the advent of large surveys using multi-object integral field unit (IFU) spectrographs (\citealt{Croom2012}: SAMI, \citealt{Sanchez2012}: CALIFA, \citealt{Bundy2015}: MaNGA) and  radio interferometer maps of galaxies with exquisite spatial resolution (e.g., ALMA). In order to provide theoretical models that can help interpret the results of these surveys, both hydrodynamical simulations and semi-analytic models of galaxy formation must extend their capabilities. In particular, they must accurately track the spatial distributions of stars, of ionized, atomic and molecular gas, of star formation and of heavy elements in galactic discs. Simultaneously, such models must remain consistent with the overall time evolution of the distribution of global galaxy properties.

These aspects of galaxy evolution will automatically be addressed by cosmological hydrodynamical simulations as their resolution and the realism of their subgrid modelling are improved \citep[see e.g.][for recent advances]{Schaye2015, Dave2016, Dubois2016, Pillepich2018} but the computational expense of such simulations makes it difficult to treat large volumes at high resolution or to survey the parameter space of their subgrid models. The dramatic speed-up afforded by semi-analytic simulations solves these last two problems at the expense of a much more schematic representation of hydrodynamic processes and galaxy structure. Some aspects of the internal structure of galaxies beyond their size, mass and bulge-to-disc ratio have been treated in previous semi-analytic models (SAMs). \cite{Fu2010, Fu2012, Fu2013} tracked the properties of galaxy discs in a set of nested rings, enabling improved treatments of the different cold gas phases and of star formation. They also tested a variety of recipes for the conversion of atomic into molecular hydrogen and of molecular hydrogen into stars. Most of this work was based on the \citet{Guo2011} version of \lgal. A similar approach has been adopted more recently by \citet{Stevens2016} in the context of the SAGE model. 

 \citet{Lagos2011a, Lagos2011b},  \citet{Popping2014} and \citet{Somerville2015} introduced recipes to follow different cold-gas phases and tested $\Htwo$-dependent star-formation laws for the GALFORM and Santa Cruz models, respectively. These studies were single-zone treatments based on a single-size estimate and an assumed  density-profile shape for each galaxy. \citet{Martindale2017} post-processed  the \citet{Henriques2015} version of \lgal\ and showed that model results for the low-mass end of the stellar and HI mass functions could be better reconciled with observations by such more elaborate treatments of the gas phases (something that had previously been found to be problematic \citealt{Lu2012, Henriques2015}).
  
Another important ingredient for realistic galaxy formation models is a detailed treatment of chemical enrichment. This needs to include the ability to follow the formation and evolution of the stars that contribute significantly to the (metallicity-dependent) release of mass, energy and heavy elements through various enrichment channels, primarily, core collapse supernovae (SNII), supernovae of type Ia (SNIa) and asymptotic giant branch (AGB) stars. Such a model was introduced into the \citet{Guo2011} version of \lgal\, by \citet{Yates2013} and into the GAEA model by \citet{DeLucia2014}. 

Incorporating a full chemical-enrichment model with the ability to track different gas phases in spatially-resolved rings represents a significant technical upgrade to standard SAMs, because the formation of each galaxy has to be resolved both temporally and spatially. This introduces a new level of complexity, but the expectation is that current and future observational constraints will require the expansion in the number of degrees of freedom and strongly constrain the associated parameters. The problem of properly sampling the high-dimensional parameter space can be tackled using robust statistical techniques initially introduced in this field by \citet{Kampakoglou2008} and \citet{Henriques2009}. These have since been extended to include a wide range of models and sampling methods \citep{Benson2010, Bower2010, Henriques2009, Lu2011b, Lu2012, Henriques2013, Mutch2013, Benson2014, Ruiz2015}. Such methods are also a crucial tool for the work presented in this paper. 

Here we present a new version of the Munich semi-analytic model of galaxy evolution, \lgal, that includes a variety of new modelling features while retaining past successes. In particular, it matches equally well to the observed evolution of galaxy abundance as a function of a wide range of global properties (stellar mass, luminosity, colour, morphology, size, star formation rate (SFR), metallicity, central black hole mass) as well as to the scaling relations between properties. Applying the new model simultaneously to two large simulations of very different mass resolution enables tests against observation over a very wide dynamic range (e.g. four orders of magnitude in stellar mass). The new features follow the structure of galaxy discs in much more detail than before, allowing for comparison with the observed masses, sizes and structure of the atomic, molecular and stellar components of discs, as well as to their observed metallicities and metallicity gradients.

In Section 2 we outline the extensions we have made to the model of
\cite{Henriques2015}. In section 3 we then show that the new model can fit the observations of abundances and passive fractions that were used to constrain parameters in that paper to about the same level of accuracy as the original model. In addition we show that the new model also fits new observational constraints for the local content of cold gas in different phases. Section 4 discusses how the more detailed treatment of galaxy structure affects model results for global galaxy properties other than those used to calibrate the model. Section 5 then concentrates on model results for the radial structure of discs, comparing with results from recent Integral Field Unit (IFU) surveys of galaxy structure. These represent the main breakthrough of the present work. Finally, Section 6 presents a summary of our results and some conclusions.

\section{Updates to galaxy formation modelling}
\label{sec:munich_model}

In this section we describe how the treatment of astrophysical processes within \lgal\ has changed since the last version of the model with publicly-released catalogues \citep{Henriques2015}. As with that version, a full and updated description of the model is presented in the supplementary material of the journal submission and is also available online on our model webpage\footnote{https://lgalaxiespublicrelease.github.io/}. Detailed predictions for this new model are made publicly available in our online database\footnote{Catalogs for the Munich models of galaxy formation can be found at http://www.mpa-garching.mpg.de/millennium.}. The same MCMC method is used to fully explore the parameter space. In addition to the stellar mass function and red fractions as a function of stellar mass at $z=0$ and 2, we also include the HI mass function at $z=0$ as a constraint. The parameter choices for our best-fit model are shown in table~\ref{table:parameters}. A detailed comparison with values from previous versions of our model, as well as marginalized 1D-likelihood distributions from the MCMC sampling, are presented in the supplementary material.

The model is self-consistently coupled to the evolution of dark-matter sub-haloes identified in dark-matter-only N-body simulations. As in \citet{Henriques2015} we use the Millennium \citep{Springel2005} and Millennium-II \citep{BoylanKolchin2008} simulations scaled to the {\it Planck} cosmology ($\sigma_8=0.829$,
$H_0=67.3\,\rm{km}\,\rm{s}^{-1}\rm{Mpc}^{-1}$, $\Omega_{\Lambda}=0.685$, $\Omega_{\rm{m}}=0.315$,
$\Omega_{\rm{b}}=0.0487$, $f_{\rm{b}}=0.155$ and $n=0.96$) following the \citet{Angulo2015} procedure.  After rescaling, the Millennium simulation box has a side-length corresponding to 714\,Mpc and a particle mass of $1.43 \times 10^9 \Msun$. The Millennium-II follows a region one fifth the linear size of the Millennium simulation, resulting in 125 times better mass resolution. Throughout the paper, the Millennium simulation is used for galaxies with $\log_{10}(M_*/\Msun)>9$ and $\log_{10}(M_{\mathrm{HI}}/\Msun)>9$ and the Millennium-II simulation for galaxies with lower masses. As in previous models, the distributions of most galaxy properties agree well in the two simulations over the mass range $9.0 < \log_{10}(M_*/\Msun) < 11.0$ where they are well determined in both\footnote{The exception are properties that strongly depend on the merger history of a galaxy, e.g. galaxy sizes and morphologies shown in Figs.~\ref{fig:sizes} and \ref{fig:morphology} for which we base our analysis mostly on Millennium-II.}.

Baryonic components are followed using a set of coupled differential equations. Primordial gas falls with the dark matter onto sufficiently massive haloes, where it is shock-heated. The efficiency of radiative cooling then determines whether it is added directly to the cold gas of the central galaxy, or resides for a while in a hot-gas atmosphere. In our new model, the properties of cold interstellar gas are followed in concentric rings where cold gas is partitioned into HI and $\Htwo$ and the latter is converted into stars, both quiescently and in merger-induced starbursts that also drive the growth of central supermassive black holes. Stellar evolution is tracked independently in each ring and not only determines the photometric appearance of the final galaxy, but also heats and enriches its gas components, in many cases driving material into the wind reservoir, from where it may later fall back into the galaxy. Accretion of hot gas onto central black holes gives rise to radio-mode feedback, regulating condensation of hot gas onto the galaxy. Environmental processes affect the gas components of galaxies (ram-pressure might strip the hot component while tidal stripping affects hot gas, cold gas and stars), as well as the partition of stars between discs, bulges and the intracluster light.

\begin{table}
\begin{center}
  \caption{Parameter choices for our best fit model, determined by MCMC sampling of the parameter space, using the evolution of the stellar mass function (Fig.~\ref{fig:SMF_evo}, at $z=0$ and $z=2$), the evolution of the red fraction as a function of stellar mass (Fig.~\ref{fig:redfraction_colorcut}, at $z=0$ and $z=2$) and the HI mass function (Fig.~\ref{fig:GMF}, at $z=0$) as observational constraints. The bottom 4 parameters are not well constrained by these observations and were fixed a priori. A full description of these parameters and of the equations that define our galaxy formation model is presented in the supplementary material of the journal submission and in our model webpage.}
\label{table:parameters}
\begin{tabular}{lccc}
\hline
\hline
&Parameter &value &units \\
\hline
&$\alpha_{\rm{SF}}$ &0.06 & \\
&$\alpha_{\rm{SF, burst}}$  &0.5 & \\
&$\beta_{\rm{SF, burst}}$ &0.38 & \\
\hline
&$k_{\rm{AGN}}$ &$2.5\times10^{-3}$ &$[\Msun\,\rm{yr^{-1}}]$ \\
&$f_{\rm{BH}}$ &0.066 & \\
&$V_{\rm{BH}}$ &700 &$[\rm{km\,s}^{-1}]$\\
\hline
&$\epsilon_{\rm{reheat}}$ &5.6 & \\
&$V_{\rm{reheat}}$ &110 &$[\rm{km\,s}^{-1}]$ \\ 
&$\beta_{\rm{reheat}}$ &2.9 & \\
&$\eta_{\rm{eject}}$ &5.5 & \\
&$V_{\rm{eject}}$ &220 &$[\rm{km\,s}^{-1}]$ \\
&$\beta_{\rm{eject}}$ &2.0 & \\
&$\gamma_{\rm{reinc}}$ &$1.2\times10^{10}$ &$[\rm{yr}^{-1}]$ \\ 
\hline
&$M_{\rm{r.p.}}$ &$5.1\times10^{4}$ &$[10^{10}\Msun]$\\
&$\alpha_{\rm{dyn. fric.}}$ &1.8 & \\
\hline
\hline
& &Fixed &\\
\hline
&$v_{\rm{inflow}}$ &1.0 &$[\rm{km\,s^{-1}\,kpc^{-1}}]$ \\
&$R_{\rm{merger}}$ &0.1 & \\
&$f_{\rm{z,hot,TypeII}}$ &0.3 & \\
&$f_{\rm{z,hot,TypeIa}}$ &0.3 & \\
\hline
\hline
 \end{tabular}
\end{center}
\end{table}

\subsection{Resolved properties of galaxies}
\label{sec:model_rings}

Semi-analytic models traditionally aim to follow the key baryonic components of galaxies in a global manner. i.e.  a single value describes the amount of cold gas in the interstellar medium (ISM), or stars in a disc, or the total amount of hot gas in a halo, with no attempt to model the spatial distribution of material inside these components. Consequently, there is a relatively simple connection between dark matter and  galaxy baryonic properties.

Despite introducing a significant additional layer of complexity, there are advantages to following the spatial distribution of material within these components. One clear benefit is the possibility of making direct comparisons with kpc-scale observations of properties such as stellar or gas surface density, SFR or metallicity within nearby galaxies. Such comparisons are becoming viable now that dedicated surveys with modern multi-object IFU spectrographs have greatly increased the number of galaxies with spatially resolved data \citep{Bacon2010, Croom2012, Sanchez2012, Bundy2015}.

The ability to track galactic discs spatially is also critical for modelling the transition from atomic to molecular gas. Molecular-gas formation is believed to depend on gas density and to occur predominantly in the densest regions near the centres of galaxies. In order to follow the formation of $\Htwo$ correctly, it is necessary to track the surface density of cold gas inside galaxies. This in turn allows for the possibility of a spatially resolved model for star formation based on $\Htwo$ rather than total ISM gas. 

In the present work, we limit our spatial tracking  to the stellar and gas discs of galaxies. Following \citet{Fu2013}, this is done by dividing the gas and stellar discs into a series of concentric annuli or `rings', with outer edges given by:  
\begin{equation} \label{eq:rings}
 r_i = 0.01 \times 2.0^{i} \,h^{-1} \rm{kpc}\, (i = 1, 2...12).
\end{equation}
We are therefore able to follow the accretion of cooling gas, the transition of atomic to molecular gas, the conversion of molecular gas into stars, and the release of energy and chemical elements in individual rings within each model galaxy. The radius of the inner rings is slightly smaller than that adopted in \citet{Fu2013} in order to resolve the properties of dwarf galaxies. The fact that the same ring structure is used for all galaxies allows us to easily add the content in each ring for different components whenever two galaxies merge (though this is, of course, an oversimplified version of what actually happens in a galaxy merger).

\subsection{Cooling, gas inflows, \texorpdfstring{H$_2$}{H2} formation and \texorpdfstring{H$_2$}{H2}-based star formation}
\label{sec:model_H2}
\subsubsection{Gas cooling}
\label{sec:cooling}

The first step in the modelling of resolved disc properties is the choice of the surface-density profile of newly-added material that cools from the hot halo. We assume that newly-accreted cold gas follows an exponential profile (with a uniform metallicity equal to that of the hot gas):

\begin{equation} \label{eq:profile}
\Sigma_{\rm{gas}} (r) = \Sigma^0_{\rm{gas}} \exp (-r/r_{\rm{infall}}),
\end{equation}
where:
\begin{equation} \label{eq:r_infall_1}
\Sigma^0_{\rm{gas}}=\frac{m_{\rm{cool}}}{2 \pi r^2_{\rm{infall}}}.
\end{equation}
Assuming angular momentum is conserved during gas cooling and infall \citep{Mo1998}:
\begin{equation} \label{eq:r_infall_2}
 r_{\rm{infall}} =  \frac{j_\mathrm{halo}}{2V_\mathrm{c}},
\end{equation}
where $j_\mathrm{halo}$ is the specific angular momentum of the halo and $V_\mathrm{c}$ is its circular speed.\footnote{For an isothermal sphere, $V_\mathrm{c}$ is independent of radius; we set it equal to the maximum circular velocity of the dark matter halo.}
Unlike in \citet{Guo2011} and \citet{Henriques2015}, we assume that the angular momenta of infalling material is aligned with that already in the disc. This, combined with having a ring structure with fixed sizes, allows us to directly add new material into pre-existing discs without having to shuffle material during accretion. The scale length of the accreted gas is determined from the spin parameter of the halo of the central galaxy and added to the existing disc. Accretion of cold gas from satellite galaxies during mergers is treated as any other gas accretion event. We note here that, if we did not assume that angular momentum is aligned whenever gas components merge, the characteristic sizes of different components in galactic discs would be significantly reduced. However, in the case of our model, that would not be sufficient to replenish the innermost gas consumed by star formation. This requires large radial inflows that we describe in the following subsection. Once these are included they also largely determine global discs sizes.

For each cooling episode, the newly-accreted gas is directly superimposed onto the pre-existing gas profile. Since the disc and halo sizes are smaller at high redshift, so is the scale length of the infalling gas. As a result, the radial extent of the infalling material is larger at later times, causing the disc to grow (see figure 1 in \citealt{Fu2010}). This naturally leads to an inside-out growth of discs, as is incorporated in many disc formation models \citep[e.g.][]{Kauffmann1996b, Dalcanton1997, Avila-Reese1998, Dutton2009, Fu2009, Pilkington2012}.

\subsubsection{Gas inflow}
\label{sec:inflow}

Whenever gas cools onto the disc we assume that it retains its angular-momentum and that there is no subsequent angular momentum loss during mergers or disc instabilities. In reality, both the differences in angular momentum between disc and infalling material \citep{Lynden-Bell1974}, the transfer of angular momentum between the disc and the dark-matter halo and galaxy interactions can lead to the radial inflow of material towards the centre of galactic discs. In addition, the gravitational interaction between gas in the disc and non-axisymmetric stellar structures such as bars and spirals can also drive the radial inflow of gas in discs \citep{Kalnajs1972}. Simple physical considerations yield estimates of flow velocities ranging from 0.1 to a few km/s \citep{Lacey1985, Bertin1996}. Observationally it is particularly challenging to measure radial-flow rates due to the irregularity of the flows in individual galaxies and the fact that discs are frequently not axisymmetric. As a result these have traditionally been modelled using simple parameterized inflow prescriptions \citep[e.g.][]{Lacey1985, Portinari2000, Schonrich2009, Spitoni2011}.

\citet{Fu2013} tested the simplest prescription in which the radial inflow velocity of gas is constant for the whole disk (model A in \citealt{Lacey1985}), but concluded that moving enough gas from the outer to the inner disc would lead to the excessive pile-up of gas in the very inner disc. Instead, the authors found that a prescription in which the rate of change of the angular momentum is proportional to the angular momentum yields results that agree best with observational data. Here, we adopt the same prescription, as described below. 

\begin{equation} \label{eq:am_1}
\frac{dL_{\rm{gas}}}{dt}\propto L_{\rm{gas}},
\end{equation}
which leads to the velocity scaling with distance from the centre:
\begin{equation} \label{eq:am_2}
v_{\rm{inflow}}=\alpha_v r=\frac{r}{t_v},
\end{equation}
assuming that $L_{\rm{gas}} = m_{\rm{gas}}r_{\rm{gas}}v_{\rm{cir}}$. 

We do not include $\alpha_v$ in the MCMC sampling of the parameter space since we do not include observations of spatially resolved-properties as constraints. Nevertheless, we adopt a larger value than in \citet{Fu2013} ($\alpha_v=1.0$ instead of 0.7 $\mathrm{km}\,\mathrm{s}^{-1}\mathrm{kpc}^{-1}$, corresponding to $t_v\approx1.1$ instead of 1.4\,Gyr) in order to reach a compromise between having extended enough gas profiles for Milky-Way like galaxies and ensuring that gas in the outer rings of very massive galaxies flows to the centre and forms stars within a few Gyrs after cooling has stopped. This is necessary to reproduce the predominantly red colours of these systems at $z=0$.

\begin{figure*}
\centering
\includegraphics[width=17cm]{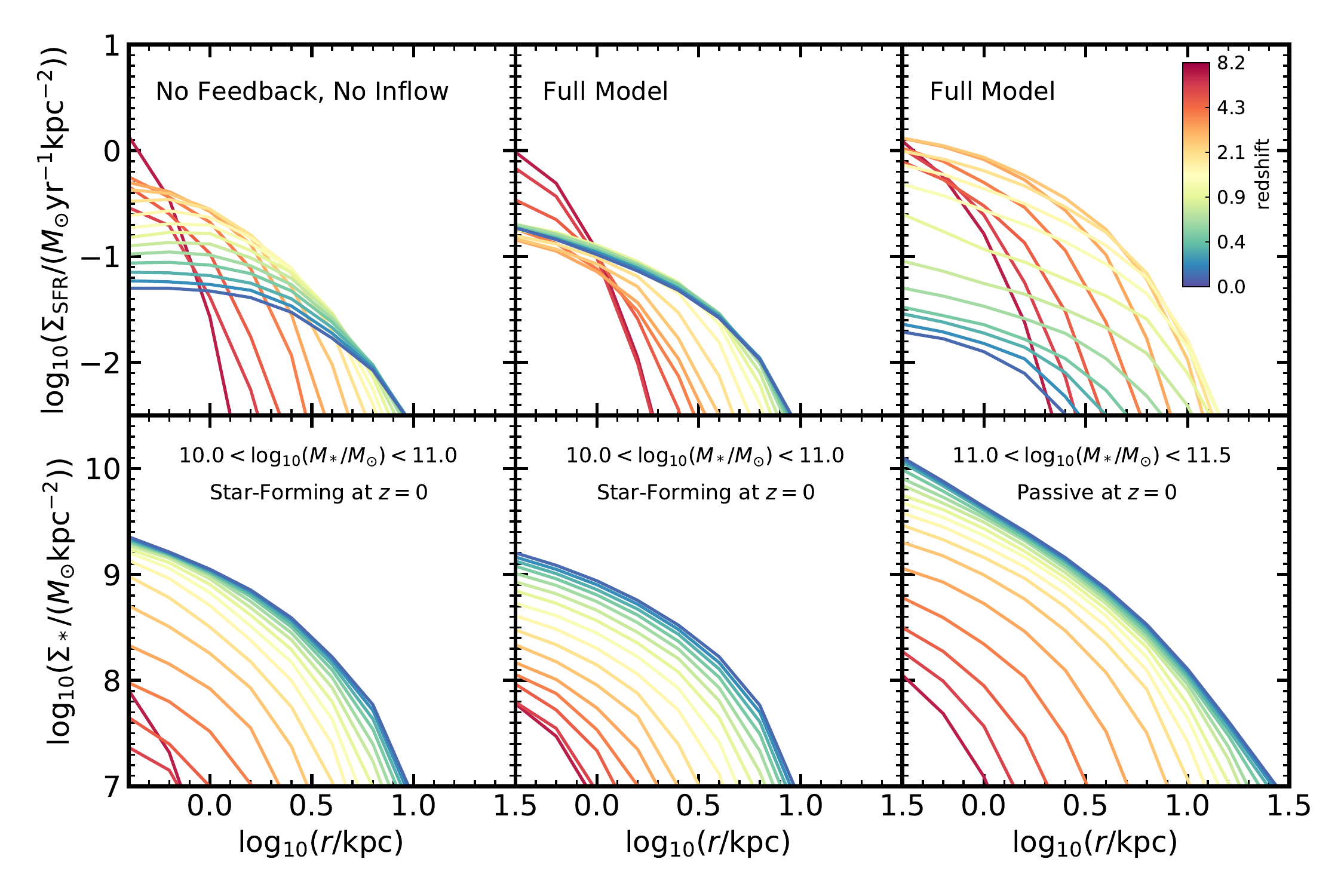}
\caption{Visualization of the build-up of the radial properties of discs in our new scheme. Top and bottom panels show the evolution of the mean radial profiles of star formation surface density and of stellar mass, respectively, for the progenitors of galaxies selected by stellar mass at $z=0$. The left column shows the evolution of profiles for the progenitors of star-forming, intermediate mass galaxies for a simplified model without SN and AGN feedback and without radial inflow of cold gas. The middle and right columns show the evolution of profiles for galaxies in the full model for two distinct bins in mass and star formation rate. The middle panels represent the evolution of star-forming, intermediate mass galaxies, while the right panels represent the evolution of passive, massive galaxies.}
\label{fig:gradients_evo}
\end{figure*}

As possible alternatives to this radial flow formalism, we have considered dropping the assumption that angular momentum is aligned when gas components merge, or assuming that as much as 70\% of the angular momentum can be lost during cooling, but we find that neither can replace our new prescription for the radial inflow of gas. The former moves insufficient material towards galactic centres, while the latter moves predominantly metal-poor gas that is inefficient at forming molecular gas and stars. A further alternative, assuming that strong disc instabilities can transfer large amounts of gas inwards in galactic discs, could mimic the effect of our new radial-inflow model. This approach was indeed adopted by \citet{Stevens2016}, resulting in a model that successfully reproduced the gas profiles of nearby galaxies.

The build-up of radially-resolved disc quantities in our new model, as well as the impact of radial inflows, is illustrated in Fig.~\ref{fig:gradients_evo}. Top panels and bottom panels show, respectively, the average evolution of radial profiles of star formation surface density and of stellar mass for the progenitors of galaxies selected to have different masses and star formation rates at $z=0$. The left panels show the evolution of profiles for galaxies in a simplified model without SN and AGN feedback and without radial inflows of cold gas in discs while the middle and right panels represent the evolution of galaxies in the full model (without feedback and radial inflows, galaxies of all masses display similar profiles, therefore we only show data for galaxies with intermediate mass for the simplified model). For all models we can see that galaxies tend to increase their sizes and stellar surface densities from high (red colours) to low redshift  (blue colours) as new cooling material is accreted onto discs. 

Particularly interesting trends can be seen in the top panels for the evolution of SFR density profiles. The need for radial inflows of gas is clearly illustrated in the top-left panel for progenitors of intermediate mass, star-forming galaxies, in the simplified model. Even without any feedback, if material is not moved inwards, star formation depletes the cold gas in the inner regions of the galaxy, producing flatter SFR surface-density profiles at low redshift. In other words, without radial inflows, star formation alone produces inside out quenching even in galaxies of relatively low mass. A more realistic evolution of SFR density profiles for star-forming, intermediate mass galaxies, is shown in the top-middle panel for the full model. Radial inflows ensure that newly accreted gas is spread evenly through the disc and star formation happens predominantly in the centre at all times.

Nevertheless, a flattening of SFR profiles at later times can be seen in all the top panels, as star formation increases in the outer rings. This is a direct consequence of having stars forming in molecular clouds, which are predominantly located in regions of high gas surface density and so near the centres of galaxies. When combined with a model that quenches massive galaxies at later times due to AGN feedback (top-right panel), the overall lower normalization of the SFR profiles at later times results in apparent inside-out quenching.

\subsubsection{H$_2$ formation}
\label{sec:H2}

\citet{Fu2010,Fu2012,Fu2013} have tested different prescriptions for the conversion of atomic hydrogen into molecules. In particular, two prescriptions were implemented. The first was the \citet{Krumholz2009} model, in which the $\Htwo$\ fraction is primarily a function of local cold gas surface density and metallicity. The second was the relation originating from \citet{Elmegreen1989, Elmegreen1993}, \citet{Blitz2006} and \citet{Obreschkow2009}, in which the $f_\Htwo$ is a function of the pressure in the ISM. 

As shown in \citet{Fu2010,Fu2012,Fu2013} galaxy properties are quite insensitive to this choice and depend much more critically on the adopted star formation law. Nevertheless, some differences can be seen for the metal content of cold gas, which better resemble observations when using the \citet{Krumholz2009} prescription. Therefore, this will be our default choice.

The \citet{Krumholz2009} model for $\Htwo$ formation calculates an equilibrium $\Htwo$ fraction, $f_{\rm{H_2}} = \Sigma_{\Htwo} /(\Sigma_{\Htwo} + \Sigma_{\rm{HI}})$, for a spherical cloud with a given dust content and surrounded by a photo-dissociating UV field. Following \citet{Fu2013} this prescription is updated using the \citet{McKee2010} fitting equations with the molecular-gas fraction $f_{\Htwo}$ given by:
\begin{equation} \label{eq:h2_1}
f_{\rm{H_2}}=\begin{cases}
\frac{2(2-s)}{4+s},& s<2;\\
0,& s\geq2.
\end{cases}
\end{equation}
In this prescription, $s$ is given by:
\begin{equation} \label{eq:h2_2}
s=\frac{\ln(1+0.6\chi+0.01\chi^2)}{0.6 \tau_c},
\end{equation}
in which $\chi = 3.1\, (1+3.1Z'^{0.365})/4.1$, $\tau_c=0.066\,(\Sigma_{\rm{comp}}/\Msun\rm{pc}^{-2})\,Z'$, $Z' = Z_{\rm{gas}}/Z_{\odot}$ is the gas-phase metallicity in solar units (with $Z_{\odot}=0.0134$, following \citealt{Asplund2009}) and $\Sigma_{\rm{comp}}$ is the gas surface density of the gas cloud. Since the gas surface density in the model is the azimuthally averaged value in each concentric ring, a clumping factor $c_{\rm{f}}$ is introduced to take into account the fact that the gas in real disk galaxies is not smooth.  We introduce an effective gas density $\Sigma_{\rm{comp}}$:
\begin{equation} \label{eq:clump_1}
\Sigma_{\rm{comp}}=c_{\rm{f}}\Sigma_{\rm{gas}}.
\end{equation}
Since there is observational evidence that the gas in metal poor dwarf galaxies is more clumpy than in more metal rich galaxies like our own Milky Way \citep{Lo1993, Stil2002}, we adopt a variable clumping factor that depends on gas-phase metallicity:
\begin{equation} \label{eq:clump_2}
c_{\rm{f}}=\begin{cases}
0.01^{-0.7},& Z'<0.01;\\
Z'^{-0.7},& 0.01\leq Z'<1;\\
1,& Z'\geq1.
\end{cases}
\end{equation}
\citet{Fu2013} found that the \citet{Krumholz2009} prescription can easily yield non-convergent results at very low metallicities. The reason is that molecular cloud formation can only happen after metals have been produced, while metal production requires molecular clouds and star formation. As a result, the $\Htwo$ formation rates for galaxies that have recently started forming stars are quite uncertain and we therefore assume that galaxies with $Z'<0.01$ have $Z'=0.01$. This yields clumping factors of $c_{\rm{f}}\sim{}25$, in agreement with the values of 20-30 used in simulations of high-redshift, low-metallicity systems (\eg{}\citealt{Wang2011,Kuhlen2012}). Between $Z'=0.1$  and 1 the clumping factor varies from 5 to 1, which agrees with the values suggested for normal galaxies in \citet{Krumholz2009}. An alternative to adopting this low-metallicity floor would be to simply allow for the initial enrichment to happen from other channels of star formation, e.g. instabilities and mergers \citep{Stevens2017}.

\subsubsection{H$_2$-based star-formation law}
\label{sec:h2_SF}

Once a model for the spatially-resolved formation of $\Htwo$ has been implemented, it is possible to adopt a star-formation law in which the amount of stars formed is directly related to the amount of $\Htwo$ present in a certain region of a galactic disc. As in \citet{Fu2013}, we assume that star-formation surface density is proportional to the $\Htwo$ surface density \citep[e.g.][]{Leroy2008, Bigiel2011,Schruba2011, Leroy2013}. However, we also include an inverse dependence with dynamical time, such that:
\begin{equation} \label{eq:SF}
\Sigma_{\rm{SFR}}=\alpha_{\Htwo}\Sigma_{\Htwo}/t_{\mathrm{dyn}},
\end{equation}
where $t_{\mathrm{dyn}}=R_{\mathrm{cold}}/V_{\mathrm{max}}$. This ensures that star formation is more efficient at early times where dynamical times are shorter, in accordance with recent observational findings \citep{Scoville2017, Genzel2015}. In addition, it makes more physical sense to have gas clouds collapsing on a timescale that is related to the dynamical time of the disc rather than a universal constant. Molecular clouds are likely to coagulate due to relative motions driven by shear for which this is the relevant timescale.

\subsection{Detailed star-formation histories}
\label{sec:model_SFH}

As in \citet{Henriques2015}, the current version of our model stores star-formation and metal-enrichment histories using the algorithm described in \citet{Shamshiri2015}. This aspect is essential for the implementation of the \citet{Yates2013} chemical-enrichment model in order to follow the time-dependent release of mass and energy by dying stars.  In addition, this makes it possible to compute luminosities, colours and spectral properties in post-processing using any stellar-population-synthesis model.

\begin{figure*}
\centering
\includegraphics[width=15.5cm]{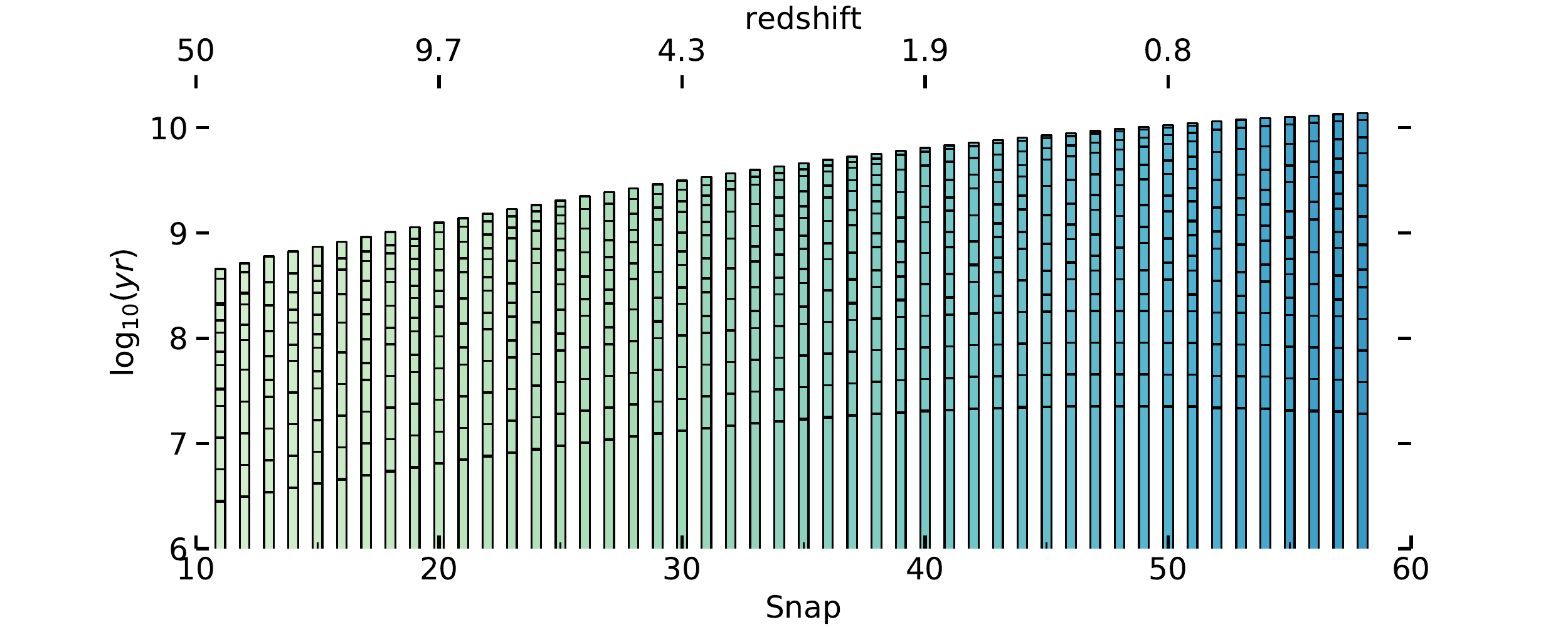}
\caption{The binning structure in which star formation histories are recorded at different snapshots/redshifts. The most recent bins, at the bottom of the plot, have the best resolution, with a median age corresponding to a single internal time-step for the semi-analytic calculation (20 times smaller than the time between snapshots). Older stellar populations are binned together resulting in widths ranging from a few to ten million years for recent bins to a few Gyrs for the oldest stars.}
\label{fig:sfh_bins}
\end{figure*}

An important difference relevant for the present version of our model is that since the detailed chemical-enrichment scheme is implemented at the level of the spatially-resolved rings in galactic discs, the recorded star-formation and metal-enrichment histories must also be recorded separately for each ring. We store these history bins in each ring with a time resolution that degrades for older stellar populations. The most recent activity is always stored with the maximum resolution, which is set to be equal to a single substep of the main timestep of the semi-analytic model. As the computation progresses, older bins are merged together logarithmically. \citet{Shamshiri2015} showed that with each timestep split into $\sim$20 substeps, one can recover the UV luminosities of
galaxies in post-processing with less than 10\% scatter for more than 90\% of the galaxies at any given time (and  with much lower scatter at longer wavelengths).

The resulting bin structure is shown as a function of time in Fig.~\ref{fig:sfh_bins}. As can be seen, there is a relatively high resolution for stellar populations with ages less than $\sim$1 Gyr. Typically, these stellar populations are tracked with more than 5 bins with the highest-resolution timesteps varying from a few Myr at high redshifts  to just over 10 Myr at low redshifts. Populations older than 1 Gyr are represented by $\sim$ half a dozen bins at $z=0$. 

\subsection{Detailed chemical enrichment on local scales}
\label{sec:model_DCE}

With the present update of \lgal, we also incorporate a galactic-chemical-enrichment (GCE) model that tracks how much enriched material is returned to the interstellar medium and circumgalactic medium by each stellar population at any given time. This GCE scheme, introduced by \citet{Yates2013}, directly follows the delayed enrichment of eleven individual chemical elements (H, He, C, N, O, Ne, Mg, Si, S, Ca, and Fe) produced by SNe-Ia, SNe-II, and winds from AGB stars, adopting mass and metallicity-dependent stellar yields and lifetimes. This scheme also includes a reformulation of the associated SN feedback, so that energy and heavy elements are released into the ISM and CGM when stars die, rather than when they are born (i.e. we eliminate the instantaneous-recycling approximation). 

The bulk of the GCE set-up used in this new version of \lgal{} is the same as described in \citet{Yates2013}. In brief, the total ejection rate of chemical element X by a simple stellar population (SSP) at time $t$ is given by,

\begin{equation}
    e\sub{X}(t) = \int^{M_{U}}_{M_{L}} M\sub{X}(M,Z_{0}) \  \psi(t-\tau_{\textnormal{M}}) \  \phi(M) \  \textnormal{dM} \;\;,
\end{equation}
where $M$ is the initial mass of a star, $\tau\sub{M}$ is its lifetime, $M_{L}$ is the lowest mass of star to eject material at time $t$ (\ie{}one with a lifetime of $\tau\sub{M} = t$), $M_{U}$ is the maximum star mass considered ($120 \Msun$ in this work), and $M\sub{X}$ is the mass of element X ejected per star, which depends on the initial mass $M$ and initial metallicity $Z_{0}$. This ejecta mass comprises both the yield, $y\sub{X}$, and the mass of element X that passes through a star unprocessed before being ejected. Finally, $\psi(t-\tau_{\textnormal{M}})$ is the SFR at a star's birth, and $\phi(M)$\,d$M$ is the number of stars in the mass range $M\mapsto M+$d$M$ per unit mass of star formation. For more details, we refer the reader to section 4 of \citet{Yates2013}.

Mass- and metallicity-dependent yields are taken from \citet{Marigo2001} for AGB stars, from \citet{Thielemann2003} for SNe-Ia (not metallicity dependent), and from \citet{Portinari1998} for SNe-II. Mass- and metallicity-dependent stellar lifetimes are also taken from the calculations of \citet{Portinari1998}. A \citet{Chabrier2003} IMF with a constant high-mass-end slope of $\alpha\sub{IMF} = -2.3$ is always assumed. When modelling the lifetimes of SNe-Ia, a constant power-law delay-time distribution (DTD) with a slope of -1.12 is assumed, following \citet{Maoz2012}.

Two modifications to the parameters governing GCE have been made here, in comparison to those chosen by \citet{Yates2013}. First, the fraction of objects between 3 and $16\Msun$ in each stellar population that are assumed to form SN-Ia progenitors has been increased from $A\sub{old} = 0.028$ to $A = 0.04$ (for our chosen IMF, this is equivalent to a fraction of \textit{all} objects in a stellar population that are assumed to form SN-Ia progenitors of $A' = 0.00154$). This new value is still well within the range inferred from observations of the SN-Ia rate, which suggest $0.024 \lesssim A \lesssim 0.081$ \citep{Maoz2012}.

Second, the amount of direct enrichment of the hot circumgalactic medium (CGM) by supernovae has been modified. Previously, all supernovae exploding in the stellar disc were assumed to directly enrich the ISM. This material was then fully mixed with the ambient cold gas, before being reheated into the CGM, or expelled from the dark-matter halo in galactic winds. In our new version of \lgal, we instead allow 30\% of the ejecta from both SNe-II and SNe-Ia to be directly dumped into the CGM, constituting a metal-rich wind that is then allowed to cool and re-accrete onto the galaxy along with the ambient hot gas. This value was chosen so that the normalization of the gas phase metallicity relation shown in Fig.~\ref{fig:MZgR_z0} is roughly consistent with observations (although we note that similar results are obtained with values ranging from 10 to 50\%). More sophisticated prescriptions for the direct enrichment of the CGM will be investigated in future work.


\subsubsection{Spatially resolved SN feedback}
\label{sec:resolved_SN}
In addition to allowing metal-rich winds to deposit material directly into the CGM, a major modification of the present work is that all the material and energy ejected into the ISM by stellar populations is spatially resolved. In practice this means that stellar populations return mass and energy to a specific ring and that the amount of energy used for reheating the ISM, and the value at which it saturates, is computed independently for that ring. Any left-over energy from reheating the ISM in each ring is then added-up and used to eject hot gas into the external reservoir.

In detail, the energy effectively available to the gas components from supernovae
and stellar winds is taken to be:
\begin{equation} \label{eq:energy_sn1}
\Delta E_{\rm SN}=\epsilon_{\rm halo}\times\Delta
M_{\star, R}\eta_{\rm{SN}} E_{\rm{SN}},
\end{equation}
where $\Delta M_{\star, R}$ is the mass returned to the ISM by different stellar populations (as opposed to the mass of stars formed used in \citealt{Henriques2015}), $\eta_{\rm{SN}}$ is the number of supernovae expected per solar mass of stars returned to the ISM ($0.0149\Msun^{-1}$, assuming a universal, \citealt{Chabrier2003}, IMF), $E_{\rm{SN}}$ is the energy released by each supernova ($10^{51}$erg) and $\epsilon_{\rm halo}$ is a free parameter given by:
\begin{equation} \label{eq:energy_sn2}
\epsilon_{\rm halo} =\eta_{\rm eject}\times \left[0.5+\left(\frac{V_{\rm
      max}}{V_{\rm eject}}\right)^{-\beta_{\rm eject}} \right] .
\end{equation}

The mass of cold gas reheated by star formation and added to the hot atmosphere is assumed to be directly proportional to the amount of stars returned to the ISM:
\begin{equation} \label{eq:reheat}
\ \Delta M_{\mathrm{reheat},i}=\epsilon_{\rm disk}\Delta M_{\star,R_i},
\end{equation}
where the second efficiency is:
\begin{equation} \label{eq:reheat2}
\ \epsilon_{\rm disk}=\epsilon_{\rm reheat} \times \left[0.5+\left(\frac{V_{\rm max}}{V_{\rm reheat}}\right)^{-\beta_{\rm reheat}}  \right]
\end{equation}
and $\Delta M_{\mathrm{reheat},i}$ and $\Delta M_{\star,R_i}$ are computed locally for each ring. This reheating is assumed to require energy $\Delta E_{\mathrm{reheat,i}}=\frac{1}{2}\Delta M_{\mathrm{reheat},i}V^2_{\twoc}$. If $\Delta E_{\mathrm{reheat},i} > \Delta E_{\rm SN} \times \Delta M_{\star,R_i}/\Delta M_{\star,R}$, the reheated mass in each ring is assumed to saturate at $\Delta M_{\mathrm{reheat},i}=\Delta E_{\rm{SN}}\times \Delta M_{\star,R_i}/\Delta M_{\star,R}/\left(\frac{1}{2}V^2_{\twoc}\right)$. 

Any remaining SN energy is added up and used to eject a mass $\Delta M_{\rm eject}$ of hot gas into an external reservoir:
\begin{equation} \label{eq:ejection3}
\frac{1}{2}\Delta M_{\rm eject}V_{\twoc}^2=\Delta E_{\rm SN}-\Delta E_{\rm reheat},
\end{equation}
where $\Delta E_{\rm reheat}$ is the sum over all the rings.

\begin{figure}
\centering
\includegraphics[width=8.4cm]{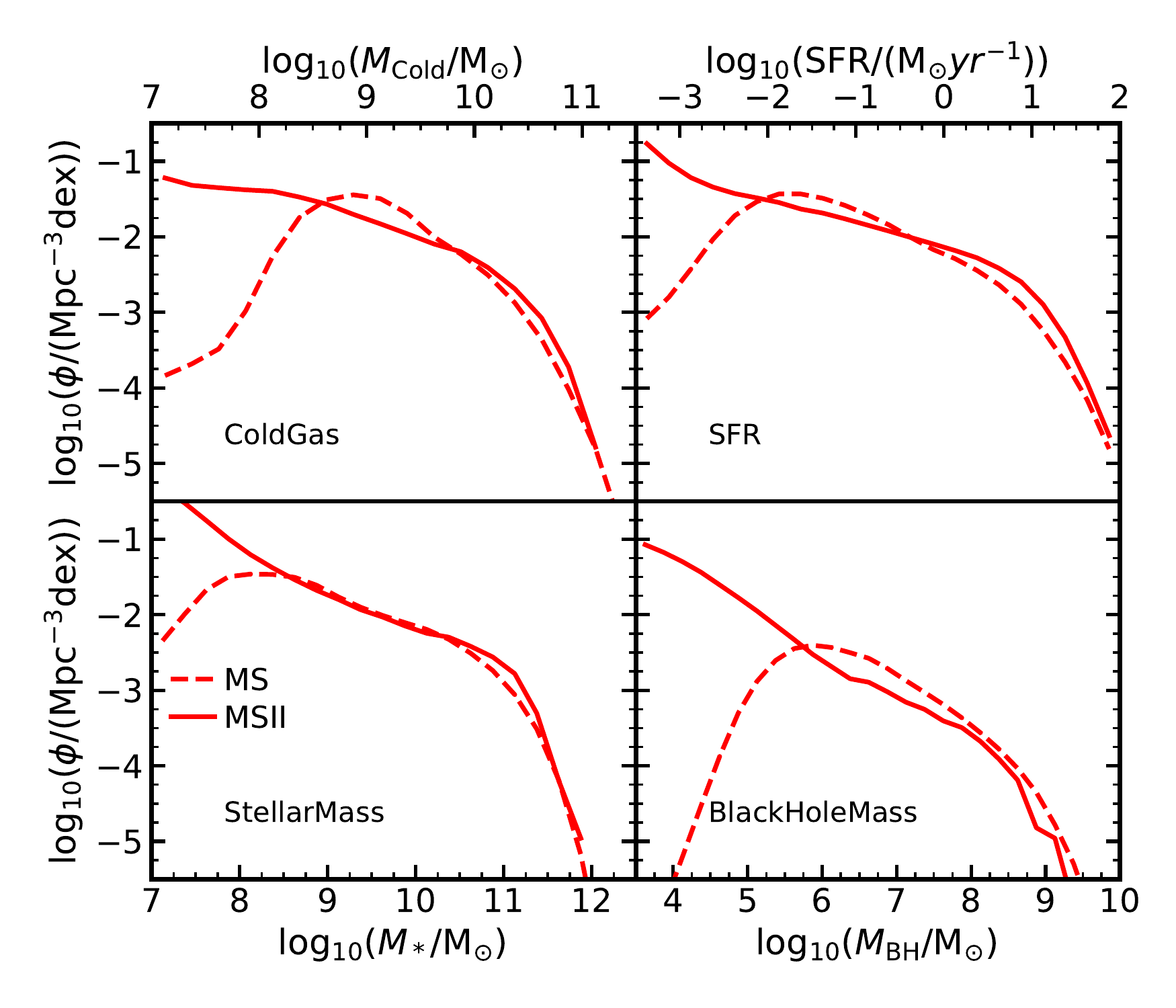}
\caption{Resolution tests using the $z=0$ ``mass'' functions for cold gas mass, SFR, stellar mass and BH mass. The dashed and solid lines show results based on the MS and MS-II simulations respectively. The larger volume/lower resolution MS simulation is incomplete at low masses, but has better statistical power for the most massive objects.}
\label{fig:resolution}
\end{figure}

\subsection{Convergence}
\label{sec:convergence}

With the significant changes introduced in the present model, in particular with the newly-implemented ability to resolve the radial structure of galaxy discs, it is important to test whether our simulation results are numerically converged, and at which mass scales resolution effects might be important. As described in \citet{Henriques2015}, the model is calibrated using a representative subset of merger trees based on the MS and MS-II simulations. Galaxies based on the former are used to compare with observations for stellar masses  $\log_{10} (M_*/\Msun) \ge 9$ and the latter for stellar masses below this value. The resulting best-fit parameters are then used to run the galaxy formation model on both simulations and produce all the plots presented in this paper. Fig.~\ref{fig:resolution} shows that mass functions based on the MS and MS-II simulations, which differ in resolution by a factor of 125, agree well for $\log_{10} (M_*/\Msun) \ge 8.5$, $\log_{10} (M_{\rm{Cold}}/\Msun) \ge 9.0$, $\log_{10} (\mathrm{SFR}/(\Msun \rm{yr}^{-1})) \ge -1.0$ and $\log_{10} (M_{\rm{BH}}/\Msun) \ge 7.5$. 


\section{Comparison with observed galaxy properties used as constraints}
\label{sec:constraints}

\begin{figure}
\centering
\includegraphics[width=8.4cm]{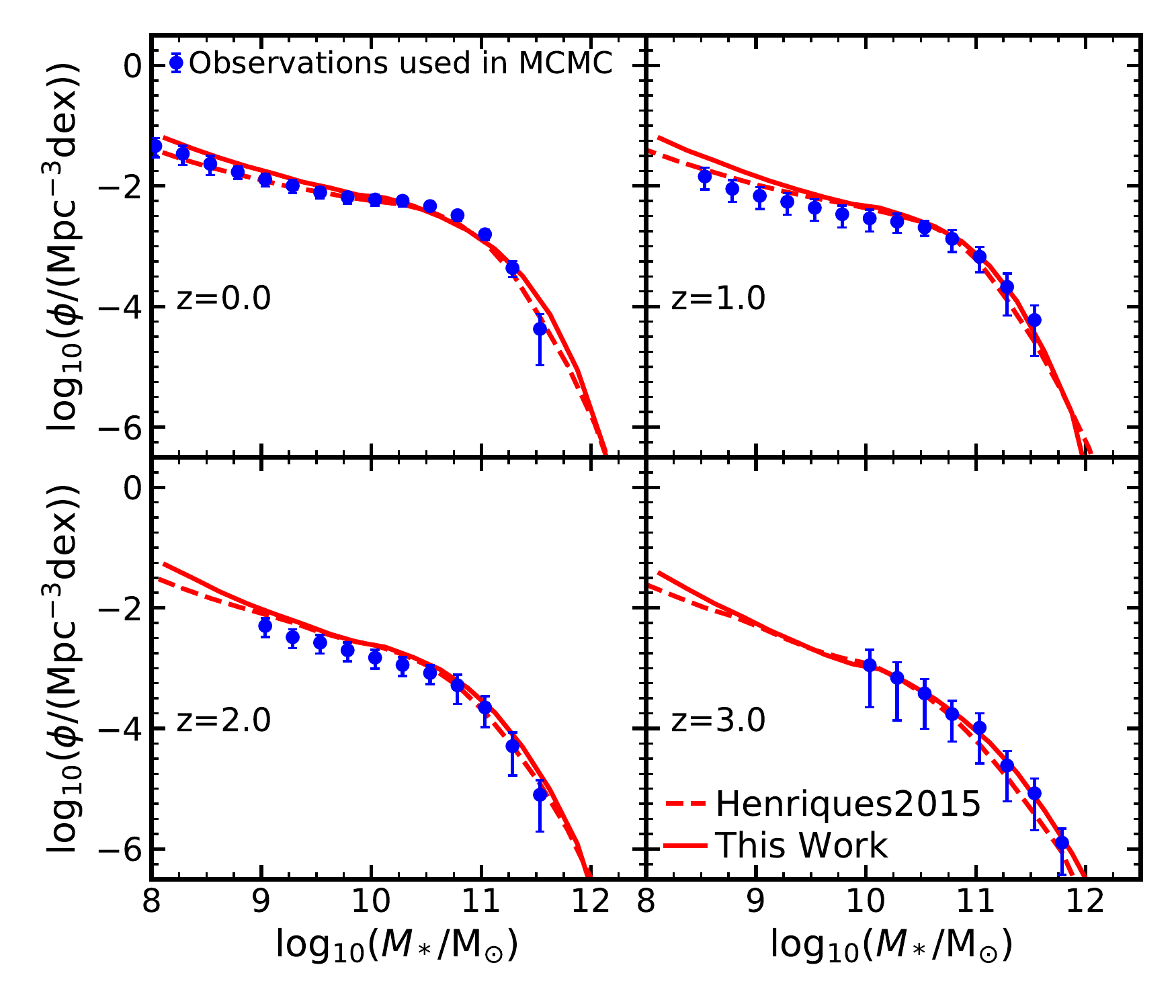}
\caption{The evolution of the stellar mass function from $z=0$ to 3. Results for the current model (solid red lines) are compared with those from \citet{Henriques2015} (dashed red lines) and with a combination of observational data sets (blue symbols) described in Appendix~A2 of \citet{Henriques2015}.}
\label{fig:SMF_evo}
\end{figure}

\begin{figure*}
\centering
\includegraphics[width=17.7cm]{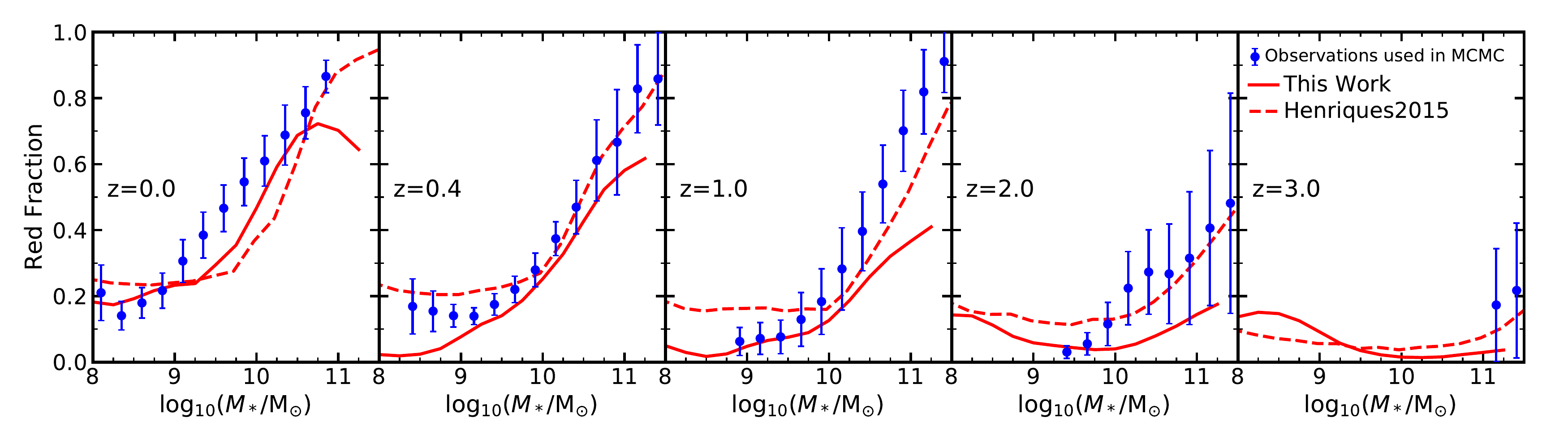}
\caption{The evolution of the red fraction as a function of stellar mass from $z=0$ to 3. Results for the current model (solid red lines) are compared with those from \citet{Henriques2015} (dashed red lines) and with a combination of observational data sets (blue symbols) derived from the stellar mass function of red and blue objects as described in Appendix~A2 of \citet{Henriques2015}.}
\label{fig:redfraction_colorcut}
\end{figure*}

\begin{figure*}
\centering
\includegraphics[width=15.9cm]{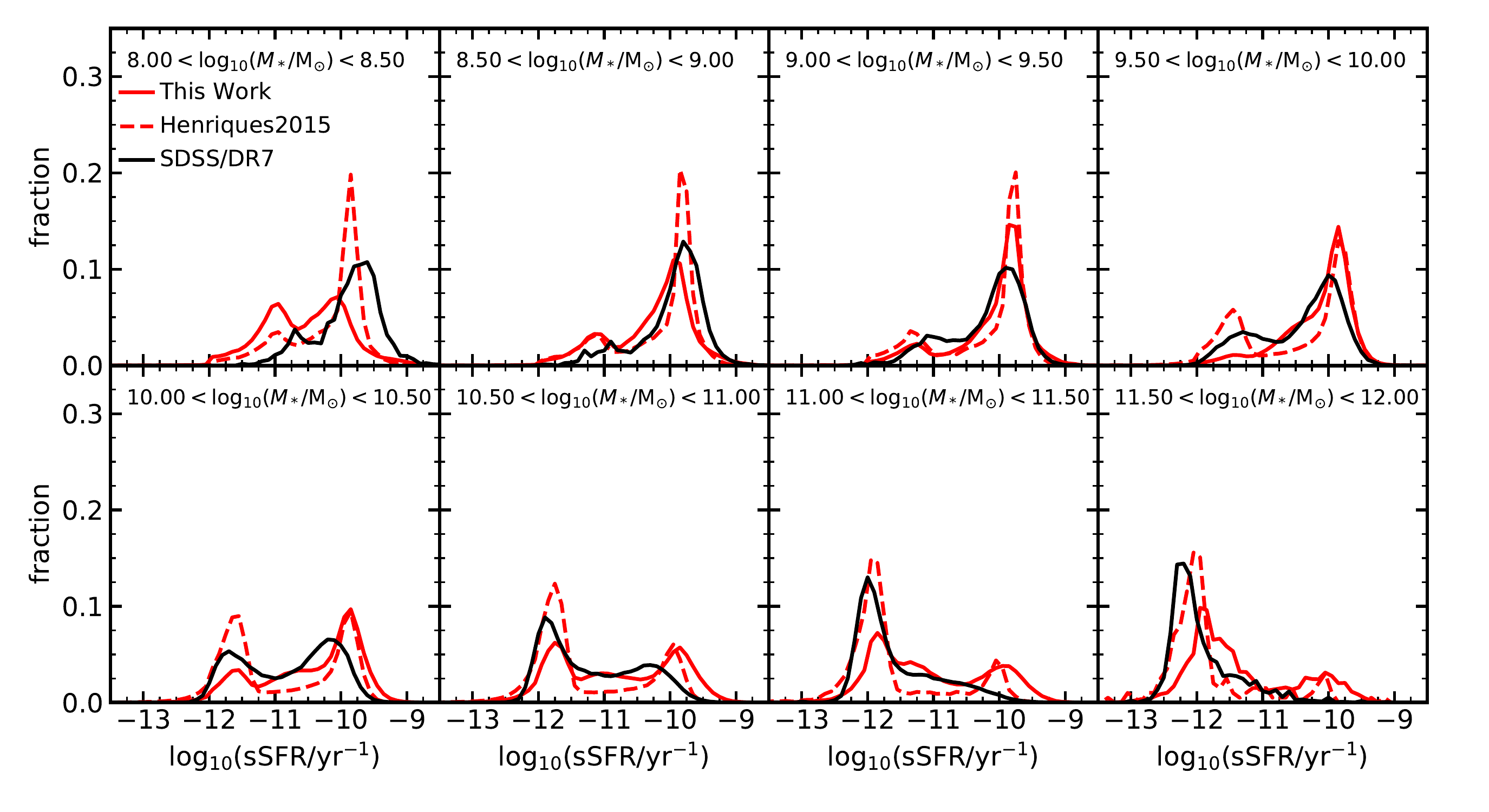}
\caption{Histograms of $z=0$ sSFR in bins of increasing stellar mass (from top left to bottom right). Our new model results (solid red lines) are compared with those from the \citet{Henriques2015} model (dashed red lines) and with observations from SDSS-DR7 (solid black lines). The latter have been weighted by $1/V_{\mathrm{max}}$ in order to produce volume-limited statistics.}
\label{fig:ssfr_hist}
\end{figure*}

\begin{figure*}
\centering
\includegraphics[width=17.7cm]{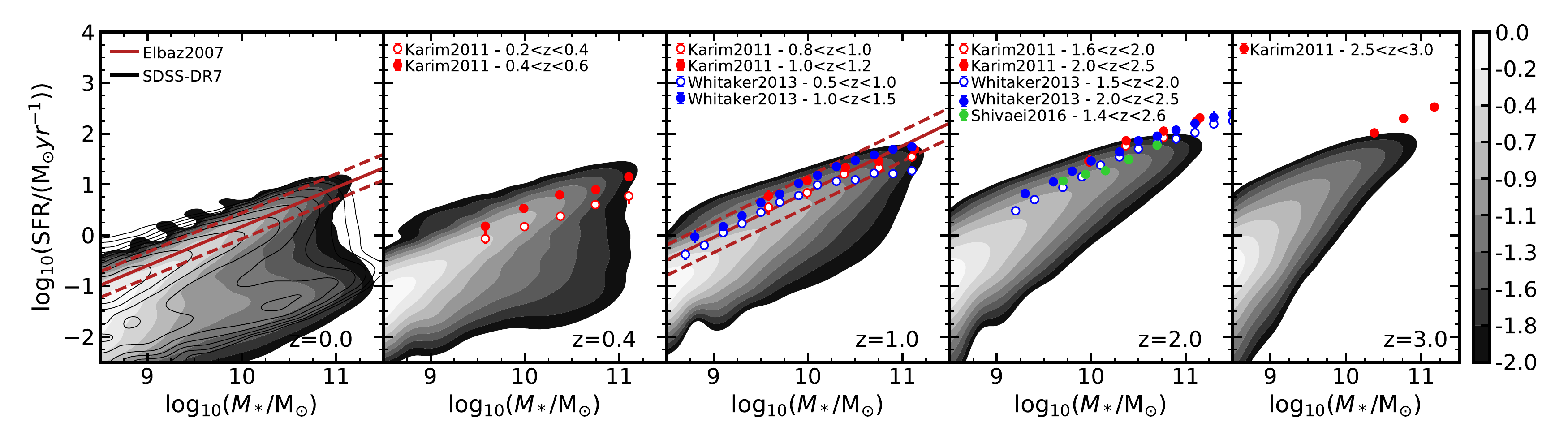}
\caption{Evolution of the SFR versus stellar mass relation from $z=0$ to 3 (the "main sequence" of star formation). The logarithmic gray contours show model results and are compared with different dobservational datasets. These are displayed as empty black contours (SDSS-DR7 at $z=0$), solid brown lines (\citealt{Elbaz2007} at $z=0$ and 1), red circles (\citealt{Karim2011} at $z=1$ to 3), blue circles (\citealt{Whitaker2012} at $z=1$ and 2) and green circles (\citealt{Shivaei2016} at $z=2$).}
\label{fig:main_sequence}
\end{figure*}

In this section we start our comparison between model results and observations for properties either used as constraints in our MCMC sampling, or directly related to those. The former are the stellar mass functions shown in Fig.~\ref{fig:SMF_evo}, the red fractions as a function of stellar mass shown in Fig.~\ref{fig:redfraction_colorcut} and the HI mass function shown in Fig.~\ref{fig:GMF}. In addition, we compare our results with observations of (i) star formation rates (Figs.~\ref{fig:ssfr_hist}, \ref{fig:main_sequence} and \ref{fig:sfrd}), which are directly related to red fractions, and (ii) different cold gas phases (Figs.~\ref{fig:HI_bins}, \ref{fig:gas_fraction} and \ref{fig:h2d}), related to the HI content. In this and the next section we will analyse how our new model, with a self-consistent treatment of multiphase ISM, star formation, feedback and chemical enrichment on local scales, compares with observed properties on galactic scales. After showing that it produces a realistic population of galaxies we will then fully exploit its new capabilities by comparing model results with observed radial profiles in Section~\ref{sec:gradients}. Multiple observational data sets were used to constrain both the evolution of the stellar mass function and red fraction as a function of stellar mass. These were presented in \citet{Henriques2015} and we refer the reader to that paper, and in particular to Appendix 2, for a full description of the methodology used to combine them.

\subsection{Evolution of the stellar mass function and red galaxy fraction}
\label{sec:mass_color}

In this section, we compare our model with observations of the evolution of the stellar mass function and the red fraction as a function of stellar mass. These properties were the main constraints used to build the \citet{Henriques2015} model. 

Fig.~\ref{fig:SMF_evo} compares the stellar mass functions from our new model over the redshift range $0\leq z \leq 3$ to the observational data that were used to constrain its parameters. The latter were obtained by combining datasets from SDSS at $z=0$ \citep{Baldry2008, Li2009, Baldry2012} and ULTRAVISTA at higher redshifts \citep{Ilbert2013, Muzzin2013}. As described in \citet{Henriques2015} we convolve our theoretical estimates with a Gaussian in $\log_{10}M_{*}$ with width increasing with redshift, in order to mimic the uncertainties in observational determinations of stellar mass ($\sigma=0.08\times(1+z)$). 

As was found in our previous model, the level of agreement between model and observations shown in Fig.~\ref{fig:SMF_evo} requires very strong SN feedback at early times, and a relatively long timescale for reincorporation of ejected material, in order to suppress the build-up of low-mass galaxies at $z \ge 2$ \citep{Henriques2013}. The latter ensures significant return of material onto intermediate-mass objects at $z \le 2$ and enhances the amplitude of the mass function around $M_*$ to fit observation. Similar trends have been identified in other semi-analytic models by \citet{Mitchell2014}, \citet{Hirschmann2016} and \citet{Lagos2018}. In the case of the present model, we need to use all the energy available from SN to reheat and eject gas from low-mass galaxies in order to reduce their number density at $\log_{10} (M_*/\Msun) \sim 9$. Nevertheless there is still an hint of an excess below this mass. This indicates that our model could still be missing some important source of heating for which cosmic rays would be a likely candidate.

Fig.~\ref{fig:redfraction_colorcut} shows the evolution of the red fraction as a function of stellar mass. Observational estimates are obtained by dividing the number density of red objects by the number density of all galaxies using the stellar mass functions of red and blue galaxies. The $z=0$ data points are based on SDSS \citep{Baldry2004} while high-redshift data points are based on ULTRAVISTA \citep{Ilbert2013, Muzzin2013}. Passive and star-forming galaxies in the model are selected using a cut in $u-r$ versus $r$ at $z=0$ and in sSFR versus stellar mass at higher redshifts (see Appendix~\ref{app:cuts}). This cut is meant to adequately separate the two populations in the model and, although normally close to it, is not necessarily the one used in interpreting observations.

The model roughly reproduces the observed trends, with massive galaxies being predominantly quenched and low-mass galaxies being predominantly star-forming. In detail, the red fraction approaches 0.7 at $\log_{10} (M_*/\Msun) \sim 11.0$ and 0.1 at $\log_{10} (M_*/\Msun) \sim 8$. As shown in \citet{Henriques2017}, low-mass red galaxies are predominantly satellites that have their star formation quenched by environmental processes. Since the fraction of satellites is $\sim$30\% at any stellar mass, and approximately half of the satellites are quenched in our model at $\log_{10} (M_*/\Msun) \sim 9$, satellites alone account for $\sim$15\% of galaxies being quenched in this mass range. At higher masses,  when the host haloes of galaxies reach a mass of $\log_{10} (M_{\mathrm{vir}}/\Msun) \sim 12$, they switch from the cold to the hot mode of accretion. SN then become inefficient in removing gas and galaxies experience a phase of enhanced star formation and black-hole growth. Shortly thereafter, black holes become massive enough to produce sufficient AGN feedback to completely shut down cooling and quench galaxies \citep{Henriques2019}.

In comparison to \citet{Henriques2015} and to observation, there is a deficit of red galaxies in our new model at high mass ($\log_{10} (M_*/\Msun) > 10.5$) and high redshift ($z \ge 2$). This deficit seems to be caused by the more detailed treatment of cold gas and star formation in our new model. Since the latter now requires the formation of molecular gas, a relatively inefficient process particularly in the outer rings of massive galaxies, these objects now have relatively long gas depletion timescales once cooling is suppressed by AGN feedback. As we shall see, the new radially resolved treatment of galactic discs does, however, result in significantly more realistic model results for their gas properties. The selection in colour done at $z=0$ (left panel) leads to a downturn in the fraction of red galaxies at high mass, consistent with the deficit in metallicity seen in Fig.~\ref{fig:MZsR_z0}.

\subsection{Star-formation rates}
\label{sec:sfr}

\subsubsection{Local star-formation rates}
\label{sec:sfr_local}

Fig.~\ref{fig:ssfr_hist}, shows histograms of specific star-formation rate (sSFR) at $z=0$, for bins of increasing stellar mass (from top left to bottom right). Observational data from \citet{Brinchmann2004} that includes the \citet{Salim2007} updates are plotted as solid black lines after applying a $V_{\mathrm{max}}$ correction (every galaxy is weighted by $1/d_{\mathrm{max}}^3$, where $d_{\mathrm{max}}$ represents the maximum distance out to which the galaxy could be observed considering the flux limit of the survey). Results from the new model are shown as solid red lines. Focusing on low-mass galaxies (the top panels) we see that the model correctly captures their star-forming peak (except for the lowest mass bin, on the top left, where resolution may affect the results). In the bottom panels of Fig.~\ref{fig:ssfr_hist} we see that the fraction of quenched galaxies significantly increases towards higher mass --- almost all galaxies with $\log_{10} (M_*/\Msun) > 11.0$ have $\log_{10} (\mathrm{sSFR} /\rm{yr}^{-1}) < -11.0$ (two bottom-right panels). Our model also correctly captures the observed trend for an increasingly dominant passive peak with increasing stellar mass. Nevertheless, a small population of star-forming galaxies persists above $\log_{10} (M_*/\Msun) \sim 11.0$ in our model, in disagreement with observation. As previously described, this seems to be caused by some massive galaxies retaining a residual amount of cold gas and star formation for many Gyr after cooling is shutdown by AGN.

\begin{figure}
\centering
\includegraphics[width=8.4cm]{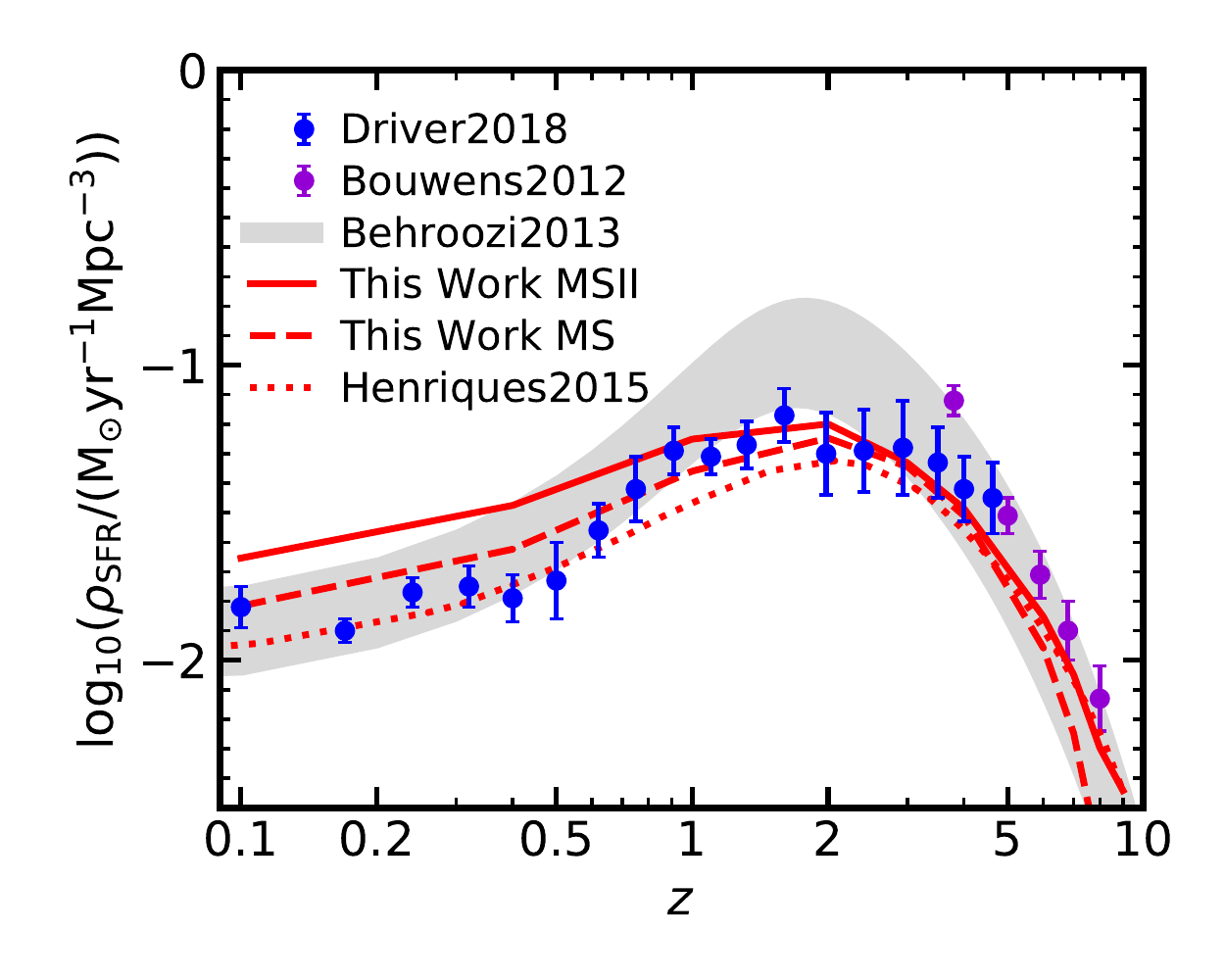}
\caption{Evolution with redshift of the comoving density of cosmic star formation. The new model (based on the MS and MS-II simulations, respective dashed and solid red lines) and that of \citet{Henriques2015} (dashed red line) are compared with observational data from \citet{Bouwens2012}, \citet{Behroozi2013} and \citet{Driver2018}.}
\label{fig:sfrd}
\end{figure}

\subsubsection{Evolution of star-formation rates}
\label{sec:sfr_highz}

As is the case for their $z=0$ analogues, star-forming galaxies at high redshift are observed to lie on a tight sequence of SFR versus stellar mass, with a scatter of only $\sim0.3$ dex, at least out to $z=2$ \citep{Elbaz2007, Noeske2007}. In Fig.~\ref{fig:main_sequence} we show this so-called "main sequence" of star formation for our model as filled grey contours at logarithmically spaced intervals. At $z=0$ the model is compared with a stellar mass-complete sample from SDSS (empty solid contours). Additional datasets, shown as coloured circles and brown lines, are for star-forming objects only. At $z \leq 1.0$, there is a clear build-up of a passive population in the model (three left panels).

In all panels, the theoretical distribution of star-forming galaxies is particularly tight, with a similar scatter to that observed for $\log_{10} (M_*/\Msun) \geq 9.5$, similar slope and increasing normalization with increasing redshift. In our model, this simply reflects the scaling of cooling rate and feedback efficiency with halo mass for $\log_{10} (M_{\mathrm{vir}}/\Msun) \lesssim 12.0$. The star-formation process is largely determined by the availability of baryons in the cold phase, which in turn is controlled by a balance between accretion of halo gas and ejection of the ISM by SNe. This also explains the similarity between the properties of star-forming objects in the new model and in \citet{Henriques2015}. 

In Fig.~\ref{fig:sfrd} we show the evolution of the integrated star formation rate density (SFRD). Our new model results based on the MS and MS-II simulations (respectively, dashed and solid red lines) are compared with a compilation of data from \citet{Behroozi2013} (gray region) and with independent derivations from \citet{Bouwens2012} and \citet{Driver2018}. While there is agreement between new model and data for MS based results, the normalization of the theoretical SFRD based on MS-II lies above the observational data, and that of \citet{Henriques2015}, at $z \sim 0$ by $\sim$ a factor of 2. This is likely a combined effect of the slight excess in the number density of low-mass galaxies seen in Fig.~\ref{fig:SMF_evo}, the fact that the sSFR distribution at fixed $M_*$ is shifted towards larger values for $\log_{10} (M_*/\Msun) \geq 10.0$ in Fig.~\ref{fig:ssfr_hist} and the slight excess in the MS-II-based SFR function at $\log_{10}(\mathrm{SFR}/\Msun yr^{-1}) \sim 1$ seen in Fig.~\ref{fig:resolution}.

\subsection{Gas properties}
\label{sec:gas}

\begin{figure}
\centering
\includegraphics[width=8.4cm]{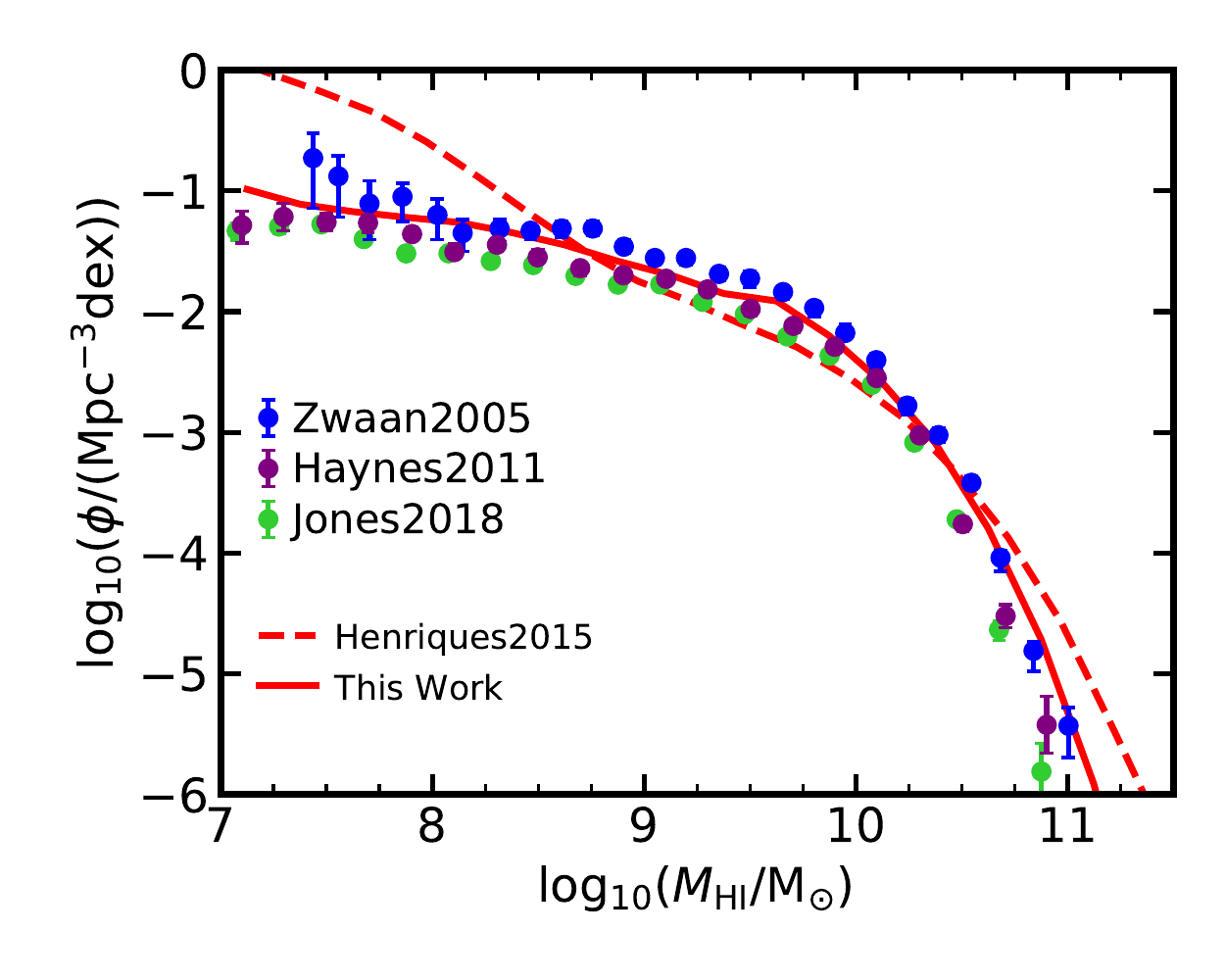}
\caption{Comparison of model results for the mass function of atomic gas at $z = 0$ with observational data from \citet{Zwaan2005}, \citet{Haynes2011} and \citet{Jones2018}. In the current model HI is tracked directly, while a conversion factor of $0.54\times M_{\rm{cold}}$ is used to plot results from \citet{Henriques2015} (dashed red line).}
\label{fig:GMF}
\end{figure}

This subsection examines our new model's results for the HI mass function, the distribution of the ratio HI/$L_r$ in bins of HI mass and the fractions of galactic mass in HI and $\Htwo$ as functions of stellar mass (all at $z=0$), as well as the evolution over redshift of the integrated cosmic density in $\Htwo$. All these quantities can be compared with  recent observational estimates. 

In Fig.~\ref{fig:GMF} we compare our new model (the solid red line) and that of \citet{Henriques2015} (the dashed red line) to observed $z=0$ HI mass functions, in particular, to HIPASS data as analysed by \citet{Zwaan2005} and to ALFALFA data as analysed by \citet{Haynes2011} and \citet{Jones2018}. Note that the HI mass function was one of the constraints used in the MCMC determination of the parameters of our new model, whereas it was not used as a constraint by \citet{Henriques2015}. The amount of  HI is tracked directly in our current modeling, but for \citet{Henriques2015} we have multiplied total cold gas masses by $M_{\rm{HI}}/M_{\rm{cold}}=0.54$ to make this plot\footnote{We note that \citet{Martindale2017} partitioned the HI and $\Htwo$ content of \citet{Henriques2015} in post-processing using an analytic surface-density profile and observed a similar behaviour at low mass to that obtained with our simple conversion.}. The figure shows that agreement is achieved across the entire observed mass range used to constrain the model \citep{Zwaan2005, Haynes2011}. This represents a significant success, since it has been particularly challenging for $\Lambda$CDM-based models to match simultaneously the abundances of low-mass galaxies as a function of stellar and HI mass \citep{Lu2015}. Our findings are consistent with recent literature, with models that adopt strong SN-driven winds and a multiphase treatment of gas on sub-kpc scales producing results that are consistent with observation \citep{Stevens2016, Crain2017, Xie2017, Lagos2018}.

\begin{figure}
\centering
\includegraphics[width=8.4cm]{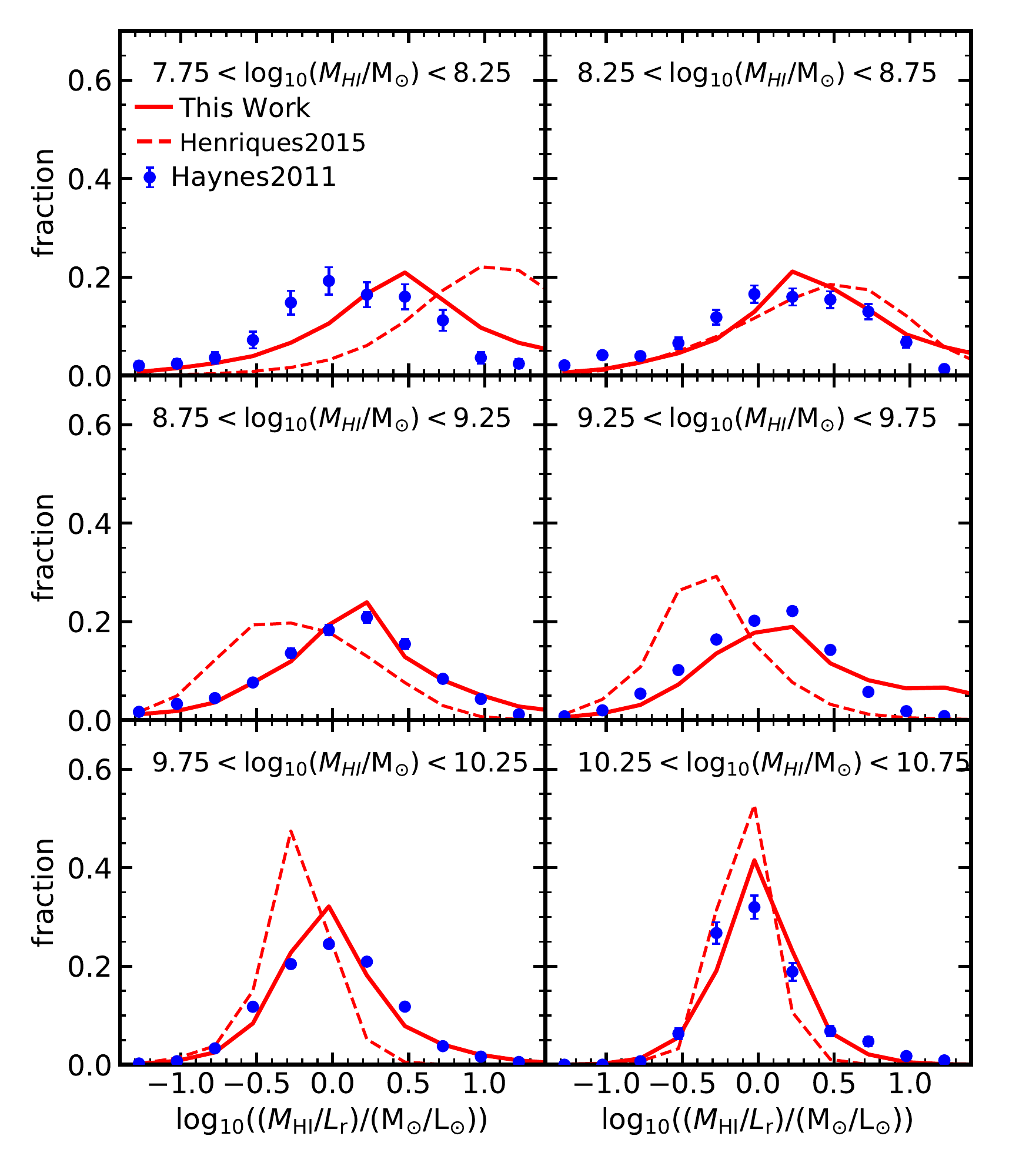}
\caption{Histograms of the ratio of cold gas mass to r band luminosity in different bins of atomic gas mass. The current model (solid red line) and that of \citet{Henriques2015} (dashed red lines) are compared with the HI-flux-limited sample of \citet{Haynes2011} from ALFALFA (blue circles).}
\label{fig:HI_bins}
\end{figure}

A stringent test of  our more detailed treatment of cold gas is provided by the {\it distribution} of gas-to-stellar mass ratio as a function of galactic mass. For large HI-flux-limited surveys such as ALFALFA, for which optical imaging is available for all detected objects, it is possible to construct complete, volume-limited versions of this distribution in the form of distributions of the ${\rm HI}/L_r$ flux ratio for galaxies binned by their HI mass. Such distributions are shown in Fig.~\ref{fig:HI_bins} for the \citet{Haynes2011} ALFALFA data using 0.5 dex bins in  $\log_{10} M_{\rm{HI}}$ (increasing from top left to bottom right) and are compared to results from the \citet{Henriques2015} model (dashed red lines) and our new model (solid red lines). We remind the reader that the cold gas content of galaxies in the local Universe is used as a constraint in our MCMC sampling of the parameter space of our new model (Fig.~\ref{fig:GMF}). As a result, unlike the \citet{Henriques2015} case that exhibits a significant deficit of HI at intermediate stellar mass and an excess at low-mass, the new model agrees well with the observations in all HI mass bins except the lowest, where HI fractions are too high by about a factor of two. Our new model produces realistic HI gas fractions over three orders of magnitude in HI mass.

\begin{figure}
\centering
\includegraphics[width=8.4cm]{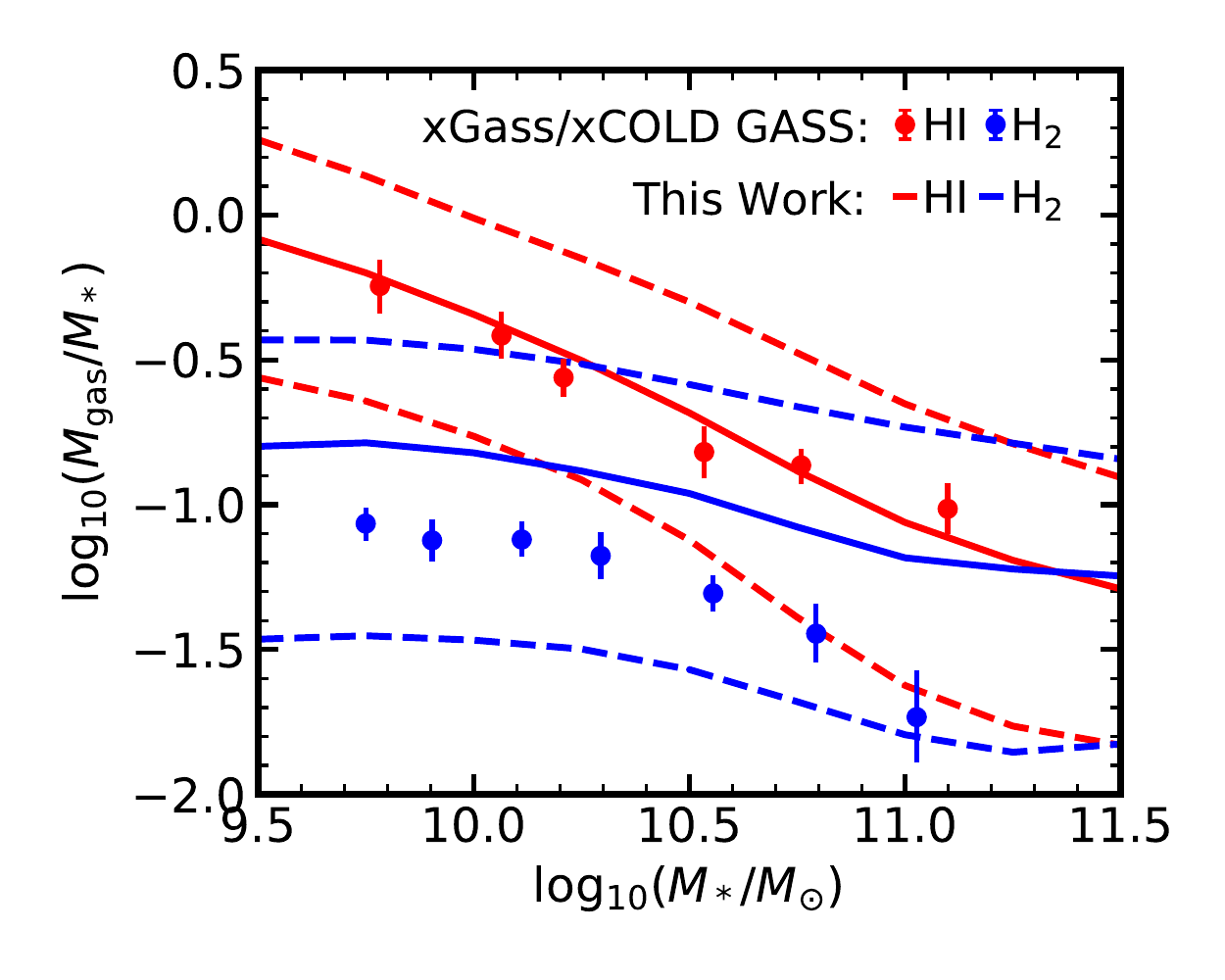}
\caption{The mass fractions in HI (red) and $\Htwo$ (blue). Model medians and 16th+84th percentiles (solid and dashed lines) are compared with xCOLD GASS \citep{Saintonge2017} and xGASS \citep{Catinella2018} data. The comparison is restricted to star-forming galaxies (defined as galaxies not more than 1 dex below the main sequence).}
\label{fig:gas_fraction}
\end{figure}

\subsubsection{$\Htwo$ content}
\label{sec:H2d}

Fig.~\ref{fig:gas_fraction} compares theoretical HI (red) and $\Htwo$ (blue) over stellar mass fractions as a function of stellar mass with xCOLD GASS \citep{Saintonge2017} and xGASS \citep{Catinella2018} data for star-forming galaxies. xCOLD GASS builds upon the original COLD GASS survey, providing a census of the molecular-gas content of a mass-selected sample ($\log_{10} (M_*/\Msun)>9.0$) at $0.01<z<0.05$. These were obtained using CO(1-0) measurements from the IRAM-30m telescope complemented by observations of the CO(2-1) line with both the IRAM-30m and APEX telescopes. Both surveys were accompanied by sister programs at the Arecibo telescope aimed at obtaining HI-mass estimates for a similar sample of galaxies. The observational data points correspond to weighted medians and the errors take into account statistical uncertainties associated with the IRAM calibration and aperture corrections and the sampling error determined from bootstrapping. It should be noted that while model results are presented for individual galaxies, observational mass estimates are derived using a beam size of 3.5 arcmin (corresponding to $\sim25$kpc at z=0.015) and might include contributions from multiple objects (see \citealt{Stevens2019} for an extensive discussion on the topic).

The distribution of HI mass fractions in our new model closely match those observed showing a strong decline with increasing stellar mass. The agreement shown in Fig.~\ref{fig:gas_fraction} is significantly better than for any of the models shown in \citet{Saintonge2017} --- which included the MUFASA and EAGLE hydro simulations and the GALFORM and Santa Cruz semi-analytic models \citep{Gonzalez2014, Popping2014, Lagos2015, Dave2016, Lacey2016} --- and is consistent with the recent results from IllustrisTNG presented in \citet{Stevens2019}. In contrast, $\Htwo$ fractions in the model are significantly higher than those observed, particularly at high-mass. While the $\Htwo$ sharply drops above $\log_{10} (M_*/\Msun) = 10.5$ in observations, our model shows a much 
less marked decrease. These objects host the most powerful AGN and it is plausible that their radiation will have some impact on the gas of their host galaxies, in particular, that it can lead to $\Htwo$ destruction.

\begin{figure}
\centering
\includegraphics[width=8.4cm]{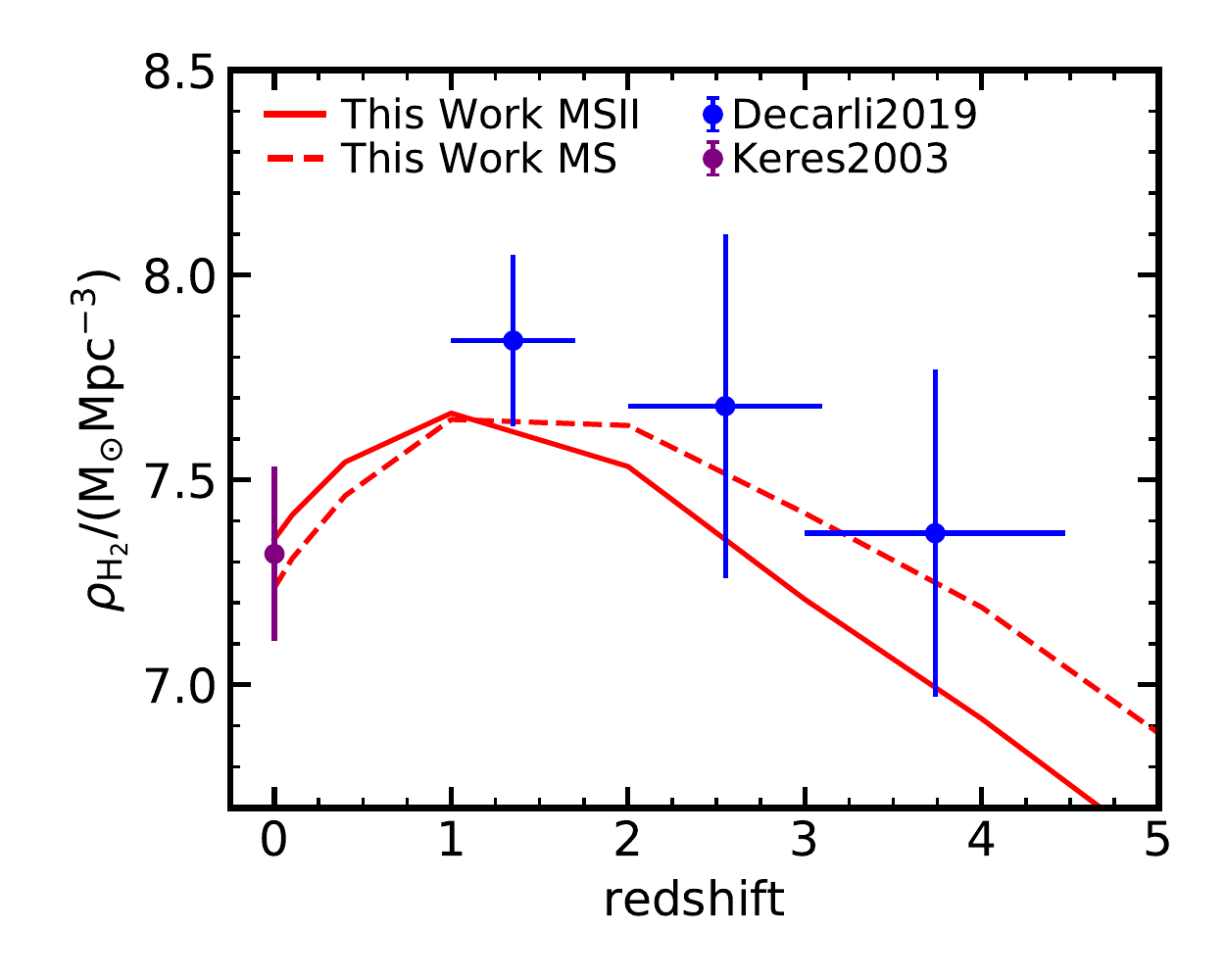}
\caption{Evolution with redshift of the comoving density of molecular hydrogen. Our model results based on the Millennium (dashed red line) and Millennium-II (solid red line) simulations are compared with observational data from \citet{Keres2003} and \citet{Decarli2019}.}
\label{fig:h2d}
\end{figure}

We end this section on gas properties by extending the comparison between model and observation to high redshift. Fig~\ref{fig:h2d} shows model results for the evolution of the cosmic mean density in  $\Htwo$ from $z=0$ to 5 in the MS (dashed red line) and MS-II (solid red line) simulations and compares it with observational data from \citet{Keres2003} (purple symbol) and \citet{Decarli2019} (blue symbols). The local derivation from \citet{Keres2003} was obtained using the FCRAO Extragalactic CO Survey with sampling corrections derived by matching it to FIR-selected data from the IRAS Bright Galaxy Surveys. The normalized CO luminosty functions were then converted to a molecular-gas density using a fixed conversion factor: $N(H2)/I(CO) = 3 \times 10^{20}\;\mathrm{cm}^{-2}[ \mathrm{K\;km\;s}^{-1}]^{-1}$.

The \citet{Decarli2019} data was obtained using the ALMA large program ASPECS, the spectroscopic survey in the Hubble Ultra Deep Field (HUDF), covering a total area of $4.6\; \mathrm{arcmin}^2$. The broad frequency range covered allows the identification of CO emission lines of different rotational transitions at $z>1$ which are corrected for CO excitation to infer the corresponding CO(1-0) luminosities. The authors then construct CO luminosity functions by splitting the line candidates by CO transitions in bins of luminosity and dividing their number counts by the comoving volume (given by counting the area with sensitivity >50\% of the peak sensitivity obtained at the center of the mosaic in each channel. These number counts take into account uncertainties in the line flux estimates, in the line identification, in the conversion factors, as well as for the fidelity of the line candidates. These are then reflected in the vertical error bars shown in Fig~\ref{fig:h2d}, while the horizontal error bars represent the binning in redshift.

Finally, H2 masses are obtained by scaling the CO(1-0) luminosities by a fixed conversion factor, $\alpha_{\mathrm{CO}}=3.6 \Msun (\mathrm{K\;km\;s^{-1}\;pc^2})^{-1}$ \citep{Daddi2010}, consistent with the galactic value \citep{Bolatto2013}. The completeness-corrected $\Htwo$ masses of each line candidate that passes the fidelity threshold are then added up in bins of redshift and divided by the comoving volume in order to derive the cosmic molecular-gas density. These densities do not take into account any extrapolation of the mass function below the detection limit. However, the true detections account for between 70 and 90\% of the luminosity-weighted integral of the fitted luminosity functions in the redshift range considered, so that the $\Htwo$ comoving densities should, at maximum, underestimate the true value by 10-30\%.

Our model correctly captures the overall observed trend with redshift, although with an apparent deficit of $\Htwo$ at the highest redshifts and an excess at $z\sim 0$ (particularly for MS-II based results). The latter is consistent with the excess in star formation seen in Fig.~\ref{fig:sfrd} and both seem to be a consequence of the excessive $\Htwo$ content, particularly of high-mass galaxies, seen in Fig.~\ref{fig:gas_fraction}. A similar excess was identified in \citet{Saintonge2017} for a variety of hydro simulations and semi-analytic models except when post-processing EAGLE with the \citet{Krumholz2013} prescription. The latter takes into account the HI column density when partitioning the ISM.

\begin{figure}
\centering
\includegraphics[width=8.4cm]{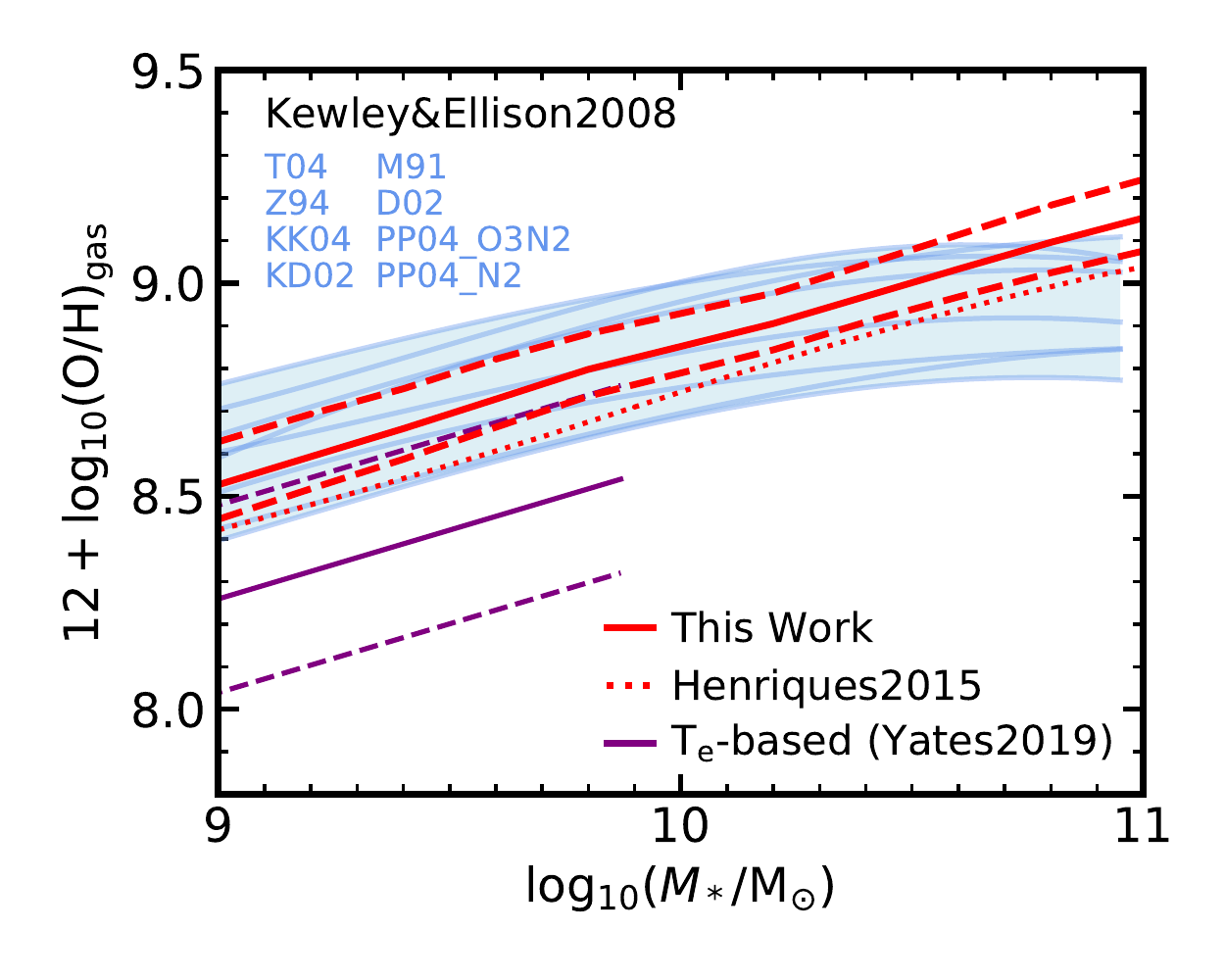}
\caption{Oxygen to hydrogen abundance ratio versus stellar mass. Our new model results (solid and dashed red lines indicating the mean, 16th and 84th percentiles of the distribution) are compared with those from \citet{Henriques2015} (dotted red line), with the compilation of observational estimates in \citet{Kewley2008} and with more recent T$_e$-based measurements from \citet{Yates2019} (Solid and dashed purple lines indicating the mean and 1$\sigma$ scatter).}
\label{fig:MZgR_z0}
\end{figure}

\section{Comparison with observed galaxy properties not used as constraints}
\label{sec:predictions}

\subsection{Metals}
\label{sec:metals}

In the following subsections, we outline our results on the metal content in the cold gas, stars, and hot gas of our model galaxies at $z=0$. A more detailed investigation into chemical enrichment, including the chemical evolution of metallicities over cosmic time, will be presented in a follow-up paper (Yates et al., in prep.). In order to mimic the observational selection, we restrict our comparison to star-forming galaxies for both cold gas and stellar metallicities (defined as galaxies not more than 1 dex below the main sequence, see Appendix~\ref{app:cuts} for details). 

\subsubsection{Gas-phase mass-metallicity relation}
\label{sec:MZgR}

Fig.~\ref{fig:MZgR_z0} shows the relation between galaxy stellar mass and the SFR-weighted oxygen abundance\footnote{Due to the independent tracking of eleven different chemical elements in \lgal, the ISM oxygen abundance can be directly estimated as the number density of oxygen atoms to hydrogen atoms in the cold gas phase, 12+log(O/H).} in the cold gas (the MZ$\sub{g}$R) at $z=0$. For our new model (solid and dashed red lines), we calculate oxygen abundances by first measuring the number-density ratio of oxygen to hydrogen atoms in each of the gas disc rings, and then taking the SFR-weighted mean of these `local' oxygen abundances as the overall metallicity of the galaxy. This is done in order to mimic the emission-weighting of observed metallicities, which are dominated by bright, more metal-rich H\textsc{ii} regions. We also impose an aperture limit of 9 kpc to replicate that of SDSS measurements in the redshift range $0.04 \lesssim z \lesssim 0.1$ used to form the \citet{Kewley2008} MZ$\sub{g}$Rs. However, this aperture correction does not significantly affect our SFR-weighted oxygen abundances (see also \citealt{Tremonti2004}), as the majority of the star-forming gas in these systems is found at smaller radii (see Fig.~\ref{fig:milkyway}). For the \citet{Henriques2015} model relation (dotted red line), we simply convert the cold gas metallicity to an oxygen abundance using $12+\log_{10}(\mathrm{O/H})_{\mathrm{gas}}=\log_{10}(Z_{\mathrm{gas}}/\mathrm{Z_{\odot}})+8.69$.

Model results are compared with observational data from \citet{Kewley2008}, who used a variety of strong-line-based calibrations to derive Z$\sub{g}$, and with measurements based on electron temperature ($T\sub{e}$) from \citet{Yates2019}. It is well known that these two methods can yield MZ$\sub{g}$Rs that are offset in normalisation by up to $\sim{}0.7$ dex, while the scatter among the strong-line-based relations alone is $\sim{}0.4$ dex (giving an indication of the broad range of metallicities observed in local extragalactic H\textsc{ii} regions)\footnote{Although the $T\sub{e}$ method is believed to be a more direct probe of the true oxygen abundance, it is currently limited to lower-mass systems, primarily because the auroral emission lines required are very faint and their strength is inversely proportional to metallicity (see \eg{}\citealt{Peimbert1967,Bresolin2008})}. The distribution of gas-phase metallicities in our new model, as in \citet{Henriques2015}, is within the observational range spanned by the \citet{Kewley2008} compilation. The slope of the relation is also well matched at low masses but it seems steeper than observed at $\log_{10} (M_*/\Msun) > 10.0$. If the flattening of the observed relation at high mass were caused by recent accretion of a small amount of low-metallicity gas (see also \citealt{Yates2014}), this would be consistent with the levels of star formation in our massive galaxies being too high.

\begin{figure}
\centering
\includegraphics[width=8.4cm]{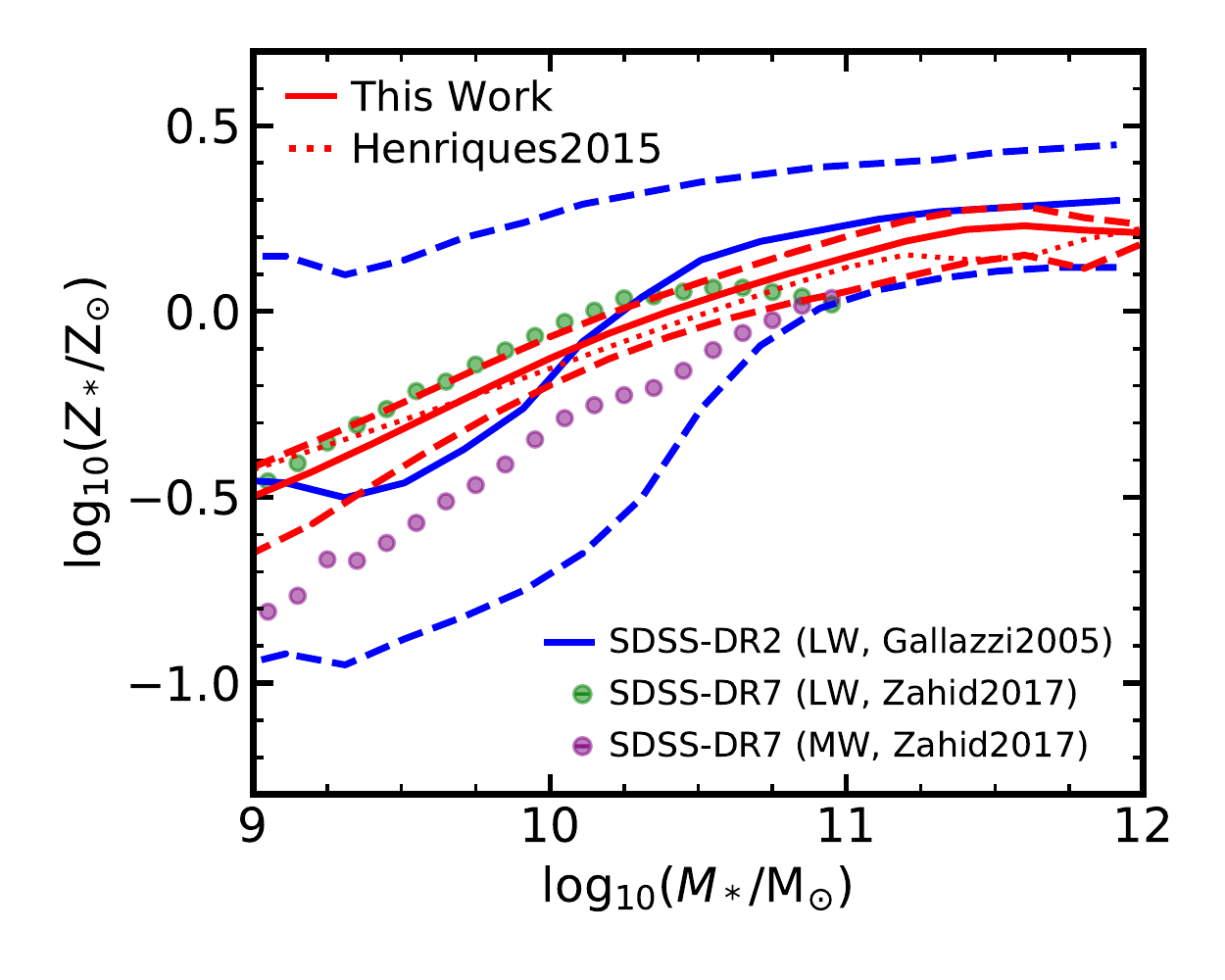}
\caption{Stellar metallicity as a function of stellar mass. Results from our new model (solid and dashed red lines indicating the mean, 16th and 84th percentiles of the distribution) are compared with those from \citet{Henriques2015} (dotted red line) and with observational data from \citet{Gallazzi2005} (light-weighted, solid and dashed blue lines) and \citet{Zahid2017} (purple and green symbols, respectively for mass- and light-weighted measurements).}
\label{fig:MZsR_z0}
\end{figure}




\subsubsection{Stellar mass-metallicity relation}
\label{sec:MZsR}

Fig.~\ref{fig:MZsR_z0} shows the relation between stellar mass and stellar metallicity (the MZ$_{*}$R) at $z=0$ for star-forming galaxies in our new model (the solid and dashed red lines) and for that of \citet{Henriques2015} (the red dotted line). For comparison, observational results from \citet{Zahid2017}, are shown as green and purple circles (for luminosity-weighted and mass-weighted measurements, respectively). These are derived using full spectral fitting of $M_{*}$-binned stacks of $\sim{}200,000$ SDSS-DR7 and utilise stellar absorption lines to determine $Z_{*}$. In addition, the figure shows measurements based on Lick indices from \citet{Gallazzi2005} (solid and dashed blue lines) for $\sim{}175,000$ SDSS-DR2 galaxies. All stellar metallicities are normalised here to the proto-solar metal abundance as measured by \citet{Asplund2009}, $\Zsun = 0.0142$.

The \citet{Gallazzi2005} dataset incorporates galaxies of all SFRs, including passive systems (especially at higher mass) that tend to have higher $Z_{*}$ than star-forming galaxies of the same mass (see \eg{}\citealt{Okamoto2017}). The data provided by \citet{Zahid2017}, which only consider star-forming systems, are therefore likely a fairer comparison to our model relation.

Overall, we find a good correspondence between our model results and the SDSS data over the three orders of magnitude in stellar mass considered. At low masses, the slope of our model MZ$_{*}$R matches that of the \citet{Zahid2017} relations fairly well, although the normalisation is a factor of $\sim$2 above the observed mass-weighted relation. Nonetheless, it is encouraging to see convergence to a similar value of $Z_{*}$ at higher mass, where star-formation histories, weighting differences, and spectral fitting issues are all likely to be better constrained in observations.

Finally, we note that the large metallicity deficit previously noted by \citet{Hirschmann2016} is not present in any of our models, despite the strong SN feedback removing metals in low-mass galaxies.

\begin{figure}
\centering
\includegraphics[width=8.4cm]{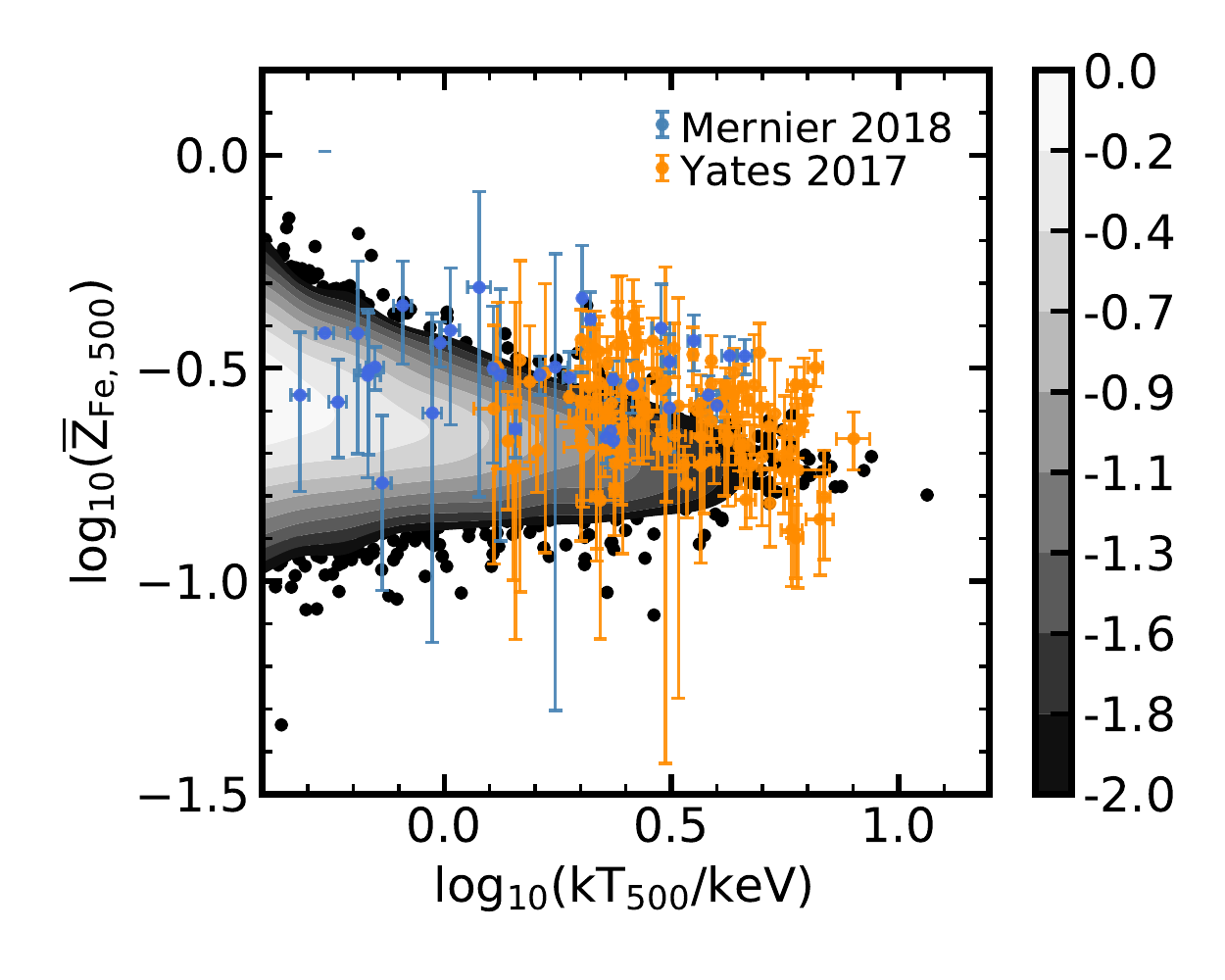}
\caption{Mean, mass-weighted ICM iron abundance within $r_{500}$ for galaxy groups and clusters, as a function of the ICM temperature at $r_{500}$. Model systems (normalised number density of galaxies in logarithmic grey-scale contours) are plotted alongside observational galaxy cluster data from \citeauthor{Yates2017} (2017, orange points) and galaxy group \& cluster data from \citeauthor{Mernier2018b} (2018b, light blue points).}
\label{fig:T-Z_ICM}
\end{figure}

\subsubsection{Metal content in the ICM}
\label{sec:ICM metals}
The iron abundance in the hot, X-ray emitting gas surrounding galaxy groups and clusters is well reproduced by our new model. Fig.~\ref{fig:T-Z_ICM} shows model results for the relation between the temperature of the hot ICM at $r_{500}$, $T_{500}$ (a proxy for the gravitating mass), and the mean iron abundance within $r_{500}$ (the black points). Also shown are the observational dataset from \citet{Yates2017} for clusters (orange points), and from \citet{Mernier2018a} for clusters and groups (light blue points). 

Observationally, $T_{500}$ is obtained from measurements of the mean emission-weighted ICM temperature within some aperture, while in the model it is derived from  a standard calculation of $T_{200}$. Both are then scaled to $r_{500}$ assuming either the \citet{Vikhlinin2006} (for clusters) or \citet{Rasmussen2007} (for groups) temperature profile (see \citealt{Yates2017} section 2.3, for details). The theoretical value for the hot-gas temperature is given by $T_{200} = \mu m_{\mathrm{p}} \sigma_{200}^2/k$, where $\mu m_{\mathrm{p}}$ is the average mass of the particles in the ICM ($m_{\mathrm{p}}$ is the proton mass and $\mu$=0.58), $k=8.6173\times 10^{-8}$KeV/K is Boltzman's constant and $\sigma_{200}$ is calculated dynamically from the mass of the cluster assuming an NFW profile.

As in the \citet{Yates2017} analysis, we have homogenised and re-scaled all observed temperature and iron abundance measurements to provide a consistent comparison between the model and the various observations. Here, we further refine this technique by including two modifications. First, we update the $Z\sub{Fe}$ radial profile assumed when re-scaling all measurements to that presented by \citet{Mernier2017}, which is based on a set of 44 groups and clusters from the CHEERS sample \citep{dePlaa2017} observed with \textit{XMM-Newton}. This CHEERS analysis found very similar profile shapes for groups and clusters. We therefore fit a beta-model to the \citet{Mernier2017} `full sample' stacked data (their table 2) and apply this  $Z\sub{Fe}$ profile to all systems, allowing the normalisation to vary for each, depending on the total iron abundance measured. The same procedure is also applied to our model systems, which are assumed to have a  \citet{Mernier2017} profile when re-scaling the iron abundance from within $r_{200}$ to within $r_{500}$ (see Section 5 of \citealt{Yates2017} for details.)

Second, we take account of recent improvements to the way iron abundances are obtained from the X-ray spectra of galaxy groups \citep{Mernier2018a} by excluding from our comparison the observational data for galaxy groups compiled by \citet{Yates2017} that relied on $Z\sub{Fe}$ estimates obtained from older spectral-fitting codes. For clusters, \citet{Mernier2018a} found little difference in the $Z\sub{Fe}$ estimates obtained from older and newer versions of SPEXACT. Therefore, we choose to include both the CHEERS dataset and that of \citet{Yates2017} in this higher-temperature regime. For a more detailed investigation into the differences these two improvements make to the modelling of the ICM metallicity in \lgal{}, see our follow-up paper, Yates et al., in prep.

Our analysis of the iron abundance of the intragroup and intracluster medium within $r_{500}$ suggests that the mean $Z\sub{Fe}$ varies little with temperature over around 2 orders of magnitude in gravitating mass (see also \citealt{Biffi2018,Mernier2018b}). We also find an increasing scatter around in $Z\sub{Fe}$ with decreasing temperature in the model. Both the slope and the normalisation of the observed relation are well reproduced by \lgal{}, without the need for a variable or top-heavy IMF, for excessive amounts of SNe-Ia per stellar population, or for enhanced iron yields from supernova nucleosynthesis.

\begin{figure}
\centering
\includegraphics[width=8.4cm]{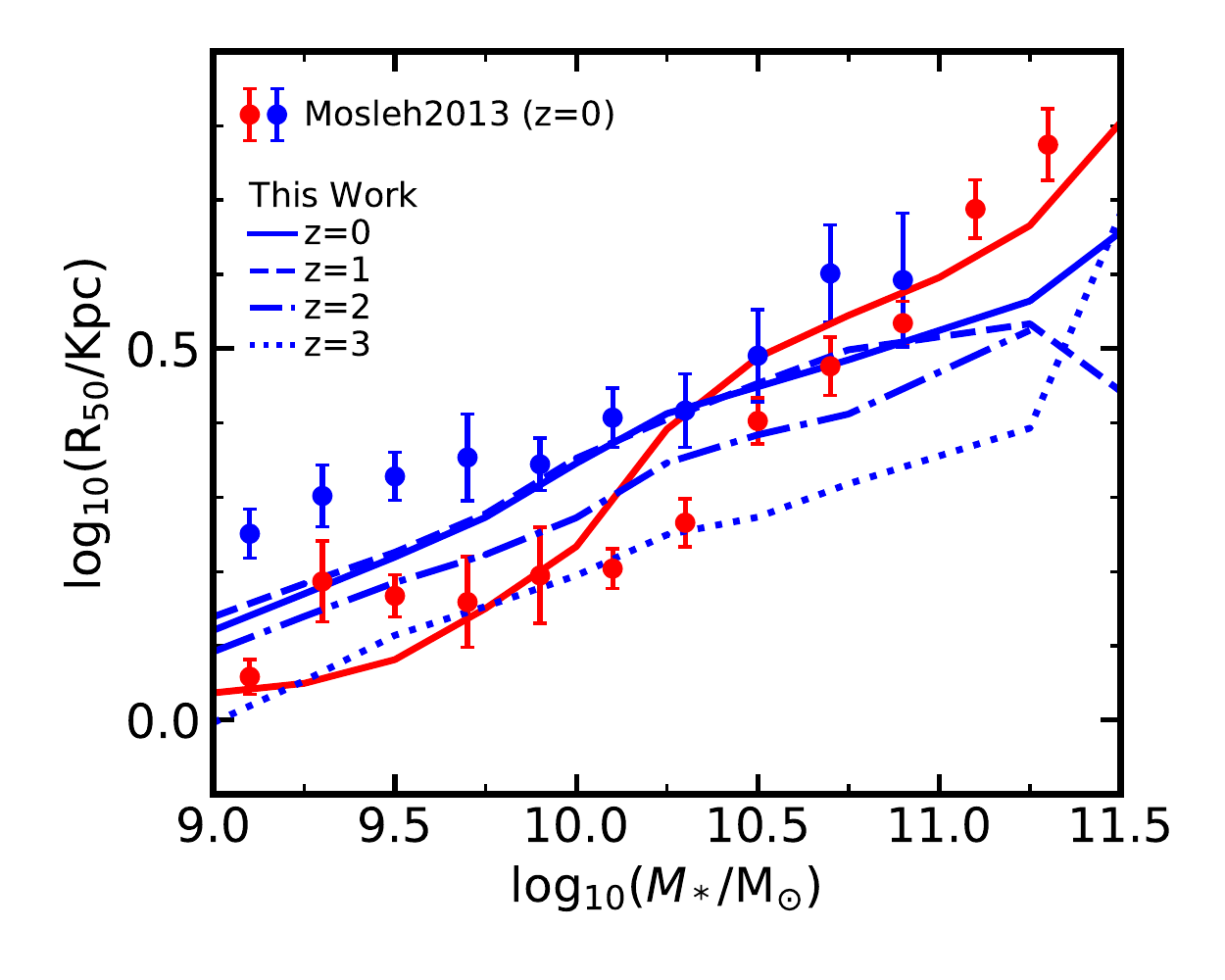}
\caption{Half-mass radius versus stellar mass for disc-dominated (blue) and bulge-dominated (red) galaxies. Model results based only on the Millennium-II simulation (solid, dashed and dotted lines at $z=0$, 1, 2 and 3) are compared with $z\sim 0$ data from \citet{Mosleh2013}}.
\label{fig:sizes}
\end{figure}

\subsection{Sizes and morphology}
\label{sec:sizes}
Although we treat disc instabilities and mergers in a similar way to \citet{Henriques2015}, we now track the evolution of the radial structure of discs explicitly and separately for their stellar and cold gas components. As a result, the sizes of galaxies may be expected to differ significantly from those found in earlier models. In this context, it is particularly interesting to look at size differences between star-forming and quenched galaxies. At given halo mass, dark matter haloes are more compact at earlier times because of the higher mean cosmic density. For standard assumptions about angular momentum conservation, the galaxy discs within them are then expected to be smaller on average \citep[e.g.][]{Mo1998}, suggesting that quenched galaxies should be smaller than star-forming ones of the same stellar mass, since their sizes will be similar to those of  star-forming galaxies at the time their stars were formed \citep{Lilly2016}. 

In Fig.~\ref{fig:sizes} we plot the median half-mass radii for star-forming galaxies in our new model (defined as galaxies not more than 1 dex below the main sequence, see Appendix~\ref{app:cuts} for details) at redshifts $z=0,$ 1, 2 and 3 as a function of their stellar mass. As expected, galaxies of given stellar mass are smaller at higher redshift, but the trend is quite weak, indeed, almost absent between the present and $z=1$. The red solid line shows the theoretical size-mass relation at $z=0$ for quenched galaxies (galaxies more than 1 dex below the main sequence). At low stellar masses, quenched galaxies in the model are smaller than star-forming ones, but the relation is steeper for quenched than for star-forming galaxies, so that above about $10^{10}\Msun$ quenched galaxies are typically larger.  For comparison, Fig.~\ref{fig:sizes} also shows observational data at $z\sim 0$ from \cite{Mosleh2013} --- for both types of galaxy the agreement with the model is relatively good over the full stellar mass range. 


\begin{figure}
\centering
\includegraphics[width=8.4cm]{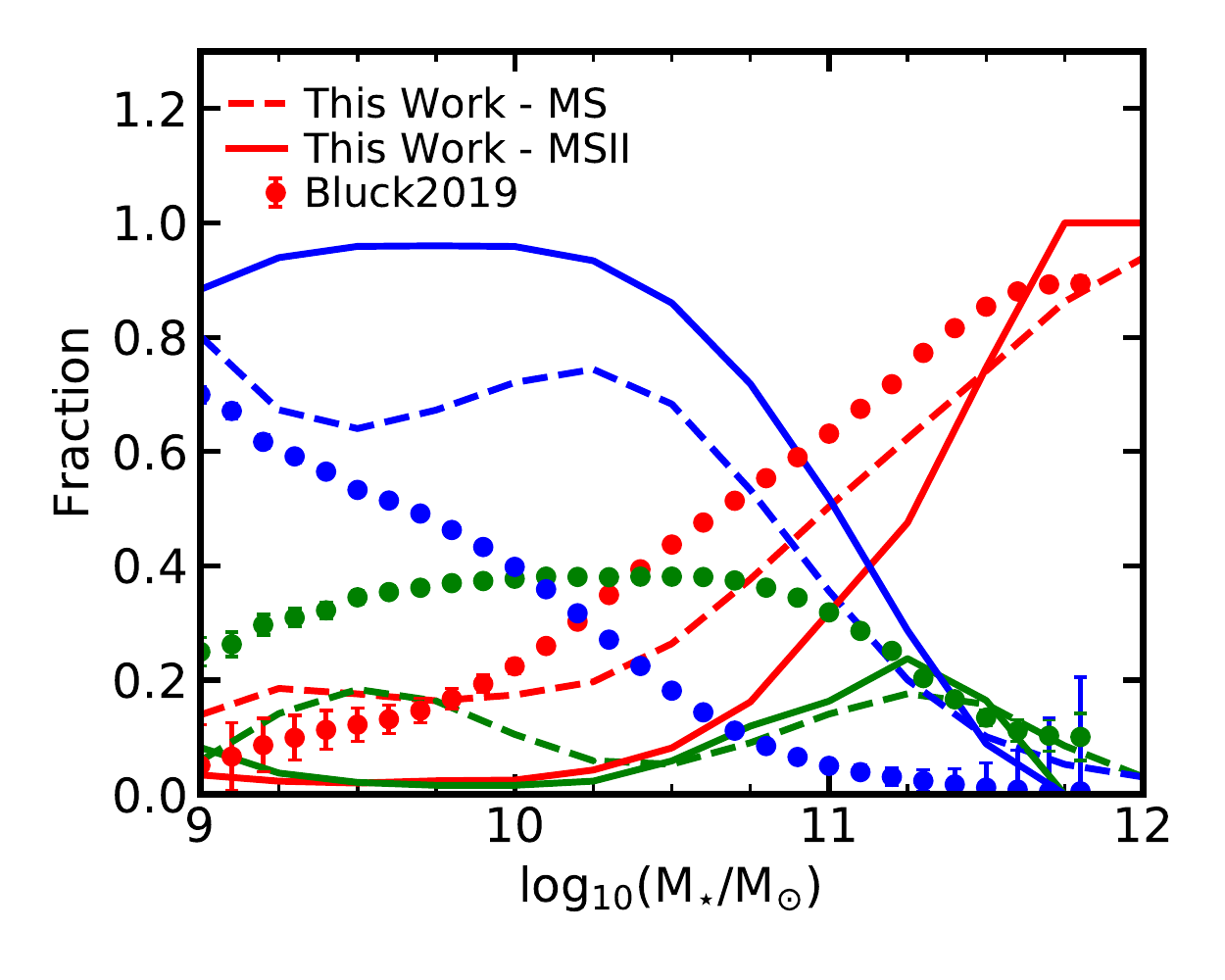}
\caption{Fraction of disc-dominated ($B/T<0.3$, blue), 
intermediate ($0.3\leq B/T\leq 0.7$, green) and bulge-dominated ($0.7<B/T$, red) galaxies as a function of stellar mass. Our new model results based on the Millennium (dashed lines) and Millennium-II simulations (solid lines) are compared with observational data from \citet{Bluck2019}.}
\label{fig:morphology}
\end{figure}

In addition to sizes, the relative importance of bulges and discs is also an important constraint on our new model. Fig.~\ref{fig:morphology} shows the fraction of bulge-dominated ($B/T > 0.7$, red), intermediate ($0.3 \leq B/T \leq 0.7$, green) and disc-dominated ($B/T < 0.3$, blue) galaxies as a function of stellar mass. Model results based on the MS-II (solid lines) and MS (dashed lines) are compared with observational data from \citet{Bluck2019}. As seen in observational data, bulge-dominated galaxies are rare at low mass and dominant at high mass in the model, but the latter severely underestimates the number of objects with intermediate B/T ratios (this was also clear in the \citealt{Bluck2019} analysis). This is particularly clear in the Millennium-II simulation, which has high enough resolution to accurately track the assembly histories of low-mass galaxies. A similar problem was already present in the models of \citet{Guo2011} and \citet{Henriques2015}. A related discrepancy is that the fraction of bulge-dominated galaxies exceeds 50\% only for $\log_{10} (M_*/\Msun) > 11.25$ in the model, whereas the observational data suggest that this should be the case for $\log_{10} (M_*/\Msun) > 10.6$. Although significant uncertainties are introduced into this comparison by the observational difficulty in measuring $B/T$ from photometric data, it seems clear that the bulge-formation mechanisms in our model need to be revised (currently bulges are produced only by mergers with stellar mass ratios exceeding 0.1). 

A more realistic implementation of disc instabilities, for example accounting for the impact of gas, will likely result in larger spheroidal components and alleviate this tension. The phenomenon has been independently studied in the context of the \citet{Henriques2015} version of our model by \citet{Irodotou2019} and \citet{Izquierdo-Villalba2019}; both papers find a significant population of pseudo-bulges at intermediate masses.  We plan to incorporate these developments in future version of our model. Similar approaches have been adopted by other models \citep{Tonini2016, Stevens2016, Lagos2018} also finding improved agreement with observed galaxy morphologies. 

\begin{figure}
\centering
\includegraphics[width=8.4cm]{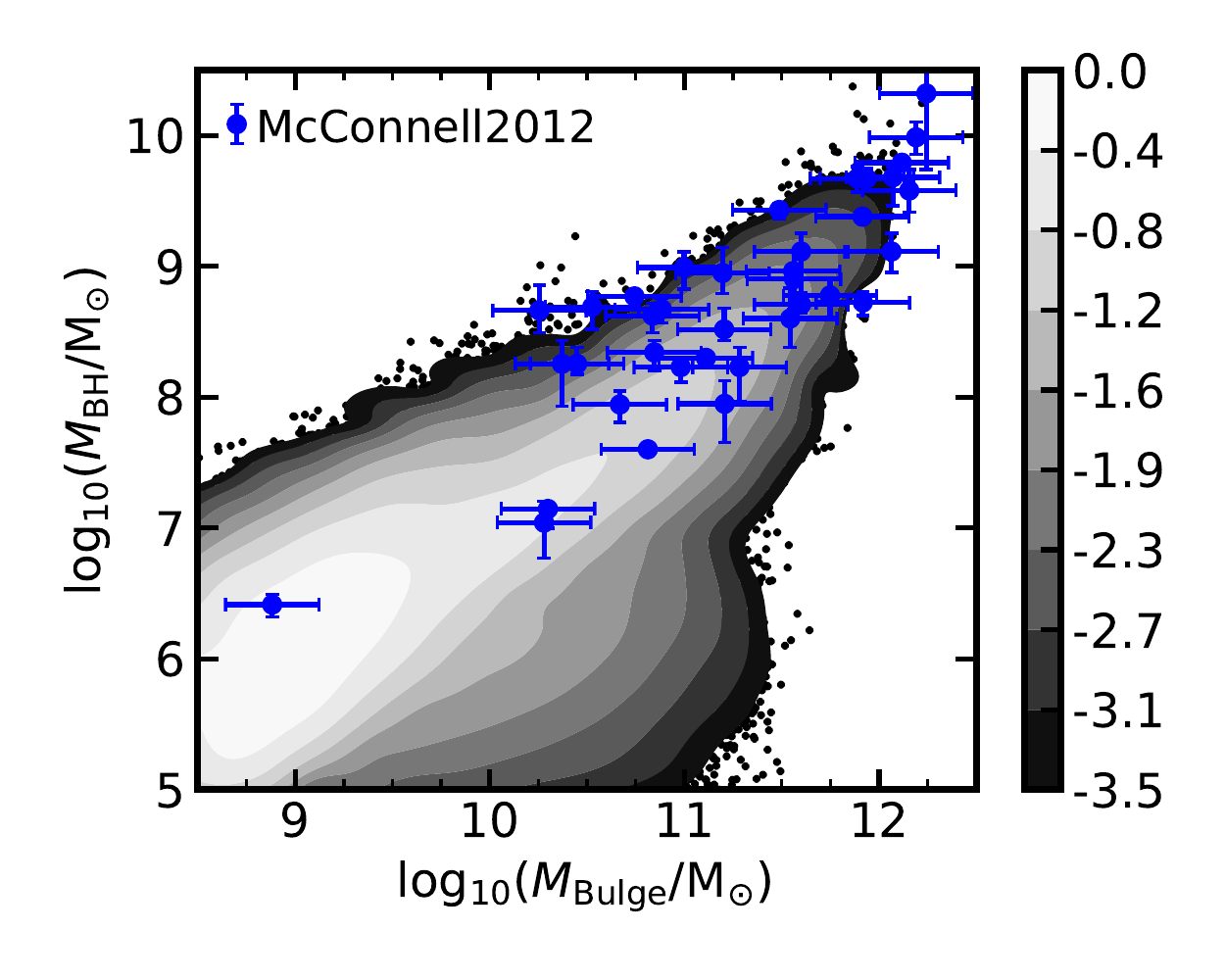}
\caption{The black hole mass --  bulge mass relation at $z=0$. Results from the current model (normalised number density of galaxies shown as logarithmic grey-scale contours) are compared with observational data from \citet{McConnell2013}. Model galaxies in regions with density below the lowest level contour are shown as black circles.}
\label{fig:bhbm}
\end{figure}

\begin{figure*}
\centering
\includegraphics[width=12cm]{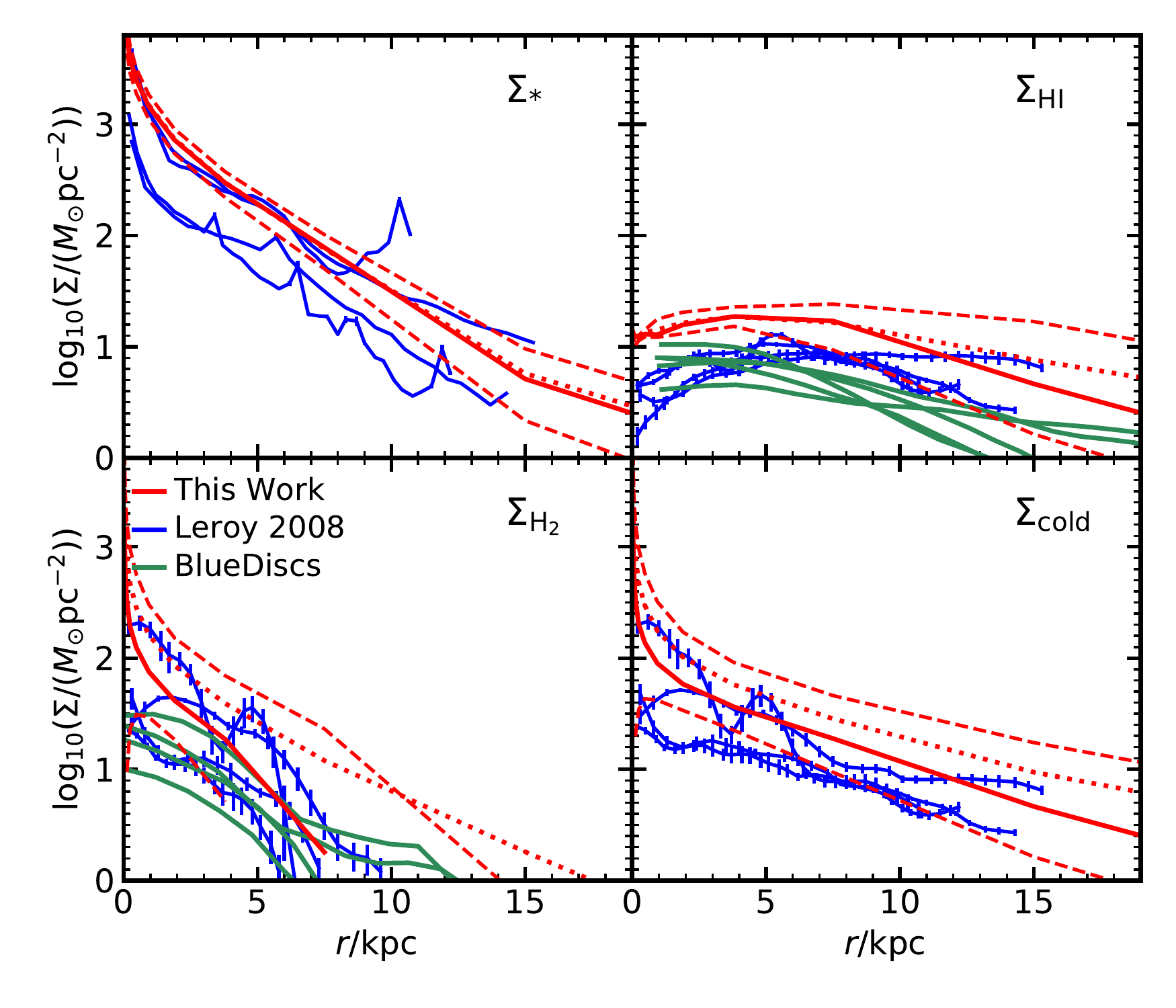}
\caption{Radial profiles of stellar (top left), HI (top right), $\Htwo$ (bottom left) and cold gas (bottom right) surface densities for Milky Way-like galaxies. The median (solid red lines), 16th and 84th percentiles (dashed red lines) and mean (dotted red lines) of the distribution for our new model are compared with observational data from \citet{Leroy2008} (solid blue lines) and the \textsc{bluedisk} survey (solid green lines).}
\label{fig:milkyway}
\end{figure*}

\subsection{Black Holes and AGN}
\label{sec:BH}

The final global galaxy property we will analyze is black hole mass, in particular, its relation with bulge mass. Black holes are a crucial part of our galaxy formation model since they provide the necessary feedback to suppress cooling in massive haloes that leads to the quenching of  star formation in the galaxies at their centres. 
This relation is shown in Fig.~\ref{fig:bhbm} where model results (grey contours) are compared with observational data from \citet{McConnell2013} at $z=0$. The tight relation between black hole and bulge mass means that very massive, bulge-dominated galaxies will host the most massive black holes. In the case of our model, and as shown in \citet{Henriques2019}, it is, in fact, the strong feedback from the central black hole that results in suppression of star formation in discs, allowing massive galaxies to grow primarily by mergers and become bulge-dominated. As in previous versions of the L-Galaxies model, the observed relation  between black hole mass and bulge mass is quite well reproduced.



\section{Spatially resolved properties}
\label{sec:gradients}

In previous sections we have demonstrated that our spatially resolved model --- of cold gas partition, star formation and delayed mass return form stars and SNe --- still provides a fair representation of the observable Universe in terms of global galaxy properties. We will now fully exploit its capabilities by extending our comparison to the radial variation of properties in observed discs. This is particularly relevant at the moment, since a new generation of multi-object integral-field units is producing complete and well defined samples of $\sim 10^4$ galaxies with spatially-resolved spectroscopic data. In this paper we will focus on stellar and cold gas density profiles as well as stellar and cold gas metallicity profiles. The latter will be analyzed in significantly more detail in a companion paper (Yates et al., in prep.).

When plotting quantities for entire galaxies (and not just their discs), we assume that mass in bulges is distributed according to a Jaffe density profile \citep{Jaffe1983} of the form: $1/x^2(1+x)^2$, where $x=r/r_b$ (and $r_b$ is the half-mass radius of the bulge), resulting in $\Mstar\propto x/(1+x)$. This form accurately reproduces a de Vaucouleurs profile when projected onto the sky.

\subsection{Profiles for Milky-Way like galaxies}
\label{sec:Milky-Way}

Fig.~\ref{fig:milkyway} shows stellar surface density (top left), HI surface density (top right), $\Htwo$ surface density (bottom left) and total cold gas surface density (bottom right) profiles for Milky-Way-like galaxies. These are defined as star-forming objects,  with disc-like morphologies ($M_{\rm{Bulge}}/M_*<0.15$), virial velocities in the range $200<V_{\rm{vir}}/$km\,s$^{-1}<235$ and stellar masses in the range $10.3<\log_{10} (M_*/\Msun)<10.7$. Since these are the type of objects where most of the star formation in the Universe occurs, their radial gas profiles represent a critical test to our new spatial-resolved model. Solid and dashed red lines show the median and 16th+84th percentiles, respectively, while dotted red lines show the mean for the simulated galaxies. Observations from \citet{Leroy2008} are shown as blue lines with error bars while data from the \textsc{bluedisk} survey are shown as solid green lines \citep{Wang2014b, Cormier2016}.

\begin{figure*}
\centering
\includegraphics[width=14cm]{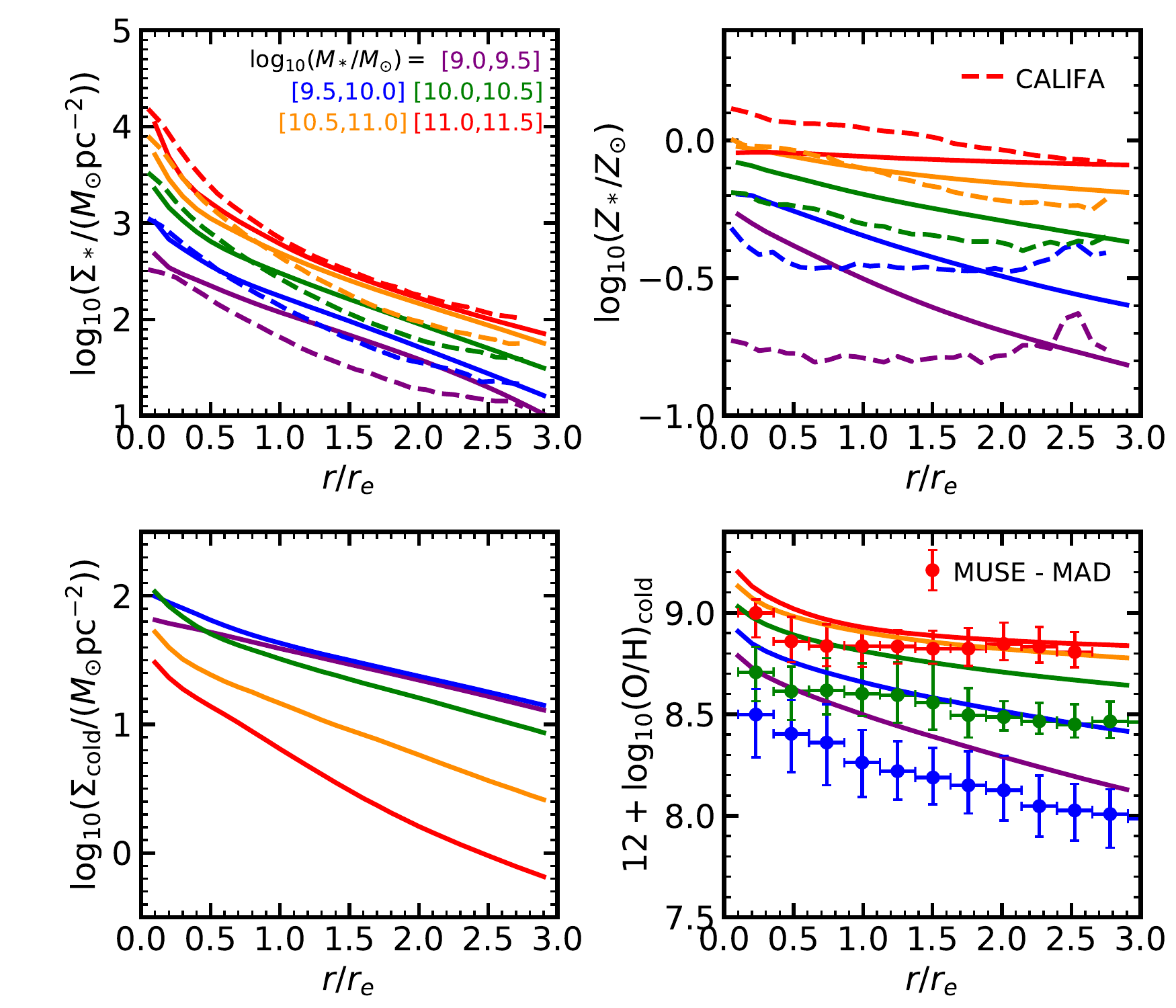}
\caption{Radial distribution of stellar (top left) and cold-gas (bottom left) surface densities and stellar (top right) and gas-phase (bottom right) metallicities in our new model for galaxies in different stellar mass bins (solid coloured lines). Stellar surface densities and metallicities are compared with observational data from CALIFA \citep{Delgado2015} and gas-phase metallicities are compared with observations from the MUSE Atlas of Discs \citep{Ferrer2019}.}
\label{fig:all_gradients}
\end{figure*}

Overall, results from our new model match the slope of the observed profiles for Milky-Way-like galaxies reasonably well. In particular, our new model displays centrally concentrated $\Htwo$ profiles and relatively flat HI distributions. This is a direct consequence of having cold gas predominantly in the form of $\Htwo$ in the highest surface-density regions. Nevertheless, the overall normalization of both stellar and cold gas density profiles seems excessive in the new model. This is consistent with the excessive levels of star formation found for these relatively massive galaxies in Fig.~\ref{fig:ssfr_hist} and seems to indicate that the complete shutdown of cooling by AGN feedback is not sufficient to reduce the cold gas content of these objects to the observed level.

While the scale length of gas profiles is strongly affected by the parameter controlling the velocity of inflowing gas in discs, the latter is only weakly constrained by the properties used to calibrate our new model. This means that its value will leave properties like the stellar mass function and red fractions relatively unchanged while dramatically affecting the radial extent of different gas components --- notwithstanding the fact that excessively decreasing this inflow velocity can reduce the amount of dense cold gas in galactic centres and suppress star formation. We therefore do not include this parameter in our MCMC exploration and fix it {\it a priori} at a value of $v_{\mathrm{inflow}}=1.0\,$km$\,$s$^{-1}$kpc$^{-1}$. This ensures that the cold gas content is extended enough in Milky-Way-like galaxies, while still allowing enough material to flow into the centres of massive galaxies and to be transformed into stars at early enough times. \citet{Stevens2016} implemented a local gas partition recipe in their semi-analytic model and found good agreement with similar radial profiles to those shown in Fig.~\ref{fig:milkyway}. Instead of relying on a constant radial inflow of gas, the authors implement a parametrised disc-instability recipe that effectively moves material inwards when too much gas is accumulated in a given ring.

\subsection{Comparison with MaNGA and CALIFA population gradients}
\label{sec:manga}

Fig.~\ref{fig:all_gradients} shows results from our new model for the radial distribution of stellar and gas surface densities and for stellar and gas-phase metallicities in different stellar mass bins (solid coloured lines). Stellar surface densities and metallicities are compared with observations data from CALIFA \citep{Delgado2015} and gas-phase metallicities are compared with observations from the MUSE Atlas of Discs \citep{Ferrer2019}. 

The comparison in the top left panel shows that our new model roughly captures the observational trend of increasing stellar surface density with decreasing radius and increasing stellar mass. Nevertheless, at large radii the model overestimates the surface density when compared to the CALIFA sample, particularly at low masses. The radial distribution of cold gas surface density in the bottom-left panel shows an interesting pattern of increasing normalization up to $\log_{10} (M_*/\Msun) \sim 10.0$ and decreasing normalization for larger stellar masses. This is a direct consequence of our implementation of AGN feedback, which suppresses cooling and subsequently star formation above this characteristic mass. 

The two panels on the right show similar patterns for the theoretical distribution of stellar and gas-phase metallicites. The radial distribution in our new model is relatively flat for massive galaxies (orange and red lines) and increasingly steeper for low-mass galaxies (blue and purple lines). This is consistent with the observed trends for gas-phase metallicities seen in MUSE (bottom-right panel), but not with the flatter stellar metallicity gradients found in CALIFA for low-mass galaxies (top-right panel). In combination, the two observational studies seem to indicate that the radial distribution of metals in low-mass galaxies is very steep for the gas-phase and almost flat for stars. Although it is possible to have different trends for current gas-phase metallicities and cumulative stellar-phase metallicities, it is less likely for young, star-forming objects that dominate the low-mass population (Fig.~\ref{fig:ssfr_hist}). One possible explanation is that there is some preferential selection of passive low-mass objects in CALIFA or that most of the mass formed at earlier times in low-mass galaxies followed a flat or negative gradient.

\section{Summary and conclusions}
\label{sec:conclusions}

We have updated the \citet{Henriques2015} version of the Munich galaxy formation model, \lgal, in order to follow the spatial variation of the properties of galactic discs. We explicitly track different cold gas phases, we include a H$_2$ based star formation law (all following \citealt{Fu2013}), and we incorporate the detailed chemical-enrichment model of \citet{Yates2013}. The new model stores radial information  about the stellar, molecular and atomic gas content of discs, as well as star formation histories, feedback and chemical-enrichment histories for 11 different elements in 12 concentric rings.

By introducing a significant additional layer of complexity, our new modelling approach allows direct comparison with the kpc- and sub-kpc-scale observations of nearby galaxies being provided by modern multi-object integral field unit (IFU) spectrographs \citep{Bacon2010, Croom2012, Sanchez2012, Bundy2015}. In addition, the ability to track the spatial structure of galactic discs is critical for modelling the transition between atomic and molecular gas, as well as the effects of the energy released from stars. Our new model uses the surface density of cold gas to calculate the relative abundance of atomic and molecular hydrogen, with the latter determining the rate of star formation. The spatially resolved star formation history is then used to compute the radial distribution of the mass, energy and nucleosynthesis products released at later times. 

Despite incorporating a significantly more detailed treatment of the physical processes shaping galactic discs, our new model maintains the successes of its predecessors in representing the observed evolution of key global galaxy properties. These include the stellar mass function, the red fraction as a function of stellar mass; the  evolution of the star forming main sequence (from $z=0$ to 3); the cosmic SFRD from $z=0$ to 10; the sSFR distribution as a function of stellar mass at $z=0$; the metal content of the cold gas, stellar and hot gas components at $z=0$; the sizes and morphologies of local galaxies as a function of their stellar mass; and the black hole-bulge mass relation. As in previous models, the observed $z=0$ distribution of bulge fractions is not well reproduced with the model producing too few objects with intermediate $B/T$ values. In addition, the new model has more frequent residual star formation in massive galaxies than previous models, and more than is observed.

Our more detailed and spatially resolved treatment of star formation and feedback leads to a significant improvement in simulated cold gas properties in the new model. These include the local HI mass function, the mass fractions in total cold gas, HI and $\Htwo$ as a function of stellar mass, the HI/$L_r$ distribution as a function of HI mass, and the evolution of the cosmic $\Htwo$ density from $z=0$ to 5. Nevertheless, model galaxies are still excessively rich in $\Htwo$, and therefore in SFR, particularly at $z=0$ and high stellar mass.

Finally, we fully exploit our new model capabilities by analysing results for the radial profiles of stellar and cold gas surface density, stellar metallicities and cold gas oxygen abundances. Our new model agrees reasonably well with the radial profiles of stellar, cold gas, HI and $\Htwo$ surface density observed for Milky-Way-like galaxies. When comparing with CALIFA and MUSE data for stellar and cold gas metallicity profiles we find good agreement for gas-phase metallicities and stellar-phase metallicity at high-mass, but excessively steep gradients for low-mass objects. 

In this paper we present a significant augmentation of the traditional semi-analytic modelling approach. This is based on introducing a spatially resolved model --- of both the cold gas partitioning and star formation and the mass, energy and elements return --- while maintaining the successes of previous work in matching the observed global properties of galaxies. This technique offers new opportunities, particularly for interpreting the results of ongoing and upcoming surveys with  high-resolution, multi-object IFUs. The analysis presented in this paper will soon be extended to include a detailed study of metallicity and star formation gradients.

\section*{Acknowledgements}

This work used the DiRAC Data Centric system at Durham University,
operated by the Institute for Computational Cosmology on behalf of the
STFC DiRAC HPC Facility (www.dirac.ac.uk). This equipment was funded
by BIS National E-infrastructure capital grant ST/K00042X/1, STFC
capital grant ST/H008519/1, and STFC DiRAC Operations grant
ST/K003267/1 and Durham University. DiRAC is part of the National
E-Infrastructure. 

BMBH acknowledges support from a Zwicky Prize fellowship. PAT (ORCID 0000-0001-6888-6483) acknowledges support from the Science and Technology Facilities Council (grant number ST/P000525/1). Jian Fu acknowledges support from National Natural Science Foundation of China (NSFC U1531123), Youth innovation Promotion Association CAS and Shanghai Committee of Science and Technology grant No.19ZR1466700. Chaichalit Srisawat acknowledges support by a Grant of Excellence from the Icelandic Research Fund (grant number 173929-051).

\bibliographystyle{mnras} \bibliography{paper_Hen17}

\begin{thebibliography}{}
\makeatletter
\relax
\def\mn@urlcharsother{\let\do\@makeother \do\$\do\&\do\#\do\^\do\_\do\%\do\~}
\def\mn@doi{\begingroup\mn@urlcharsother \@ifnextchar [ {\mn@doi@}
  {\mn@doi@[]}}
\def\mn@doi@[#1]#2{\def\@tempa{#1}\ifx\@tempa\@empty \href
  {http://dx.doi.org/#2} {doi:#2}\else \href {http://dx.doi.org/#2} {#1}\fi
  \endgroup}
\def\mn@eprint#1#2{\mn@eprint@#1:#2::\@nil}
\def\mn@eprint@arXiv#1{\href {http://arxiv.org/abs/#1} {{\tt arXiv:#1}}}
\def\mn@eprint@dblp#1{\href {http://dblp.uni-trier.de/rec/bibtex/#1.xml}
  {dblp:#1}}
\def\mn@eprint@#1:#2:#3:#4\@nil{\def\@tempa {#1}\def\@tempb {#2}\def\@tempc
  {#3}\ifx \@tempc \@empty \let \@tempc \@tempb \let \@tempb \@tempa \fi \ifx
  \@tempb \@empty \def\@tempb {arXiv}\fi \@ifundefined
  {mn@eprint@\@tempb}{\@tempb:\@tempc}{\expandafter \expandafter \csname
  mn@eprint@\@tempb\endcsname \expandafter{\@tempc}}}

\bibitem[\protect\citeauthoryear{{Angulo} \& {Hilbert}}{{Angulo} \&
  {Hilbert}}{2015}]{Angulo2015}
{Angulo} R.~E.,  {Hilbert} S.,  2015, \mn@doi [MNRAS] {10.1093/mnras/stv050},
  \href {http://adsabs.harvard.edu/abs/2015MNRAS.448..364A} {448, 364}

\bibitem[\protect\citeauthoryear{{Asplund}, {Grevesse}, {Sauval}  \&
  {Scott}}{{Asplund} et~al.}{2009}]{Asplund2009}
{Asplund} M.,  {Grevesse} N.,  {Sauval} A.~J.,   {Scott} P.,  2009, \mn@doi
  [\araa] {10.1146/annurev.astro.46.060407.145222}, \href
  {http://adsabs.harvard.edu/abs/2009ARA%26A..47..481A} {47, 481}

\bibitem[\protect\citeauthoryear{{Avila-Reese}, {Firmani}  \&
  {Hern{\'a}ndez}}{{Avila-Reese} et~al.}{1998}]{Avila-Reese1998}
{Avila-Reese} V.,  {Firmani} C.,   {Hern{\'a}ndez} X.,  1998, \mn@doi [ApJ]
  {10.1086/306136}, \href {http://adsabs.harvard.edu/abs/1998ApJ...505...37A}
  {505, 37}

\bibitem[\protect\citeauthoryear{{Bacon}, {Accardo}, {Adjali}  et~al.}{{Bacon}
  et~al.}{2010}]{Bacon2010}
{Bacon} R.,  {Accardo} M.,  {Adjali} L.,   et~al., 2010, in Ground-based and
  Airborne Instrumentation for Astronomy III. p. 773508,
  \mn@doi{10.1117/12.856027}

\bibitem[\protect\citeauthoryear{{Baldry}, {Glazebrook}, {Brinkmann},
  {Ivezi{\'c}}, {Lupton}, {Nichol}  \& {Szalay}}{{Baldry}
  et~al.}{2004}]{Baldry2004}
{Baldry} I.~K.,  {Glazebrook} K.,  {Brinkmann} J.,  {Ivezi{\'c}} {\v Z}.,
  {Lupton} R.~H.,  {Nichol} R.~C.,   {Szalay} A.~S.,  2004, \mn@doi [ApJ]
  {10.1086/380092}, \href {http://adsabs.harvard.edu/abs/2004ApJ...600..681B}
  {600, 681}

\bibitem[\protect\citeauthoryear{{Baldry}, {Glazebrook}  \& {Driver}}{{Baldry}
  et~al.}{2008}]{Baldry2008}
{Baldry} I.~K.,  {Glazebrook} K.,   {Driver} S.~P.,  2008, \mn@doi [MNRAS]
  {10.1111/j.1365-2966.2008.13348.x}, \href
  {http://adsabs.harvard.edu/abs/2008MNRAS.388..945B} {388, 945}

\bibitem[\protect\citeauthoryear{{Baldry}, {Driver}, {Loveday}
  et~al.}{{Baldry} et~al.}{2012}]{Baldry2012}
{Baldry} I.~K.,  {Driver} S.~P.,  {Loveday} J.,   et~al., 2012, \mn@doi [MNRAS]
  {10.1111/j.1365-2966.2012.20340.x}, \href
  {http://adsabs.harvard.edu/abs/2012MNRAS.421..621B} {421, 621}

\bibitem[\protect\citeauthoryear{{Behroozi}, {Wechsler}  \&
  {Conroy}}{{Behroozi} et~al.}{2013}]{Behroozi2013}
{Behroozi} P.~S.,  {Wechsler} R.~H.,   {Conroy} C.,  2013, \mn@doi [ApJ]
  {10.1088/0004-637X/770/1/57}, \href
  {http://adsabs.harvard.edu/abs/2013ApJ...770...57B} {770, 57}

\bibitem[\protect\citeauthoryear{{Bell}, {McIntosh}, {Katz}  \&
  {Weinberg}}{{Bell} et~al.}{2003}]{Bell2003}
{Bell} E.~F.,  {McIntosh} D.~H.,  {Katz} N.,   {Weinberg} M.~D.,  2003, \mn@doi
  [ApJ Supp.] {10.1086/378847}, \href
  {http://adsabs.harvard.edu/abs/2003ApJS..149..289B} {149, 289}

\bibitem[\protect\citeauthoryear{{Benson}}{{Benson}}{2014}]{Benson2014}
{Benson} A.~J.,  2014, \mn@doi [MNRAS] {10.1093/mnras/stu1630}, \href
  {http://adsabs.harvard.edu/abs/2014MNRAS.444.2599B} {444, 2599}

\bibitem[\protect\citeauthoryear{{Benson} \& {Bower}}{{Benson} \&
  {Bower}}{2010}]{Benson2010}
{Benson} A.~J.,  {Bower} R.,  2010, \mn@doi [MNRAS]
  {10.1111/j.1365-2966.2010.16592.x}, \href
  {http://adsabs.harvard.edu/abs/2010MNRAS.405.1573B} {405, 1573}

\bibitem[\protect\citeauthoryear{{Bertin} \& {Lin}}{{Bertin} \&
  {Lin}}{1996}]{Bertin1996}
{Bertin} G.,  {Lin} C.~C.,  1996, {Spiral structure in galaxies a density wave
  theory}

\bibitem[\protect\citeauthoryear{{Biffi}, {Mernier}  \& {Medvedev}}{{Biffi}
  et~al.}{2018}]{Biffi2018}
{Biffi} V.,  {Mernier} F.,   {Medvedev} P.,  2018, \mn@doi [\ssr]
  {10.1007/s11214-018-0557-7}, \href
  {http://adsabs.harvard.edu/abs/2018SSRv..214..123B} {214, 123}

\bibitem[\protect\citeauthoryear{{Bigiel}, {Leroy}, {Walter}, {Brinks}
  et~al.}{{Bigiel} et~al.}{2011}]{Bigiel2011}
{Bigiel} F.,  {Leroy} A.~K.,  {Walter} F.,  {Brinks} E.,   et~al., 2011,
  \mn@doi [ApJ] {10.1088/2041-8205/730/2/L13}, \href
  {http://adsabs.harvard.edu/abs/2011ApJ...730L..13B} {730, L13}

\bibitem[\protect\citeauthoryear{{Blitz} \& {Rosolowsky}}{{Blitz} \&
  {Rosolowsky}}{2006}]{Blitz2006}
{Blitz} L.,  {Rosolowsky} E.,  2006, \mn@doi [ApJ] {10.1086/505417}, \href
  {http://adsabs.harvard.edu/abs/2006ApJ...650..933B} {650, 933}

\bibitem[\protect\citeauthoryear{{Bluck}, {Bottrell}, {Teimoorinia}
  et~al.}{{Bluck} et~al.}{2019}]{Bluck2019}
{Bluck} A. F.~L.,  {Bottrell} C.,  {Teimoorinia} H.,   et~al., 2019, \mn@doi
  [\mnras] {10.1093/mnras/stz363}, \href
  {https://ui.adsabs.harvard.edu/abs/2019MNRAS.485..666B} {485, 666}

\bibitem[\protect\citeauthoryear{{Bolatto}, {Wolfire}  \& {Leroy}}{{Bolatto}
  et~al.}{2013}]{Bolatto2013}
{Bolatto} A.~D.,  {Wolfire} M.,   {Leroy} A.~K.,  2013, \mn@doi [\araa]
  {10.1146/annurev-astro-082812-140944}, \href
  {https://ui.adsabs.harvard.edu/abs/2013ARA&A..51..207B} {51, 207}

\bibitem[\protect\citeauthoryear{{Bouwens}, {Illingworth}, {Oesch}
  et~al.}{{Bouwens} et~al.}{2012}]{Bouwens2012}
{Bouwens} R.~J.,  {Illingworth} G.~D.,  {Oesch} P.~A.,   et~al., 2012, \mn@doi
  [ApJ] {10.1088/0004-637X/754/2/83}, \href
  {http://adsabs.harvard.edu/abs/2012ApJ...754...83B} {754, 83}

\bibitem[\protect\citeauthoryear{{Bower}, {Vernon}, {Goldstein}, {Benson},
  {Lacey}, {Baugh}, {Cole}  \& {Frenk}}{{Bower} et~al.}{2010}]{Bower2010}
{Bower} R.~G.,  {Vernon} I.,  {Goldstein} M.,  {Benson} A.~J.,  {Lacey} C.~G.,
  {Baugh} C.~M.,  {Cole} S.,   {Frenk} C.~S.,  2010, \mn@doi [MNRAS]
  {10.1111/j.1365-2966.2010.16991.x}, \href
  {http://adsabs.harvard.edu/abs/2010MNRAS.407.2017B} {407, 2017}

\bibitem[\protect\citeauthoryear{{Boylan-Kolchin}, {Ma}  \&
  {Quataert}}{{Boylan-Kolchin} et~al.}{2008}]{BoylanKolchin2008}
{Boylan-Kolchin} M.,  {Ma} C.-P.,   {Quataert} E.,  2008, \mn@doi [MNRAS]
  {10.1111/j.1365-2966.2007.12530.x}, \href
  {http://adsabs.harvard.edu/abs/2008MNRAS.383...93B} {383, 93}

\bibitem[\protect\citeauthoryear{{Bresolin}}{{Bresolin}}{2008}]{Bresolin2008}
{Bresolin} F.,  2008, in {Israelian} G.,  {Meynet} G.,  eds, The Metal-Rich
  Universe. p.~155 (\mn@eprint {} {astro-ph/0608410})

\bibitem[\protect\citeauthoryear{{Brinchmann}, {Charlot}, {White}, {Tremonti},
  {Kauffmann}, {Heckman}  \& {Brinkmann}}{{Brinchmann}
  et~al.}{2004}]{Brinchmann2004}
{Brinchmann} J.,  {Charlot} S.,  {White} S.~D.~M.,  {Tremonti} C.,  {Kauffmann}
  G.,  {Heckman} T.,   {Brinkmann} J.,  2004, \mn@doi [MNRAS]
  {10.1111/j.1365-2966.2004.07881.x}, \href
  {http://adsabs.harvard.edu/abs/2004MNRAS.351.1151B} {351, 1151}

\bibitem[\protect\citeauthoryear{{Bundy}, {Ellis}  \& {Conselice}}{{Bundy}
  et~al.}{2005}]{Bundy2005}
{Bundy} K.,  {Ellis} R.~S.,   {Conselice} C.~J.,  2005, \mn@doi [ApJ]
  {10.1086/429549}, \href {http://adsabs.harvard.edu/abs/2005ApJ...625..621B}
  {625, 621}

\bibitem[\protect\citeauthoryear{{Bundy}, {Bershady}, {Law}  et~al.}{{Bundy}
  et~al.}{2015}]{Bundy2015}
{Bundy} K.,  {Bershady} M.~A.,  {Law} D.~R.,   et~al., 2015, \mn@doi [\apj]
  {10.1088/0004-637X/798/1/7}, \href
  {http://adsabs.harvard.edu/abs/2015ApJ...798....7B} {798, 7}

\bibitem[\protect\citeauthoryear{{Catinella} et~al.,}{{Catinella}
  et~al.}{2018}]{Catinella2018}
{Catinella} B.,  et~al., 2018, \mn@doi [MNRAS] {10.1093/mnras/sty089}, \href
  {https://ui.adsabs.harvard.edu/abs/2018MNRAS.476..875C} {476, 875}

\bibitem[\protect\citeauthoryear{{Chabrier}}{{Chabrier}}{2003}]{Chabrier2003}
{Chabrier} G.,  2003, PASP, \href
  {http://adsabs.harvard.edu/abs/2003PASP..115..763C} {115, 763}

\bibitem[\protect\citeauthoryear{{Cormier}, {Bigiel}, {Wang}  et~al.}{{Cormier}
  et~al.}{2016}]{Cormier2016}
{Cormier} D.,  {Bigiel} F.,  {Wang} J.,   et~al., 2016, \mn@doi [\mnras]
  {10.1093/mnras/stw2097}, \href
  {https://ui.adsabs.harvard.edu/abs/2016MNRAS.463.1724C} {463, 1724}

\bibitem[\protect\citeauthoryear{{Crain} et~al.,}{{Crain}
  et~al.}{2017}]{Crain2017}
{Crain} R.~A.,  et~al., 2017, \mn@doi [MNRAS] {10.1093/mnras/stw2586}, \href
  {https://ui.adsabs.harvard.edu/abs/2017MNRAS.464.4204C} {464, 4204}

\bibitem[\protect\citeauthoryear{{Croom}, {Lawrence}, {Bland-Hawthorn}
  et~al.}{{Croom} et~al.}{2012}]{Croom2012}
{Croom} S.~M.,  {Lawrence} J.~S.,  {Bland-Hawthorn} J.,   et~al., 2012, \mn@doi
  [MNRAS] {10.1111/j.1365-2966.2011.20365.x}, \href
  {http://adsabs.harvard.edu/abs/2012MNRAS.421..872C} {421, 872}

\bibitem[\protect\citeauthoryear{{Daddi} et~al.,}{{Daddi}
  et~al.}{2010}]{Daddi2010}
{Daddi} E.,  et~al., 2010, \mn@doi [ApJ] {10.1088/0004-637X/713/1/686}, \href
  {https://ui.adsabs.harvard.edu/abs/2010ApJ...713..686D} {713, 686}

\bibitem[\protect\citeauthoryear{{Dalcanton}, {Spergel}  \&
  {Summers}}{{Dalcanton} et~al.}{1997}]{Dalcanton1997}
{Dalcanton} J.~J.,  {Spergel} D.~N.,   {Summers} F.~J.,  1997, ApJ, \href
  {http://adsabs.harvard.edu/abs/1997ApJ...482..659D} {482, 659}

\bibitem[\protect\citeauthoryear{{Dav{\'e}}, {Thompson}  \&
  {Hopkins}}{{Dav{\'e}} et~al.}{2016}]{Dave2016}
{Dav{\'e}} R.,  {Thompson} R.,   {Hopkins} P.~F.,  2016, \mn@doi [\mnras]
  {10.1093/mnras/stw1862}, \href
  {http://adsabs.harvard.edu/abs/2016MNRAS.462.3265D} {462, 3265}

\bibitem[\protect\citeauthoryear{{De Lucia}, {Tornatore}, {Frenk}, {Helmi},
  {Navarro}  \& {White}}{{De Lucia} et~al.}{2014}]{DeLucia2014}
{De Lucia} G.,  {Tornatore} L.,  {Frenk} C.~S.,  {Helmi} A.,  {Navarro} J.~F.,
   {White} S.~D.~M.,  2014, \mn@doi [\mnras] {10.1093/mnras/stu1752}, \href
  {http://adsabs.harvard.edu/abs/2014MNRAS.445..970D} {445, 970}

\bibitem[\protect\citeauthoryear{{Decarli}, {Walter}, {G{\'o}nzalez-L{\'o}pez}
  et~al.}{{Decarli} et~al.}{2019}]{Decarli2019}
{Decarli} R.,  {Walter} F.,  {G{\'o}nzalez-L{\'o}pez} J.,   et~al., 2019, arXiv
  e-prints, \href {http://adsabs.harvard.edu/abs/2019arXiv190309164D} {}

\bibitem[\protect\citeauthoryear{{Driver}, {Andrews}, {da Cunha}
  et~al.}{{Driver} et~al.}{2018}]{Driver2018}
{Driver} S.~P.,  {Andrews} S.~K.,  {da Cunha} E.,   et~al., 2018, \mn@doi
  [MNRAS] {10.1093/mnras/stx2728}, \href
  {http://adsabs.harvard.edu/abs/2018MNRAS.475.2891D} {475, 2891}

\bibitem[\protect\citeauthoryear{{Dubois}, {Peirani}, {Pichon}, {Devriendt},
  {Gavazzi}, {Welker}  \& {Volonteri}}{{Dubois} et~al.}{2016}]{Dubois2016}
{Dubois} Y.,  {Peirani} S.,  {Pichon} C.,  {Devriendt} J.,  {Gavazzi} R.,
  {Welker} C.,   {Volonteri} M.,  2016, \mn@doi [\mnras]
  {10.1093/mnras/stw2265}, \href
  {https://ui.adsabs.harvard.edu/abs/2016MNRAS.463.3948D} {463, 3948}

\bibitem[\protect\citeauthoryear{{Dutton}}{{Dutton}}{2009}]{Dutton2009}
{Dutton} A.~A.,  2009, \mn@doi [MNRAS] {10.1111/j.1365-2966.2009.14741.x},
  \href {http://adsabs.harvard.edu/abs/2009MNRAS.396..121D} {396, 121}

\bibitem[\protect\citeauthoryear{{Elbaz} et~al.,}{{Elbaz}
  et~al.}{2007}]{Elbaz2007}
{Elbaz} D.,  et~al., 2007, \mn@doi [Astronomy and Astrophysics Supplement
  Series] {10.1051/0004-6361:20077525}, \href
  {http://adsabs.harvard.edu/abs/2007A%26A...468...33E} {468, 33}

\bibitem[\protect\citeauthoryear{{Elmegreen}}{{Elmegreen}}{1989}]{Elmegreen1989}
{Elmegreen} B.~G.,  1989, \mn@doi [ApJ] {10.1086/167192}, \href
  {http://adsabs.harvard.edu/abs/1989ApJ...338..178E} {338, 178}

\bibitem[\protect\citeauthoryear{{Elmegreen}}{{Elmegreen}}{1993}]{Elmegreen1993}
{Elmegreen} B.~G.,  1993, \mn@doi [ApJ] {10.1086/172816}, \href
  {http://adsabs.harvard.edu/abs/1993ApJ...411..170E} {411, 170}

\bibitem[\protect\citeauthoryear{{Erroz-Ferrer}, {Carollo}, {den Brok}
  et~al.}{{Erroz-Ferrer} et~al.}{2019}]{Ferrer2019}
{Erroz-Ferrer} S.,  {Carollo} C.~M.,  {den Brok} M.,   et~al., 2019, arXiv
  e-prints, \href {https://ui.adsabs.harvard.edu/\#abs/2019arXiv190104493E} {p.
  arXiv:1901.04493}

\bibitem[\protect\citeauthoryear{{Faber}, {Willmer}, {Wolf}  et~al.}{{Faber}
  et~al.}{2007}]{Faber2007}
{Faber} S.~M.,  {Willmer} C.~N.~A.,  {Wolf} C.,   et~al., 2007, \mn@doi [ApJ]
  {10.1086/519294}, \href {http://adsabs.harvard.edu/abs/2007ApJ...665..265F}
  {665, 265}

\bibitem[\protect\citeauthoryear{{Fu}, {Hou}, {Yin}  \& {Chang}}{{Fu}
  et~al.}{2009}]{Fu2009}
{Fu} J.,  {Hou} J.~L.,  {Yin} J.,   {Chang} R.~X.,  2009, \mn@doi [ApJ]
  {10.1088/0004-637X/696/1/668}, \href
  {http://adsabs.harvard.edu/abs/2009ApJ...696..668F} {696, 668}

\bibitem[\protect\citeauthoryear{{Fu}, {Guo}, {Kauffmann}  \& {Krumholz}}{{Fu}
  et~al.}{2010}]{Fu2010}
{Fu} J.,  {Guo} Q.,  {Kauffmann} G.,   {Krumholz} M.~R.,  2010, \mn@doi [MNRAS]
  {10.1111/j.1365-2966.2010.17342.x}, \href
  {http://adsabs.harvard.edu/abs/2010MNRAS.409..515F} {409, 515}

\bibitem[\protect\citeauthoryear{{Fu}, {Kauffmann}, {Li}  \& {Guo}}{{Fu}
  et~al.}{2012}]{Fu2012}
{Fu} J.,  {Kauffmann} G.,  {Li} C.,   {Guo} Q.,  2012, \mn@doi [MNRAS]
  {10.1111/j.1365-2966.2012.21356.x}, \href
  {http://adsabs.harvard.edu/abs/2012MNRAS.424.2701F} {424, 2701}

\bibitem[\protect\citeauthoryear{{Fu} et~al.,}{{Fu} et~al.}{2013}]{Fu2013}
{Fu} J.,  et~al., 2013, \mn@doi [MNRAS] {10.1093/mnras/stt1117}, \href
  {http://adsabs.harvard.edu/abs/2013MNRAS.434.1531F} {434, 1531}

\bibitem[\protect\citeauthoryear{{Gallazzi}, {Charlot}, {Brinchmann}, {White}
  \& {Tremonti}}{{Gallazzi} et~al.}{2005}]{Gallazzi2005}
{Gallazzi} A.,  {Charlot} S.,  {Brinchmann} J.,  {White} S.~D.~M.,   {Tremonti}
  C.~A.,  2005, \mn@doi [MNRAS] {10.1111/j.1365-2966.2005.09321.x}, \href
  {http://adsabs.harvard.edu/abs/2005MNRAS.362...41G} {362, 41}

\bibitem[\protect\citeauthoryear{{Genzel}, {Tacconi}, {Lutz}  et~al.}{{Genzel}
  et~al.}{2015}]{Genzel2015}
{Genzel} R.,  {Tacconi} L.~J.,  {Lutz} D.,   et~al., 2015, \mn@doi [ApJ]
  {10.1088/0004-637X/800/1/20}, \href
  {http://adsabs.harvard.edu/abs/2015ApJ...800...20G} {800, 20}

\bibitem[\protect\citeauthoryear{{Gonz{\'a}lez Delgado},
  {Garc{\'{\i}}a-Benito}, {P{\'e}rez}  et~al.}{{Gonz{\'a}lez Delgado}
  et~al.}{2015}]{Delgado2015}
{Gonz{\'a}lez Delgado} R.~M.,  {Garc{\'{\i}}a-Benito} R.,  {P{\'e}rez} E.,
  et~al., 2015, \mn@doi [\aap] {10.1051/0004-6361/201525938}, \href
  {http://adsabs.harvard.edu/abs/2015A%26A...581A.103G} {581, A103}

\bibitem[\protect\citeauthoryear{{Gonzalez-Perez}, {Lacey}, {Baugh}, {Lagos},
  {Helly}, {Campbell}  \& {Mitchell}}{{Gonzalez-Perez}
  et~al.}{2014}]{Gonzalez2014}
{Gonzalez-Perez} V.,  {Lacey} C.~G.,  {Baugh} C.~M.,  {Lagos} C.~D.~P.,
  {Helly} J.,  {Campbell} D.~J.~R.,   {Mitchell} P.~D.,  2014, \mn@doi [MNRAS]
  {10.1093/mnras/stt2410}, \href
  {http://adsabs.harvard.edu/abs/2014MNRAS.tmp..191G} {}

\bibitem[\protect\citeauthoryear{{Guo} et~al.,}{{Guo} et~al.}{2011}]{Guo2011}
{Guo} Q.,  et~al., 2011, \mn@doi [MNRAS] {10.1111/j.1365-2966.2010.18114.x},
  \href {http://adsabs.harvard.edu/abs/2011MNRAS.413..101G} {413, 101}

\bibitem[\protect\citeauthoryear{{Haynes}, {Giovanelli}, {Martin}
  et~al.}{{Haynes} et~al.}{2011}]{Haynes2011}
{Haynes} M.~P.,  {Giovanelli} R.,  {Martin} A.~M.,   et~al., 2011, \mn@doi [AJ]
  {10.1088/0004-6256/142/5/170}, \href
  {http://adsabs.harvard.edu/abs/2011AJ....142..170H} {142, 170}

\bibitem[\protect\citeauthoryear{{Henriques}, {Thomas}, {Oliver}  \&
  {Roseboom}}{{Henriques} et~al.}{2009}]{Henriques2009}
{Henriques} B.~M.~B.,  {Thomas} P.~A.,  {Oliver} S.,   {Roseboom} I.,  2009,
  \mn@doi [MNRAS] {10.1111/j.1365-2966.2009.14730.x}, \href
  {http://adsabs.harvard.edu/abs/2009MNRAS.396..535H} {396, 535}

\bibitem[\protect\citeauthoryear{{Henriques}, {White}, {Thomas}, {Angulo},
  {Guo}, {Lemson}  \& {Springel}}{{Henriques} et~al.}{2013}]{Henriques2013}
{Henriques} B.~M.~B.,  {White} S.~D.~M.,  {Thomas} P.~A.,  {Angulo} R.~E.,
  {Guo} Q.,  {Lemson} G.,   {Springel} V.,  2013, \mn@doi [MNRAS]
  {10.1093/mnras/stt415}, \href
  {http://adsabs.harvard.edu/abs/2013MNRAS.431.3373H} {431, 3373}

\bibitem[\protect\citeauthoryear{{Henriques}, {White}, {Thomas}, {Angulo},
  {Guo}, {Lemson}, {Springel}  \& {Overzier}}{{Henriques}
  et~al.}{2015}]{Henriques2015}
{Henriques} B.~M.~B.,  {White} S.~D.~M.,  {Thomas} P.~A.,  {Angulo} R.,  {Guo}
  Q.,  {Lemson} G.,  {Springel} V.,   {Overzier} R.,  2015, \mn@doi [MNRAS]
  {10.1093/mnras/stv705}, \href
  {http://adsabs.harvard.edu/abs/2015MNRAS.451.2663H} {451, 2663}

\bibitem[\protect\citeauthoryear{{Henriques}, {White}, {Thomas}, {Angulo},
  {Guo}, {Lemson}  \& {Wang}}{{Henriques} et~al.}{2017}]{Henriques2017}
{Henriques} B.~M.~B.,  {White} S.~D.~M.,  {Thomas} P.~A.,  {Angulo} R.~E.,
  {Guo} Q.,  {Lemson} G.,   {Wang} W.,  2017, \mn@doi [\mnras]
  {10.1093/mnras/stx1010}, \href
  {http://adsabs.harvard.edu/abs/2017MNRAS.469.2626H} {469, 2626}

\bibitem[\protect\citeauthoryear{{Henriques}, {White}, {Lilly}, {Bell}, {Bluck}
   \& {Terrazas}}{{Henriques} et~al.}{2019}]{Henriques2019}
{Henriques} B. M.~B.,  {White} S. D.~M.,  {Lilly} S.~J.,  {Bell} E.~F.,
  {Bluck} A. F.~L.,   {Terrazas} B.~A.,  2019, \mn@doi [\mnras]
  {10.1093/mnras/stz577}, \href
  {https://ui.adsabs.harvard.edu/abs/2019MNRAS.485.3446H} {485, 3446}

\bibitem[\protect\citeauthoryear{{Hirschmann}, {De Lucia}  \&
  {Fontanot}}{{Hirschmann} et~al.}{2016}]{Hirschmann2016}
{Hirschmann} M.,  {De Lucia} G.,   {Fontanot} F.,  2016, \mn@doi [MNRAS]
  {10.1093/mnras/stw1318}, \href
  {http://adsabs.harvard.edu/abs/2016MNRAS.461.1760H} {461, 1760}

\bibitem[\protect\citeauthoryear{{Ilbert}, {McCracken}, {Le F{\`e}vre}
  et~al.}{{Ilbert} et~al.}{2013}]{Ilbert2013}
{Ilbert} O.,  {McCracken} H.~J.,  {Le F{\`e}vre} O.,   et~al., 2013, \mn@doi
  [Astronomy and Astrophysics Supplement Series] {10.1051/0004-6361/201321100},
  \href {http://adsabs.harvard.edu/abs/2013A%26A...556A..55I} {556, A55}

\bibitem[\protect\citeauthoryear{{Irodotou}, {Thomas}, {Henriques}  \&
  {Sargent}}{{Irodotou} et~al.}{2018}]{Irodotou2019}
{Irodotou} D.,  {Thomas} P.~A.,  {Henriques} B.~M.,   {Sargent} M.~T.,  2018,
  arXiv e-prints, \href {http://adsabs.harvard.edu/abs/2018arXiv181005173I} {}

\bibitem[\protect\citeauthoryear{{Izquierdo-Villalba}, {Bonoli}, {Spinoso},
  {Rosas-Guevara}, {Henriques}  \& {Hernand
  ez-Monteagudo}}{{Izquierdo-Villalba} et~al.}{2019}]{Izquierdo-Villalba2019}
{Izquierdo-Villalba} D.,  {Bonoli} S.,  {Spinoso} D.,  {Rosas-Guevara} Y.,
  {Henriques} B. M.~B.,   {Hernand ez-Monteagudo} C.,  2019, arXiv e-prints,
  \href {https://ui.adsabs.harvard.edu/abs/2019arXiv190110490I} {p.
  arXiv:1901.10490}

\bibitem[\protect\citeauthoryear{{Jaffe}}{{Jaffe}}{1983}]{Jaffe1983}
{Jaffe} W.,  1983, \mn@doi [MNRAS] {10.1093/mnras/202.4.995}, \href
  {https://ui.adsabs.harvard.edu/abs/1983MNRAS.202..995J} {202, 995}

\bibitem[\protect\citeauthoryear{{Jones}, {Haynes}, {Giovanelli}  \&
  {Moorman}}{{Jones} et~al.}{2018}]{Jones2018}
{Jones} M.~G.,  {Haynes} M.~P.,  {Giovanelli} R.,   {Moorman} C.,  2018,
  \mn@doi [\mnras] {10.1093/mnras/sty521}, \href
  {https://ui.adsabs.harvard.edu/abs/2018MNRAS.477....2J} {477, 2}

\bibitem[\protect\citeauthoryear{{Kalnajs}}{{Kalnajs}}{1972}]{Kalnajs1972}
{Kalnajs} A.~J.,  1972, Astrophys.~Lett., \href
  {https://ui.adsabs.harvard.edu/abs/1972ApL....11...41K} {11, 41}

\bibitem[\protect\citeauthoryear{{Kampakoglou}, {Trotta}  \&
  {Silk}}{{Kampakoglou} et~al.}{2008}]{Kampakoglou2008}
{Kampakoglou} M.,  {Trotta} R.,   {Silk} J.,  2008, \mn@doi [MNRAS]
  {10.1111/j.1365-2966.2007.12747.x}, \href
  {http://adsabs.harvard.edu/abs/2008MNRAS.384.1414K} {384, 1414}

\bibitem[\protect\citeauthoryear{{Karim} et~al.,}{{Karim}
  et~al.}{2011}]{Karim2011}
{Karim} A.,  et~al., 2011, \mn@doi [ApJ] {10.1088/0004-637X/730/2/61}, \href
  {http://adsabs.harvard.edu/abs/2011ApJ...730...61K} {730, 61}

\bibitem[\protect\citeauthoryear{{Kauffmann}}{{Kauffmann}}{1996}]{Kauffmann1996b}
{Kauffmann} G.,  1996, MNRAS, \href
  {http://adsabs.harvard.edu/abs/1996MNRAS.281..475K} {281, 475}

\bibitem[\protect\citeauthoryear{{Keres}, {Yun}  \& {Young}}{{Keres}
  et~al.}{2003}]{Keres2003}
{Keres} D.,  {Yun} M.~S.,   {Young} J.~S.,  2003, \mn@doi [\apj]
  {10.1086/344820}, \href
  {https://ui.adsabs.harvard.edu/abs/2003ApJ...582..659K} {582, 659}

\bibitem[\protect\citeauthoryear{{Kewley} \& {Ellison}}{{Kewley} \&
  {Ellison}}{2008}]{Kewley2008}
{Kewley} L.~J.,  {Ellison} S.~L.,  2008, \mn@doi [\apj] {10.1086/587500}, \href
  {http://adsabs.harvard.edu/abs/2008ApJ...681.1183K} {681, 1183}

\bibitem[\protect\citeauthoryear{{Krumholz}}{{Krumholz}}{2013}]{Krumholz2013}
{Krumholz} M.~R.,  2013, \mn@doi [MNRAS] {10.1093/mnras/stt1780}, \href
  {https://ui.adsabs.harvard.edu/abs/2013MNRAS.436.2747K} {436, 2747}

\bibitem[\protect\citeauthoryear{{Krumholz}, {McKee}  \&
  {Tumlinson}}{{Krumholz} et~al.}{2009}]{Krumholz2009}
{Krumholz} M.~R.,  {McKee} C.~F.,   {Tumlinson} J.,  2009, \mn@doi [ApJ]
  {10.1088/0004-637X/693/1/216}, \href
  {http://adsabs.harvard.edu/abs/2009ApJ...693..216K} {693, 216}

\bibitem[\protect\citeauthoryear{{Kuhlen}, {Krumholz}, {Madau}, {Smith}  \&
  {Wise}}{{Kuhlen} et~al.}{2012}]{Kuhlen2012}
{Kuhlen} M.,  {Krumholz} M.~R.,  {Madau} P.,  {Smith} B.~D.,   {Wise} J.,
  2012, \mn@doi [\apj] {10.1088/0004-637X/749/1/36}, \href
  {http://adsabs.harvard.edu/abs/2012ApJ...749...36K} {749, 36}

\bibitem[\protect\citeauthoryear{{Lacey} \& {Fall}}{{Lacey} \&
  {Fall}}{1985}]{Lacey1985}
{Lacey} C.~G.,  {Fall} S.~M.,  1985, \mn@doi [ApJ] {10.1086/162970}, \href
  {http://adsabs.harvard.edu/abs/1985ApJ...290..154L} {290, 154}

\bibitem[\protect\citeauthoryear{{Lacey} et~al.,}{{Lacey}
  et~al.}{2016}]{Lacey2016}
{Lacey} C.~G.,  et~al., 2016, \mn@doi [MNRAS] {10.1093/mnras/stw1888}, \href
  {https://ui.adsabs.harvard.edu/abs/2016MNRAS.462.3854L} {462, 3854}

\bibitem[\protect\citeauthoryear{{Lagos}, {Lacey}, {Baugh}, {Bower}  \&
  {Benson}}{{Lagos} et~al.}{2011a}]{Lagos2011a}
{Lagos} C.~D.~P.,  {Lacey} C.~G.,  {Baugh} C.~M.,  {Bower} R.~G.,   {Benson}
  A.~J.,  2011a, \mn@doi [\mnras] {10.1111/j.1365-2966.2011.19160.x}, \href
  {http://adsabs.harvard.edu/abs/2011MNRAS.416.1566L} {416, 1566}

\bibitem[\protect\citeauthoryear{{Lagos}, {Baugh}, {Lacey}, {Benson}, {Kim}  \&
  {Power}}{{Lagos} et~al.}{2011b}]{Lagos2011b}
{Lagos} C.~D.~P.,  {Baugh} C.~M.,  {Lacey} C.~G.,  {Benson} A.~J.,  {Kim}
  H.-S.,   {Power} C.,  2011b, \mn@doi [\mnras]
  {10.1111/j.1365-2966.2011.19583.x}, \href
  {http://adsabs.harvard.edu/abs/2011MNRAS.418.1649L} {418, 1649}

\bibitem[\protect\citeauthoryear{{Lagos} et~al.,}{{Lagos}
  et~al.}{2015}]{Lagos2015}
{Lagos} C. d.~P.,  et~al., 2015, \mn@doi [\mnras] {10.1093/mnras/stv1488},
  \href {https://ui.adsabs.harvard.edu/abs/2015MNRAS.452.3815L} {452, 3815}

\bibitem[\protect\citeauthoryear{{Lagos}, {Tobar}, {Robotham}, {Obreschkow},
  {Mitchell}, {Power}  \& {Elahi}}{{Lagos} et~al.}{2018}]{Lagos2018}
{Lagos} C. d.~P.,  {Tobar} R.~J.,  {Robotham} A. S.~G.,  {Obreschkow} D.,
  {Mitchell} P.~D.,  {Power} C.,   {Elahi} P.~J.,  2018, \mn@doi [MNRAS]
  {10.1093/mnras/sty2440}, \href
  {https://ui.adsabs.harvard.edu/abs/2018MNRAS.481.3573L} {481, 3573}

\bibitem[\protect\citeauthoryear{{Leroy}, {Walter}, {Brinks}, {Bigiel}, {de
  Blok}, {Madore}  \& {Thornley}}{{Leroy} et~al.}{2008}]{Leroy2008}
{Leroy} A.~K.,  {Walter} F.,  {Brinks} E.,  {Bigiel} F.,  {de Blok} W.~J.~G.,
  {Madore} B.,   {Thornley} M.~D.,  2008, \mn@doi [AJ]
  {10.1088/0004-6256/136/6/2782}, \href
  {http://adsabs.harvard.edu/abs/2008AJ....136.2782L} {136, 2782}

\bibitem[\protect\citeauthoryear{{Leroy}, {Walter}, {Sandstrom}
  et~al.}{{Leroy} et~al.}{2013}]{Leroy2013}
{Leroy} A.~K.,  {Walter} F.,  {Sandstrom} K.,   et~al., 2013, \mn@doi [AJ]
  {10.1088/0004-6256/146/2/19}, \href
  {http://adsabs.harvard.edu/abs/2013AJ....146...19L} {146, 19}

\bibitem[\protect\citeauthoryear{{Li} \& {White}}{{Li} \&
  {White}}{2009}]{Li2009}
{Li} C.,  {White} S.~D.~M.,  2009, \mn@doi [MNRAS]
  {10.1111/j.1365-2966.2009.15268.x}, \href
  {http://adsabs.harvard.edu/abs/2009MNRAS.398.2177L} {398, 2177}

\bibitem[\protect\citeauthoryear{{Lilly} \& {Carollo}}{{Lilly} \&
  {Carollo}}{2016}]{Lilly2016}
{Lilly} S.~J.,  {Carollo} C.~M.,  2016, \mn@doi [ApJ]
  {10.3847/0004-637X/833/1/1}, \href
  {http://adsabs.harvard.edu/abs/2016ApJ...833....1L} {833, 1}

\bibitem[\protect\citeauthoryear{{Lilly}, {Le Fevre}, {Hammer}  \&
  {Crampton}}{{Lilly} et~al.}{1996}]{Lilly1996}
{Lilly} S.~J.,  {Le Fevre} O.,  {Hammer} F.,   {Crampton} D.,  1996, \mn@doi
  [ApJ] {10.1086/309975}, \href
  {http://adsabs.harvard.edu/abs/1996ApJ...460L...1L} {460, L1}

\bibitem[\protect\citeauthoryear{{Lo}, {Sargent}  \& {Young}}{{Lo}
  et~al.}{1993}]{Lo1993}
{Lo} K.~Y.,  {Sargent} W.~L.~W.,   {Young} K.,  1993, \mn@doi [\aj]
  {10.1086/116658}, \href {http://adsabs.harvard.edu/abs/1993AJ....106..507L}
  {106, 507}

\bibitem[\protect\citeauthoryear{{Lu}, {Mo}, {Weinberg}  \& {Katz}}{{Lu}
  et~al.}{2011}]{Lu2011b}
{Lu} Y.,  {Mo} H.~J.,  {Weinberg} M.~D.,   {Katz} N.,  2011, \mn@doi [MNRAS]
  {10.1111/j.1365-2966.2011.19170.x}, \href
  {http://adsabs.harvard.edu/abs/2011MNRAS.416.1949L} {416, 1949}

\bibitem[\protect\citeauthoryear{{Lu}, {Mo}, {Katz}  \& {Weinberg}}{{Lu}
  et~al.}{2012}]{Lu2012}
{Lu} Y.,  {Mo} H.~J.,  {Katz} N.,   {Weinberg} M.~D.,  2012, \mn@doi [MNRAS]
  {10.1111/j.1365-2966.2012.20435.x}, \href
  {http://adsabs.harvard.edu/abs/2012MNRAS.421.1779L} {421, 1779}

\bibitem[\protect\citeauthoryear{{Lu}, {Mo}  \& {Wechsler}}{{Lu}
  et~al.}{2015}]{Lu2015}
{Lu} Y.,  {Mo} H.~J.,   {Wechsler} R.~H.,  2015, \mn@doi [MNRAS]
  {10.1093/mnras/stu2215}, \href
  {http://adsabs.harvard.edu/abs/2015MNRAS.446.1907L} {446, 1907}

\bibitem[\protect\citeauthoryear{{Lynden-Bell} \& {Pringle}}{{Lynden-Bell} \&
  {Pringle}}{1974}]{Lynden-Bell1974}
{Lynden-Bell} D.,  {Pringle} J.~E.,  1974, \mn@doi [MNRAS]
  {10.1093/mnras/168.3.603}, \href
  {http://adsabs.harvard.edu/abs/1974MNRAS.168..603L} {168, 603}

\bibitem[\protect\citeauthoryear{{Madau} \& {Dickinson}}{{Madau} \&
  {Dickinson}}{2014}]{Madau2014}
{Madau} P.,  {Dickinson} M.,  2014, \mn@doi [ARA\&A]
  {10.1146/annurev-astro-081811-125615}, \href
  {http://adsabs.harvard.edu/abs/2014ARA%26A..52..415M} {52, 415}

\bibitem[\protect\citeauthoryear{{Madau}, {Ferguson}, {Dickinson},
  {Giavalisco}, {Steidel}  \& {Fruchter}}{{Madau} et~al.}{1996}]{Madau1996}
{Madau} P.,  {Ferguson} H.~C.,  {Dickinson} M.~E.,  {Giavalisco} M.,  {Steidel}
  C.~C.,   {Fruchter} A.,  1996, MNRAS, \href
  {http://adsabs.harvard.edu/abs/1996MNRAS.283.1388M} {283, 1388}

\bibitem[\protect\citeauthoryear{{Maoz}, {Mannucci}  \& {Brandt}}{{Maoz}
  et~al.}{2012}]{Maoz2012}
{Maoz} D.,  {Mannucci} F.,   {Brandt} T.~D.,  2012, \mn@doi [\mnras]
  {10.1111/j.1365-2966.2012.21871.x}, \href
  {http://adsabs.harvard.edu/abs/2012MNRAS.426.3282M} {426, 3282}

\bibitem[\protect\citeauthoryear{{Marigo}}{{Marigo}}{2001}]{Marigo2001}
{Marigo} P.,  2001, \mn@doi [\aap] {10.1051/0004-6361:20000247}, \href
  {http://adsabs.harvard.edu/abs/2001A%26A...370..194M} {370, 194}

\bibitem[\protect\citeauthoryear{{Martindale}, {Thomas}, {Henriques}  \&
  {Loveday}}{{Martindale} et~al.}{2017}]{Martindale2017}
{Martindale} H.,  {Thomas} P.~A.,  {Henriques} B.~M.,   {Loveday} J.,  2017,
  \mn@doi [\mnras] {10.1093/mnras/stx2131}, \href
  {http://adsabs.harvard.edu/abs/2017MNRAS.472.1981M} {472, 1981}

\bibitem[\protect\citeauthoryear{{McConnell} \& {Ma}}{{McConnell} \&
  {Ma}}{2013}]{McConnell2013}
{McConnell} N.~J.,  {Ma} C.-P.,  2013, \mn@doi [ApJ]
  {10.1088/0004-637X/764/2/184}, \href
  {http://adsabs.harvard.edu/abs/2013ApJ...764..184M} {764, 184}

\bibitem[\protect\citeauthoryear{{McKee} \& {Krumholz}}{{McKee} \&
  {Krumholz}}{2010}]{McKee2010}
{McKee} C.~F.,  {Krumholz} M.~R.,  2010, \mn@doi [ApJ]
  {10.1088/0004-637X/709/1/308}, \href
  {http://adsabs.harvard.edu/abs/2010ApJ...709..308M} {709, 308}

\bibitem[\protect\citeauthoryear{{Mernier} et~al.,}{{Mernier}
  et~al.}{2017}]{Mernier2017}
{Mernier} F.,  et~al., 2017, \mn@doi [A\&A] {10.1051/0004-6361/201630075},
  \href {http://adsabs.harvard.edu/abs/2017A%26A...603A..80M} {603, A80}

\bibitem[\protect\citeauthoryear{{Mernier} et~al.,}{{Mernier}
  et~al.}{2018a}]{Mernier2018b}
{Mernier} F.,  et~al., 2018a, \mn@doi [\ssr] {10.1007/s11214-018-0565-7}, \href
  {http://adsabs.harvard.edu/abs/2018SSRv..214..129M} {214, 129}

\bibitem[\protect\citeauthoryear{{Mernier} et~al.,}{{Mernier}
  et~al.}{2018b}]{Mernier2018a}
{Mernier} F.,  et~al., 2018b, \mn@doi [\mnras] {10.1093/mnrasl/sly080}, \href
  {http://adsabs.harvard.edu/abs/2018MNRAS.478L.116M} {478, L116}

\bibitem[\protect\citeauthoryear{{Mitchell}, {Lacey}, {Cole}  \&
  {Baugh}}{{Mitchell} et~al.}{2014}]{Mitchell2014}
{Mitchell} P.~D.,  {Lacey} C.~G.,  {Cole} S.,   {Baugh} C.~M.,  2014, \mn@doi
  [MNRAS] {10.1093/mnras/stu1639}, \href
  {https://ui.adsabs.harvard.edu/abs/2014MNRAS.444.2637M} {444, 2637}

\bibitem[\protect\citeauthoryear{{Mo}, {Mao}  \& {White}}{{Mo}
  et~al.}{1998}]{Mo1998}
{Mo} H.~J.,  {Mao} S.,   {White} S.~D.~M.,  1998, MNRAS, \href
  {http://adsabs.harvard.edu/abs/1998MNRAS.295..319M} {295, 319}

\bibitem[\protect\citeauthoryear{{Mosleh}, {Williams}  \& {Franx}}{{Mosleh}
  et~al.}{2013}]{Mosleh2013}
{Mosleh} M.,  {Williams} R.~J.,   {Franx} M.,  2013, \mn@doi [\apj]
  {10.1088/0004-637X/777/2/117}, \href
  {https://ui.adsabs.harvard.edu/abs/2013ApJ...777..117M} {777, 117}

\bibitem[\protect\citeauthoryear{{Mutch}, {Poole}  \& {Croton}}{{Mutch}
  et~al.}{2013}]{Mutch2013}
{Mutch} S.~J.,  {Poole} G.~B.,   {Croton} D.~J.,  2013, \mn@doi [MNRAS]
  {10.1093/mnras/sts182}, \href
  {http://adsabs.harvard.edu/abs/2013MNRAS.428.2001M} {428, 2001}

\bibitem[\protect\citeauthoryear{{Muzzin}, {Marchesini}, {Stefanon}
  et~al.}{{Muzzin} et~al.}{2013}]{Muzzin2013}
{Muzzin} A.,  {Marchesini} D.,  {Stefanon} M.,   et~al., 2013, \mn@doi [ApJ]
  {10.1088/0004-637X/777/1/18}, \href
  {http://adsabs.harvard.edu/abs/2013ApJ...777...18M} {777, 18}

\bibitem[\protect\citeauthoryear{{Noeske}, {Weiner}, {Faber}  et~al.}{{Noeske}
  et~al.}{2007}]{Noeske2007}
{Noeske} K.~G.,  {Weiner} B.~J.,  {Faber} S.~M.,   et~al., 2007, \mn@doi [ApJ]
  {10.1086/517926}, \href {http://adsabs.harvard.edu/abs/2007ApJ...660L..43N}
  {660, L43}

\bibitem[\protect\citeauthoryear{{Obreschkow} \& {Rawlings}}{{Obreschkow} \&
  {Rawlings}}{2009}]{Obreschkow2009}
{Obreschkow} D.,  {Rawlings} S.,  2009, \mn@doi [MNRAS]
  {10.1111/j.1365-2966.2009.14497.x}, \href
  {http://adsabs.harvard.edu/abs/2009MNRAS.394.1857O} {394, 1857}

\bibitem[\protect\citeauthoryear{{Okamoto}, {Nagashima}, {Lacey}  \&
  {Frenk}}{{Okamoto} et~al.}{2017}]{Okamoto2017}
{Okamoto} T.,  {Nagashima} M.,  {Lacey} C.~G.,   {Frenk} C.~S.,  2017, \mn@doi
  [\mnras] {10.1093/mnras/stw2729}, \href
  {http://adsabs.harvard.edu/abs/2017MNRAS.464.4866O} {464, 4866}

\bibitem[\protect\citeauthoryear{{Peimbert}}{{Peimbert}}{1967}]{Peimbert1967}
{Peimbert} M.,  1967, \mn@doi [ApJ] {10.1086/149385}, \href
  {http://adsabs.harvard.edu/abs/1967ApJ...150..825P} {150, 825}

\bibitem[\protect\citeauthoryear{{Pilkington}, {Few}, {Gibson}
  et~al.}{{Pilkington} et~al.}{2012}]{Pilkington2012}
{Pilkington} K.,  {Few} C.~G.,  {Gibson} B.~K.,   et~al., 2012, \mn@doi
  [Astronomy and Astrophysics Supplement Series] {10.1051/0004-6361/201117466},
  \href {http://adsabs.harvard.edu/abs/2012A%26A...540A..56P} {540, A56}

\bibitem[\protect\citeauthoryear{{Pillepich}, {Springel}, {Nelson}
  et~al.}{{Pillepich} et~al.}{2018}]{Pillepich2018}
{Pillepich} A.,  {Springel} V.,  {Nelson} D.,   et~al., 2018, \mn@doi [\mnras]
  {10.1093/mnras/stx2656}, \href
  {https://ui.adsabs.harvard.edu/abs/2018MNRAS.473.4077P} {473, 4077}

\bibitem[\protect\citeauthoryear{{Popping}, {Somerville}  \&
  {Trager}}{{Popping} et~al.}{2014}]{Popping2014}
{Popping} G.,  {Somerville} R.~S.,   {Trager} S.~C.,  2014, \mn@doi [\mnras]
  {10.1093/mnras/stu991}, \href
  {http://adsabs.harvard.edu/abs/2014MNRAS.442.2398P} {442, 2398}

\bibitem[\protect\citeauthoryear{{Portinari} \& {Chiosi}}{{Portinari} \&
  {Chiosi}}{2000}]{Portinari2000}
{Portinari} L.,  {Chiosi} C.,  2000, Astronomy and Astrophysics Supplement
  Series, \href {https://ui.adsabs.harvard.edu/abs/2000A&A...355..929P} {355,
  929}

\bibitem[\protect\citeauthoryear{{Portinari}, {Chiosi}  \&
  {Bressan}}{{Portinari} et~al.}{1998}]{Portinari1998}
{Portinari} L.,  {Chiosi} C.,   {Bressan} A.,  1998, \aap, \href
  {http://adsabs.harvard.edu/abs/1998A%26A...334..505P} {334, 505}

\bibitem[\protect\citeauthoryear{{Rasmussen} \& {Ponman}}{{Rasmussen} \&
  {Ponman}}{2007}]{Rasmussen2007}
{Rasmussen} J.,  {Ponman} T.~J.,  2007, \mn@doi [\mnras]
  {10.1111/j.1365-2966.2007.12191.x}, \href
  {http://adsabs.harvard.edu/abs/2007MNRAS.380.1554R} {380, 1554}

\bibitem[\protect\citeauthoryear{{Ruiz} et~al.,}{{Ruiz}
  et~al.}{2015}]{Ruiz2015}
{Ruiz} A.~N.,  et~al., 2015, \mn@doi [ApJ] {10.1088/0004-637X/801/2/139}, \href
  {http://adsabs.harvard.edu/abs/2015ApJ...801..139R} {801, 139}

\bibitem[\protect\citeauthoryear{{Saintonge}, {Catinella}, {Tacconi}
  et~al.}{{Saintonge} et~al.}{2017}]{Saintonge2017}
{Saintonge} A.,  {Catinella} B.,  {Tacconi} L.~J.,   et~al., 2017, \mn@doi [ApJ
  Supp.] {10.3847/1538-4365/aa97e0}, \href
  {http://adsabs.harvard.edu/abs/2017ApJS..233...22S} {233, 22}

\bibitem[\protect\citeauthoryear{{Salim}, {Rich}, {Charlot}  et~al.}{{Salim}
  et~al.}{2007}]{Salim2007}
{Salim} S.,  {Rich} R.~M.,  {Charlot} S.,   et~al., 2007, \mn@doi [ApJ Supp.]
  {10.1086/519218}, \href {http://adsabs.harvard.edu/abs/2007ApJS..173..267S}
  {173, 267}

\bibitem[\protect\citeauthoryear{{S{\'a}nchez}, {Kennicutt}, {Gil de Paz}
  et~al.}{{S{\'a}nchez} et~al.}{2012}]{Sanchez2012}
{S{\'a}nchez} S.~F.,  {Kennicutt} R.~C.,  {Gil de Paz} A.,   et~al., 2012,
  \mn@doi [\aap] {10.1051/0004-6361/201117353}, \href
  {http://adsabs.harvard.edu/abs/2012A%26A...538A...8S} {538, A8}

\bibitem[\protect\citeauthoryear{{Schaye} et~al.}{{Schaye}
  et~al.}{2015}]{Schaye2015}
{Schaye} J.,  et~al., 2015, \mn@doi [MNRAS] {10.1093/mnras/stu2058}, \href
  {http://adsabs.harvard.edu/abs/2015MNRAS.446..521S} {446, 521}

\bibitem[\protect\citeauthoryear{{Sch{\"o}nrich} \& {Binney}}{{Sch{\"o}nrich}
  \& {Binney}}{2009}]{Schonrich2009}
{Sch{\"o}nrich} R.,  {Binney} J.,  2009, \mn@doi [MNRAS]
  {10.1111/j.1365-2966.2009.14750.x}, \href
  {https://ui.adsabs.harvard.edu/abs/2009MNRAS.396..203S} {396, 203}

\bibitem[\protect\citeauthoryear{{Schruba} et~al.,}{{Schruba}
  et~al.}{2011}]{Schruba2011}
{Schruba} A.,  et~al., 2011, \mn@doi [AJ] {10.1088/0004-6256/142/2/37}, \href
  {http://adsabs.harvard.edu/abs/2011AJ....142...37S} {142, 37}

\bibitem[\protect\citeauthoryear{{Scoville}, {Lee}, {Vanden Bout}
  et~al.}{{Scoville} et~al.}{2017}]{Scoville2017}
{Scoville} N.,  {Lee} N.,  {Vanden Bout} P.,   et~al., 2017, \mn@doi [ApJ]
  {10.3847/1538-4357/aa61a0}, \href
  {http://adsabs.harvard.edu/abs/2017ApJ...837..150S} {837, 150}

\bibitem[\protect\citeauthoryear{{Shamshiri}, {Thomas}, {Henriques}, {Tojeiro},
  {Lemson}, {Oliver}  \& {Wilkins}}{{Shamshiri} et~al.}{2015}]{Shamshiri2015}
{Shamshiri} S.,  {Thomas} P.~A.,  {Henriques} B.~M.,  {Tojeiro} R.,  {Lemson}
  G.,  {Oliver} S.~J.,   {Wilkins} S.,  2015, \mn@doi [MNRAS]
  {10.1093/mnras/stv883}, \href
  {http://adsabs.harvard.edu/abs/2015MNRAS.451.2681S} {451, 2681}

\bibitem[\protect\citeauthoryear{{Shivaei}, {Kriek}, {Reddy}  et~al.}{{Shivaei}
  et~al.}{2016}]{Shivaei2016}
{Shivaei} I.,  {Kriek} M.,  {Reddy} N.~A.,   et~al., 2016, \mn@doi [ApJ]
  {10.3847/2041-8205/820/2/L23}, \href
  {http://adsabs.harvard.edu/abs/2016ApJ...820L..23S} {820, L23}

\bibitem[\protect\citeauthoryear{{Somerville}, {Popping}  \&
  {Trager}}{{Somerville} et~al.}{2015}]{Somerville2015}
{Somerville} R.~S.,  {Popping} G.,   {Trager} S.~C.,  2015, \mn@doi [\mnras]
  {10.1093/mnras/stv1877}, \href
  {https://ui.adsabs.harvard.edu/\#abs/2015MNRAS.453.4337S} {453, 4337}

\bibitem[\protect\citeauthoryear{{Spitoni} \& {Matteucci}}{{Spitoni} \&
  {Matteucci}}{2011}]{Spitoni2011}
{Spitoni} E.,  {Matteucci} F.,  2011, \mn@doi [Astronomy and Astrophysics
  Supplement Series] {10.1051/0004-6361/201015749}, \href
  {http://adsabs.harvard.edu/abs/2011A%26A...531A..72S} {531, A72}

\bibitem[\protect\citeauthoryear{{Springel}, {White}, {Jenkins}
  et~al.}{{Springel} et~al.}{2005}]{Springel2005}
{Springel} V.,  {White} S.~D.~M.,  {Jenkins} A.,   et~al., 2005, \mn@doi [Nat.]
  {10.1038/nature03597}, \href
  {http://adsabs.harvard.edu/abs/2005Natur.435..629S} {435, 629}

\bibitem[\protect\citeauthoryear{{Stevens} \& {Brown}}{{Stevens} \&
  {Brown}}{2017}]{Stevens2017}
{Stevens} A. R.~H.,  {Brown} T.,  2017, \mn@doi [MNRAS]
  {10.1093/mnras/stx1596}, \href
  {https://ui.adsabs.harvard.edu/abs/2017MNRAS.471..447S} {471, 447}

\bibitem[\protect\citeauthoryear{{Stevens}, {Croton}  \& {Mutch}}{{Stevens}
  et~al.}{2016}]{Stevens2016}
{Stevens} A.~R.~H.,  {Croton} D.~J.,   {Mutch} S.~J.,  2016, \mn@doi [\mnras]
  {10.1093/mnras/stw1332}, \href
  {http://adsabs.harvard.edu/abs/2016MNRAS.461..859S} {461, 859}

\bibitem[\protect\citeauthoryear{{Stevens} et~al.,}{{Stevens}
  et~al.}{2019}]{Stevens2019}
{Stevens} A. R.~H.,  et~al., 2019, \mn@doi [MNRAS] {10.1093/mnras/sty3451},
  \href {https://ui.adsabs.harvard.edu/abs/2019MNRAS.483.5334S} {483, 5334}

\bibitem[\protect\citeauthoryear{{Stil} \& {Israel}}{{Stil} \&
  {Israel}}{2002}]{Stil2002}
{Stil} J.~M.,  {Israel} F.~P.,  2002, \mn@doi [\aap]
  {10.1051/0004-6361:20020352}, \href
  {http://adsabs.harvard.edu/abs/2002A%26A...389...29S} {389, 29}

\bibitem[\protect\citeauthoryear{{Thielemann} et~al.,}{{Thielemann}
  et~al.}{2003}]{Thielemann2003}
{Thielemann} F.-K.,  et~al., 2003, in {Hillebrandt} W.,  {Leibundgut} B.,  eds,
  From Twilight to Highlight: The Physics of Supernovae. p.~331,
  \mn@doi{10.1007/10828549_46}

\bibitem[\protect\citeauthoryear{{Tonini}, {Mutch}, {Croton}  \&
  {Wyithe}}{{Tonini} et~al.}{2016}]{Tonini2016}
{Tonini} C.,  {Mutch} S.~J.,  {Croton} D.~J.,   {Wyithe} J.~S.~B.,  2016,
  \mn@doi [MNRAS] {10.1093/mnras/stw956}, \href
  {https://ui.adsabs.harvard.edu/abs/2016MNRAS.459.4109T} {459, 4109}

\bibitem[\protect\citeauthoryear{{Tremonti} et~al.,}{{Tremonti}
  et~al.}{2004}]{Tremonti2004}
{Tremonti} C.~A.,  et~al., 2004, \mn@doi [\apj] {10.1086/423264}, \href
  {http://adsabs.harvard.edu/abs/2004ApJ...613..898T} {613, 898}

\bibitem[\protect\citeauthoryear{{Vikhlinin}, {Kravtsov}, {Forman}, {Jones},
  {Markevitch}, {Murray}  \& {Van Speybroeck}}{{Vikhlinin}
  et~al.}{2006}]{Vikhlinin2006}
{Vikhlinin} A.,  {Kravtsov} A.,  {Forman} W.,  {Jones} C.,  {Markevitch} M.,
  {Murray} S.~S.,   {Van Speybroeck} L.,  2006, \mn@doi [ApJ] {10.1086/500288},
  \href {https://ui.adsabs.harvard.edu/abs/2006ApJ...640..691V} {640, 691}

\bibitem[\protect\citeauthoryear{{Vogelsberger} et~al.}{{Vogelsberger}
  et~al.}{2014}]{Vogelsberger2014}
{Vogelsberger} M.,  et~al., 2014, \mn@doi [Nat.] {10.1038/nature13316}, \href
  {http://adsabs.harvard.edu/abs/2014Natur.509..177V} {509, 177}

\bibitem[\protect\citeauthoryear{{Wang}, {Lu}, {Luo}, {Yang}, {Hua}  \&
  {Chen}}{{Wang} et~al.}{2011}]{Wang2011}
{Wang} W.,  {Lu} J.,  {Luo} Z.,  {Yang} Z.,  {Hua} H.,   {Chen} Z.,  eds, 2011,
  {Galaxy Evolution: Infrared to Millimeter Wavelength Perspective}
  Astronomical Society of the Pacific Conference Series Vol. 446

\bibitem[\protect\citeauthoryear{{Wang}, {Fu}, {Aumer}  et~al.}{{Wang}
  et~al.}{2014}]{Wang2014b}
{Wang} J.,  {Fu} J.,  {Aumer} M.,   et~al., 2014, \mn@doi [\mnras]
  {10.1093/mnras/stu649}, \href
  {https://ui.adsabs.harvard.edu/abs/2014MNRAS.441.2159W} {441, 2159}

\bibitem[\protect\citeauthoryear{{Whitaker}, {van Dokkum}, {Brammer}  \&
  {Franx}}{{Whitaker} et~al.}{2012}]{Whitaker2012}
{Whitaker} K.~E.,  {van Dokkum} P.~G.,  {Brammer} G.,   {Franx} M.,  2012,
  \mn@doi [ApJ] {10.1088/2041-8205/754/2/L29}, \href
  {http://adsabs.harvard.edu/abs/2012ApJ...754L..29W} {754, L29}

\bibitem[\protect\citeauthoryear{{Xie}, {De Lucia}, {Hirschmann}, {Fontanot}
  \& {Zoldan}}{{Xie} et~al.}{2017}]{Xie2017}
{Xie} L.,  {De Lucia} G.,  {Hirschmann} M.,  {Fontanot} F.,   {Zoldan} A.,
  2017, \mn@doi [MNRAS] {10.1093/mnras/stx889}, \href
  {https://ui.adsabs.harvard.edu/abs/2017MNRAS.469..968X} {469, 968}

\bibitem[\protect\citeauthoryear{{Yates} \& {Kauffmann}}{{Yates} \&
  {Kauffmann}}{2014}]{Yates2014}
{Yates} R.~M.,  {Kauffmann} G.,  2014, \mn@doi [\mnras] {10.1093/mnras/stu233},
  \href {https://ui.adsabs.harvard.edu/abs/2014MNRAS.439.3817Y} {439, 3817}

\bibitem[\protect\citeauthoryear{{Yates}, {Henriques}, {Thomas}, {Kauffmann},
  {Johansson}  \& {White}}{{Yates} et~al.}{2013}]{Yates2013}
{Yates} R.~M.,  {Henriques} B.,  {Thomas} P.~A.,  {Kauffmann} G.,  {Johansson}
  J.,   {White} S.~D.~M.,  2013, \mn@doi [MNRAS] {10.1093/mnras/stt1542}, \href
  {http://adsabs.harvard.edu/abs/2013MNRAS.435.3500Y} {435, 3500}

\bibitem[\protect\citeauthoryear{{Yates}, {Thomas}  \& {Henriques}}{{Yates}
  et~al.}{2017}]{Yates2017}
{Yates} R.~M.,  {Thomas} P.~A.,   {Henriques} B.~M.~B.,  2017, \mn@doi [\mnras]
  {10.1093/mnras/stw2361}, \href
  {http://adsabs.harvard.edu/abs/2017MNRAS.464.3169Y} {464, 3169}

\bibitem[\protect\citeauthoryear{{Yates}, {Schady}, {Chen}, {Schweyer}  \&
  {Wiseman}}{{Yates} et~al.}{2019}]{Yates2019}
{Yates} R.~M.,  {Schady} P.,  {Chen} T.-W.,  {Schweyer} T.,   {Wiseman} P.,
  2019, arXiv e-prints, \href
  {http://adsabs.harvard.edu/abs/2019arXiv190102890Y} {}

\bibitem[\protect\citeauthoryear{{Zahid}, {Kudritzki}, {Conroy}, {Andrews}  \&
  {Ho}}{{Zahid} et~al.}{2017}]{Zahid2017}
{Zahid} H.~J.,  {Kudritzki} R.-P.,  {Conroy} C.,  {Andrews} B.,   {Ho} I.-T.,
  2017, \mn@doi [ApJ] {10.3847/1538-4357/aa88ae}, \href
  {http://adsabs.harvard.edu/abs/2017ApJ...847...18Z} {847, 18}

\bibitem[\protect\citeauthoryear{{Zwaan}, {Meyer}, {Staveley-Smith}  \&
  {Webster}}{{Zwaan} et~al.}{2005}]{Zwaan2005}
{Zwaan} M.~A.,  {Meyer} M.~J.,  {Staveley-Smith} L.,   {Webster} R.~L.,  2005,
  \mn@doi [\mnras] {10.1111/j.1745-3933.2005.00029.x}, \href
  {http://adsabs.harvard.edu/abs/2005MNRAS.359L..30Z} {359, L30}

\bibitem[\protect\citeauthoryear{{de Plaa}, {Kaastra}  \& {Werner}}{{de Plaa}
  et~al.}{2017}]{dePlaa2017}
{de Plaa} J.,  {Kaastra} J.~S.,   {Werner} N. e.~a.,  2017, \mn@doi [A\&A]
  {10.1051/0004-6361/201629926}, \href
  {http://adsabs.harvard.edu/abs/2017A.26A\&A...607A..98D} {607, A98}

\makeatother
\end{thebibliography}

\appendix

\section{Selection of main sequence and passive galaxies}
\label{app:cuts}

\begin{figure*}
\centering
\includegraphics[width=17cm]{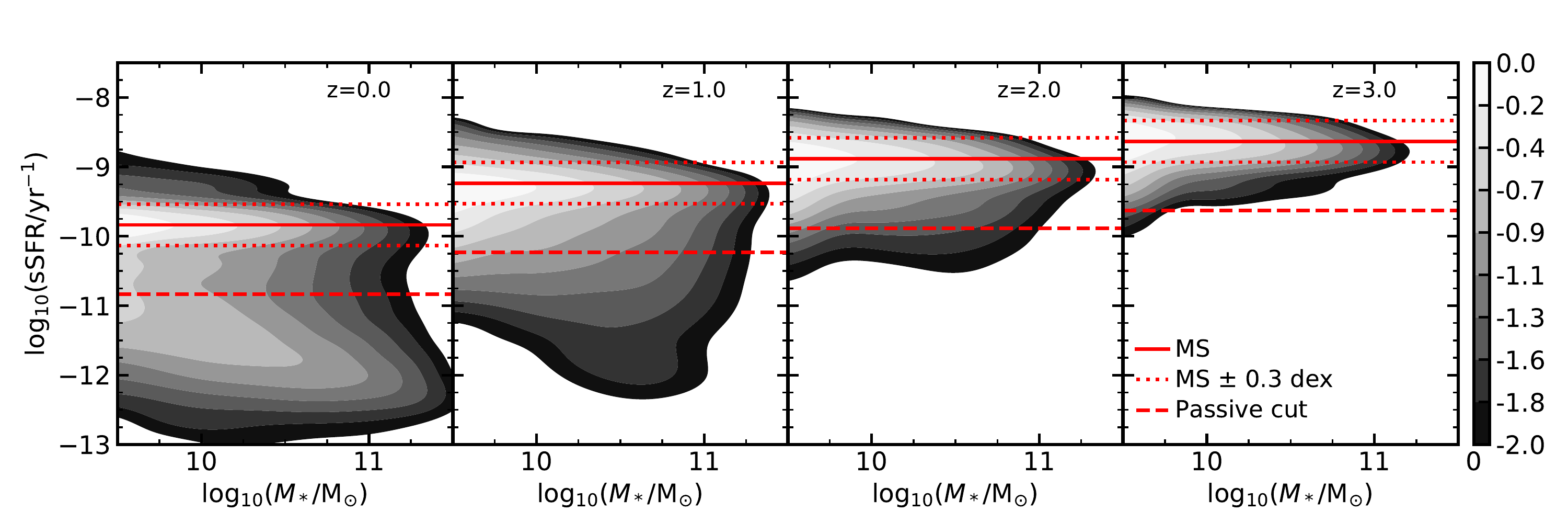}
\caption{The evolution of the sSFR vs stellar mass relation from $z=0$ to 3 in our new model. The solid red line roughly traces the evolution of the median of the main sequence of star formation and is given by $\rm{sSFR} = 2 \times (1+z)^2/t_{\rm{H}(z=0)}$. We use the dashed line,  1.0 dex below the main sequence, to distinguish between passive and star forming galaxies throughout the paper.}
\label{fig:ssfr_evo}
\end{figure*}

Throughout this paper we use a selection in sSFR versus stellar mass to distinguish between passive and star forming galaxies. As shown in Fig.~\ref{fig:ssfr_evo} the evolution of the main sequence of star formation in our model is well described by $\rm{sSFR} = 2 \times (1+z)^2/t_{\rm{H}(z=0)}$ (solid red line) out to $z=3$. Our threshold between passive and star forming galaxies is set by the dashed red line, 1 dex below the main sequence. In addition to these cuts, and in order to better match the observational selection, we use a cut in $u-r$ versus $r$ at $z=0$ in the left panel of Fig.~\ref{fig:redfraction_colorcut}, as described in Appendix A1 of \citet{Henriques2017}.


\label{lastpage}

\end{document}